%% file: NLS-Universalityrev.tex
\pdfoutput=1
\documentclass{article}
\usepackage{framed}
\usepackage{amssymb}
\usepackage{wrapfig}
\usepackage{amsbsy}
\usepackage{color}
\usepackage{graphicx}
\usepackage[colorlinks=true, pdfstartview=FitV, linkcolor=black, citecolor=blue, urlcolor=blue]{hyperref}

\definecolor{shadecolor}{rgb}{0.9, 0.9, 0.81}
\def\P{\mathbf P}
\def\pole{ v_p}
\def\KK {\mathbf K}
\def\GGamma {\mathbf \Gamma}

\def\yellow#1{\textcolor[rgb]{0.6, 0.5, 0 }{#1}}

\def\green#1{\textcolor[rgb]{0.2, 0.5,  0} {#1}}
\def\O{\Omega}
\def\S{\mathfrak S}

\def\Rscr{\mathcal R}
\def\Sscr{\mathcal S}
\def\Iscr{\mathcal I}
\def\Hscr{\mathcal H}
\def \Dscr{\mathcal D}
\def \Uscr{\mathcal U}
\def\sign{\mathrm sign}
\def \ve{\varepsilon}

\def\wt{\widetilde}
\def\wh{\widehat}
\def\ds{\displaystyle}

\renewcommand{\theequation}{\arabic{section}-\arabic{equation}}
\makeatletter
\@addtoreset{equation}{section}
\makeatother
\def\le{\left}
\def\ri{\right}
\def\QED{{\bf Q.E.D.}\par\vskip 5pt}
\def\bc{\begin{corollary}}
\def\ec{\end{corollary}}
\def\&{&{\hskip -20pt}}
\def\q{\mathbf q}
\def\ov{\overline}
\def\br{\begin{remark}\rm\small}
\def\1{{\bf 1}}
\def\er{\end{remark}}
\def\bt{\begin{theorem}}
\def\et{\end{theorem}}

\def\bx{\begin{example}}
\def\ex{\end{example}}
\def\bi{\begin{itemize}}
\def\ei{\end{itemize}}
\def\bd{\begin{definition}}
\def\ed{\end{definition}}
\def\bp{\begin{proposition}\rm}
\def\bl{\begin{lemma}\em}
\def\el{\end{lemma}}
\def\ep{\end{proposition}}
\def\bea{\begin{eqnarray}}
\def\eea{\end{eqnarray}}
\def \pa{\partial}
\def\R{{\mathbb R}}
\def\N{{\mathbb N}}
\def\Z{{\mathbb Z}}

\newenvironment{bmatrix}
{\le[\begin{array}{cc}} {\end{array}\ri]}
\renewenvironment{pmatrix}
{\le(\begin{array}{cc}} {\end{array}\ri)}
\newtheorem{problem}{Problem}[section]
\newtheorem{conjecture}{Conjecture}[section]
\newtheorem{theorem}{Theorem}[section]
\newtheorem{example}{Example}[section]
\newtheorem{coroll}{Corollary}[section]
\newtheorem{examps}{Examples}[section]

\newtheorem{lemma}{Lemma}[section]
\newtheorem{remark}{Remark}[section]
\newtheorem{remarks}[remark]{Remarks}
\newtheorem{proposition}{Proposition}[section] 
\newtheorem{definition}{Definition}[section]
\def\br{\begin{remark}}
\def\er{\end{remark}}
\def\bt{\begin{theorem}}
\def\et{\end{theorem}}
\def\bc{\begin{coroll}}
\def\ec{\end{coroll}}
\def\brs{\begin{remarks} \rm\
\begin{enumerate}}
\def\ers{\end{enumerate}\end{remarks}}
\def\bl{\begin{lemma}}
\def\el{\end{lemma}}
\def\bxs{\begin{examps}. \rm\begin{enumerate}}
\def\exs{\end{enumerate}\end{examps}}
\def\bd{\begin{definition}}
\def\ed{\end{definition}}
\def\bp{\begin{proposition}}
\def\ep{\end{proposition}}
\def\be{\begin{equation}}
\def\ee{\end{equation}}
\def\d{{\rm d}}
\def\bea{\begin{eqnarray}}
\def\eea{\end{eqnarray}}
\def\beas{\begin{eqnarray*}}
\def\eeas{\end{eqnarray*}}
\def\gt{\hat\gamma}
\def \hf{\frac{1}{2}}
\def \qt{\frac{1}{4}}
\def \pa{\partial}

\def \ra{\rightarrow}
\def\C{{\mathbb C}}
\def \D{\Delta}

\def \G{\Gamma}
\def \A{\mathbf A}
\def\a{\alpha}
\def\d{\delta}
\def\g{\gamma}
\def\k{\varkappa}
\def\l{\lambda}
\def\m{\mu}

\def\s{\sigma}
\def\sign{{\rm sign}}
\def\t{\tau}
\def\x{\xi}
\def\e{\varepsilon}
\def\z{\zeta}

\def\L{\Lambda}
\def\R{{\mathbb R}}
\def\N{{\mathbb N}}

\def\Z{{\mathbb Z}}

\def\star{*}
\date{}
\begin{document}
%                                                                  
%\fontfamily{cmss}
%\fontsize{11pt}{15pt}
%\selectfont

\baselineskip 16pt plus 1pt minus 1pt
\begin{titlepage}
\begin{flushright}
%CRM-???? (2009)\\
%nlin.SI/05xxxxxx
\end{flushright}
\vspace{0.2cm}
\begin{center}
\begin{Large}
\textbf{Universality for the focusing nonlinear Schr\"odinger equation at the gradient catastrophe point:}
\end{Large}\\
\textbf{\large Rational breathers and poles of the {\em tritronqu\'ee} solution to Painlev\'e\  I}\\
\bigskip
M. Bertola$^{\dagger\ddagger}$\footnote{Work supported in part by the Natural
  Sciences and Engineering Research Council of Canada (NSERC)}\footnote{bertola@crm.umontreal.ca},  
A. Tovbis$^{\sharp}$ 
\\
\bigskip
\begin{small}
$^{\dagger}$ {\em Centre de recherches math\'ematiques,
Universit\'e de Montr\'eal\\ C.~P.~6128, succ. centre ville, Montr\'eal,
Qu\'ebec, Canada H3C 3J7} \\
\smallskip
$^{\ddagger}$ {\em  Department of Mathematics and
Statistics, Concordia University\\ 1455 de Maisonneuve W., Montr\'eal, Qu\'ebec,
Canada H3G 1M8} \\
\smallskip
$^{\sharp}$ {\em  University of Central Florida
	Department of Mathematics\\
	4000 Central Florida Blvd.
	P.O. Box 161364
	Orlando, FL 32816-1364
} \\
\end{small}
\end{center}
\bigskip 
%%%%%%%%%%%%%%%%  Abstract  %%%%%%%%%%%%%%%%
\begin{center}{\bf Abstract}\\
\end{center} 
The semiclassical (zero-dispersion) limit of solutions $q=q(x,t,\e)$ to the  one-dimensional focusing Nonlinear  
Schr\"odinger equation (NLS) 
%\remove{\insrt{with WKB-like initial data}} \remove{with decaying potentials}
 is studied in a
 % \remove{full} 
scaling neighborhood $D$ of 
%\remove{the}
a point of gradient catastrophe ($x_0,t_0$).
We consider a certain class of solutions that decay as $|x|\ra \infty$ specified in the text.
The neighborhood $D$ contains the region of modulated  plane wave (with rapid phase oscillations), as well as the region
of fast  amplitude oscillations (spikes). 
In this paper we
establish the following {\em universal behaviors}  of the NLS solutions {$q$} near the point of gradient 
catastrophe:
i) each spike has height $3|q{_0}(x_0,t_0)|$ and  uniform shape of the rational 
breather solution to the NLS, scaled to the size $O(\ve)$; ii) the location of the spikes is determined by
the poles of the  {\em tritronqu\'ee} solution of the Painlev\'e\  I (P1) equation through an explicit map 
between $D$ and a region
of the Painlev\'e\  {independent variable}; iii) if $(x,t)\in D$ but lies away from the spikes, the asymptotics of 
the NLS solution 
$q(x,t,\e)$ is given by the plane wave approximation $q_0(x,t,\e)$, with the correction term being expressed in 
terms of the {\em tritronqu\'ee} solution of P1. 
The relation  with the conjecture of Dubrovin, Grava and Klein \cite{DubrovinGravaKlein} 
about the behavior of solutions to the focusing NLS near a point of  gradient catastrophe is discussed.
 We conjecture  that the P1 hierarchy occurs at higher 
degenerate catastrophe points and that the amplitudes of the  spikes are  odd multiples of the amplitude at the corresponding catastrophe point.
Our technique is based on the nonlinear steepest descent method for matrix Riemann-Hilbert problems and
discrete Schlesinger isomonodromic transformations.

\medskip
\bigskip
\bigskip
\bigskip
\bigskip

\end{titlepage}
\tableofcontents
\section{Introduction and main results}
\label{sectintro}
In this paper we consider the focusing Nonlinear Schr\"odinger (NLS) equation 
\be  \label{FNLS}
i\varepsilon\partial_t q + \varepsilon^2\partial_x^2 q +2 |q|^2q=0, 
\footnote {Note that this equation differs by the coefficient $2$ from the NLS equation considered
in previous papers \cite{TVZ1}-\cite{TVZ3}, whose results we use here; this is the correct equation for the particular form of 
the time evolution of the scattering data (see (\ref{rhpgam})), adopted  in \cite{TVZ1}-\cite{TVZ3}
and in the present paper. Alternatively, one can keep the ``old'' form of the NLS, but 
scale by $2$ the time variable while using the results of \cite{TVZ1}-\cite{TVZ3}.} 
\ee
where $x\in\R$ and $t\ge 0$ are space-time variable and $\e>0$.
It is  a basic model  for self-focusing and
self-modulation, for example, it governs nonlinear transmission in optical fibers; 
it can also be derived as a  modulation equation for  general nonlinear systems. 
It was first integrated (with $\e=1$) by Zakharov and Shabat \cite{ZS}
who produced a Lax pair for it and used the inverse scattering
procedure to describe general decaying solutions ($\lim_{|x|\to \infty}q(x,0)=0$) in terms of radiation  and solitons. Throughout this work, we will
use the abbreviation NLS to mean ``focusing Nonlinear Schr\"odinger equation".

Our  interest in the  semiclassical (zero-dispersion)
limit ($\varepsilon \to 0$) of NLS  stems largely from  its
{\em modulationally unstable} behavior. As shown  by  Forest 
and Lee \cite{FL}, the modulation system  for NLS can be expressed as 
a set of nonlinear PDE with  complex characteristics; 
thus, the system  is ill posed as an initial value 
problem with the initial  data (potential) in the form of a modulated plane wave. 
As a result,  this plane wave is
%\be \label{IDe}q(x,0,\varepsilon)=A(x)e^{i\Phi(x)/\varepsilon},~~~~~~~\Phi(0)=0,~~~~~~~~~~~~~\lim_{|x|\ra\infty}A(x)=0,
%\label{q0pure}\ee
%i.e., a plane wave with amplitude modulated by $A(x)$ and phase modulated by $\Phi(x)$,  are 
expected to  break immediately into some other, 
presumably disordered,  wave form when the amplitude and the phase of the potential 
possess no special properties.

\begin{figure}[t]
\begin{center}
\includegraphics[width=0.8\textwidth]{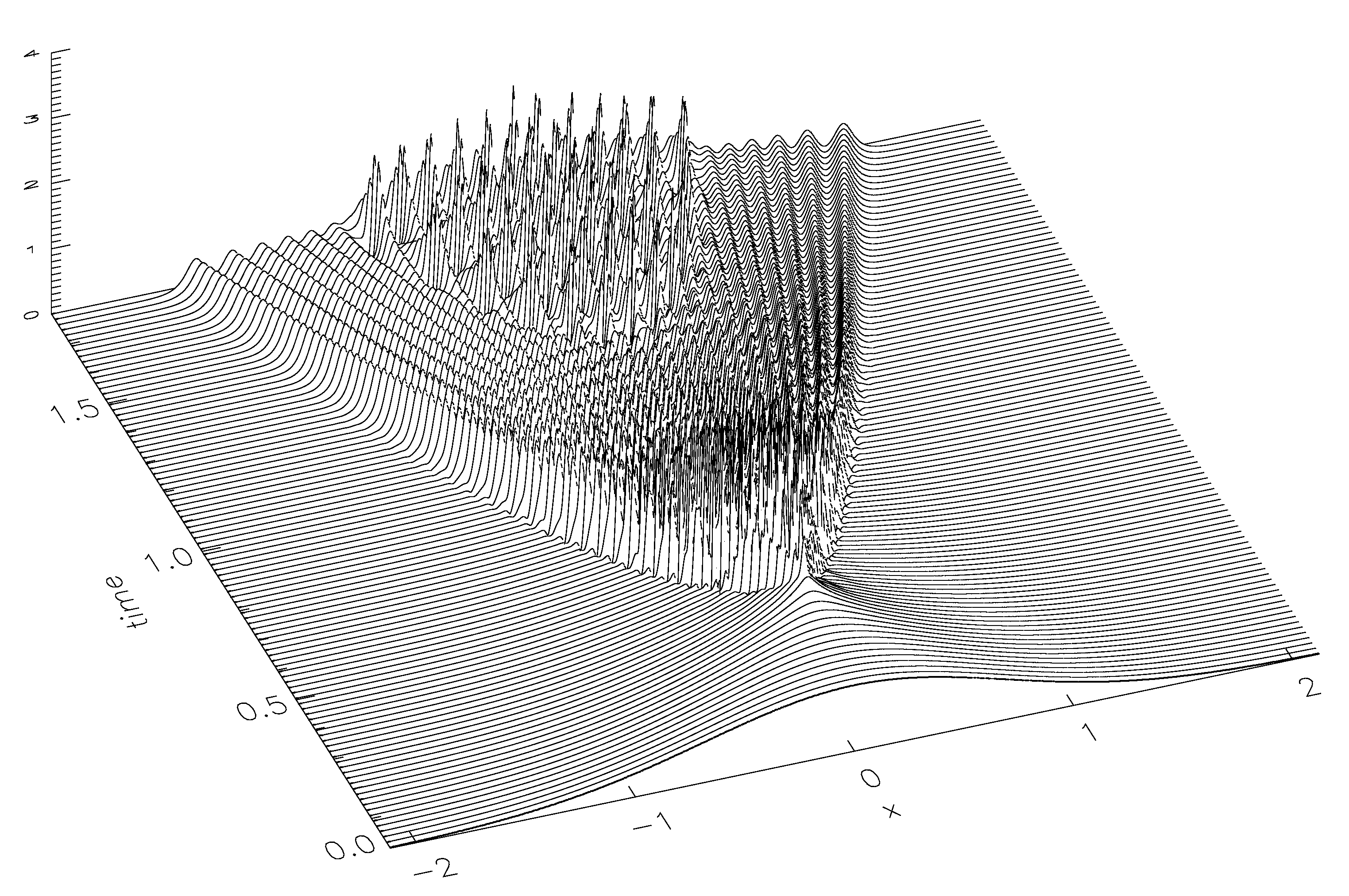}
\caption{Absolute value $|q(x,t,\varepsilon)|$ of a solution
    $q(x,t,\varepsilon)$ to the focusing NLS (\ref{FNLS}) versus the space
    $x$ and the time $t$ coordinates from \cite{CMM}. Here the potential $q(x,0,\e)=A(x)e^{\frac i\e \Phi(x)}$ with
    $A(x)=e^{-{x^2}}$, $\Phi'(x)=-\tanh x$, and
    $\e=0.02$.}
\label{Caiorig}
\end{center}
\end{figure}

In the case of an {\em analytic} initial data, the NLS evolution  
displays
some  ordered structure 
instead of the disorder suggested by the modulational
instability (see \cite{MillerKamvissis},
\cite{CT} and \cite{CMM}), as can be seen on the well-known Figure \ref{Caiorig} (from  \cite{CMM}).
This figure depicts
 numerical simulations (obtained by D. Cai) for the absolute value of the 
solution $q(x,t,\e)$ of the focusing NLS (\ref{FNLS}) with the initial data of a modulated plane wave
$q(x,0,\e)=A(x)e^{\frac i\e \Phi(x)}$, where $A(x)=e^{-{x^2}}$ and $\Phi'(x)=-\tanh x$,~~ $\Phi(0)=0$.

\begin{figure}[t]
\begin{center}
\resizebox{0.7\textwidth}{!}{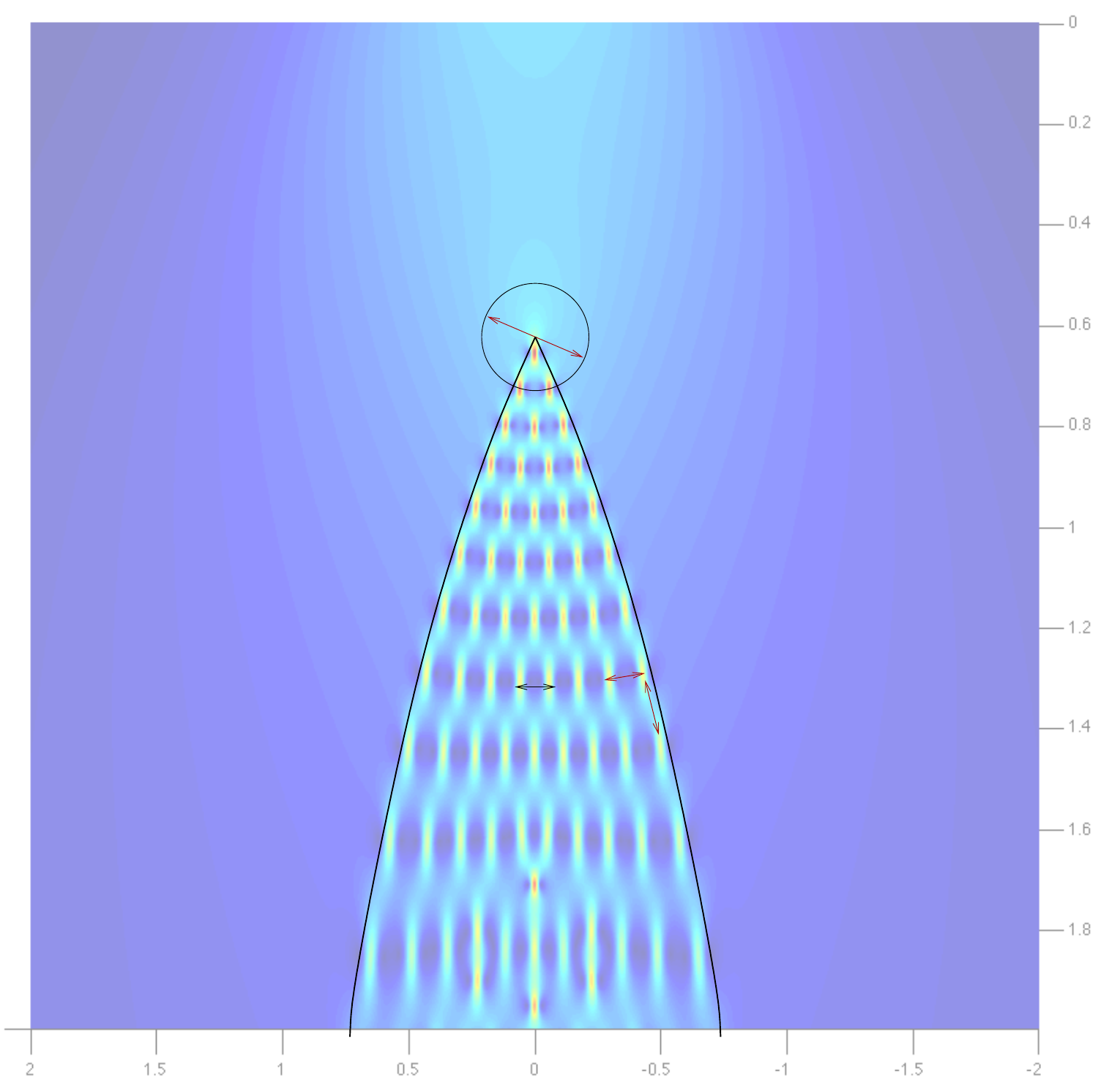}
\caption{Absolute value $|q(x,t,\varepsilon)|$ of a solution  $q(x,t,\varepsilon)$ to the focusing NLS (\ref{FNLS})
versus $x,t$ coordinates. Here $q(x,0,\e)=A(x)e^{\frac i\e \Phi(x)}$ with $A(x)=e^{-{x^2}}$, $\Phi'(x)=\tanh x$ and $\ve=0.03$.}
\label{Cai}
\end{center}
\end{figure}

\begin{figure}[t]
\begin{center}
\includegraphics[width=0.8\textwidth]{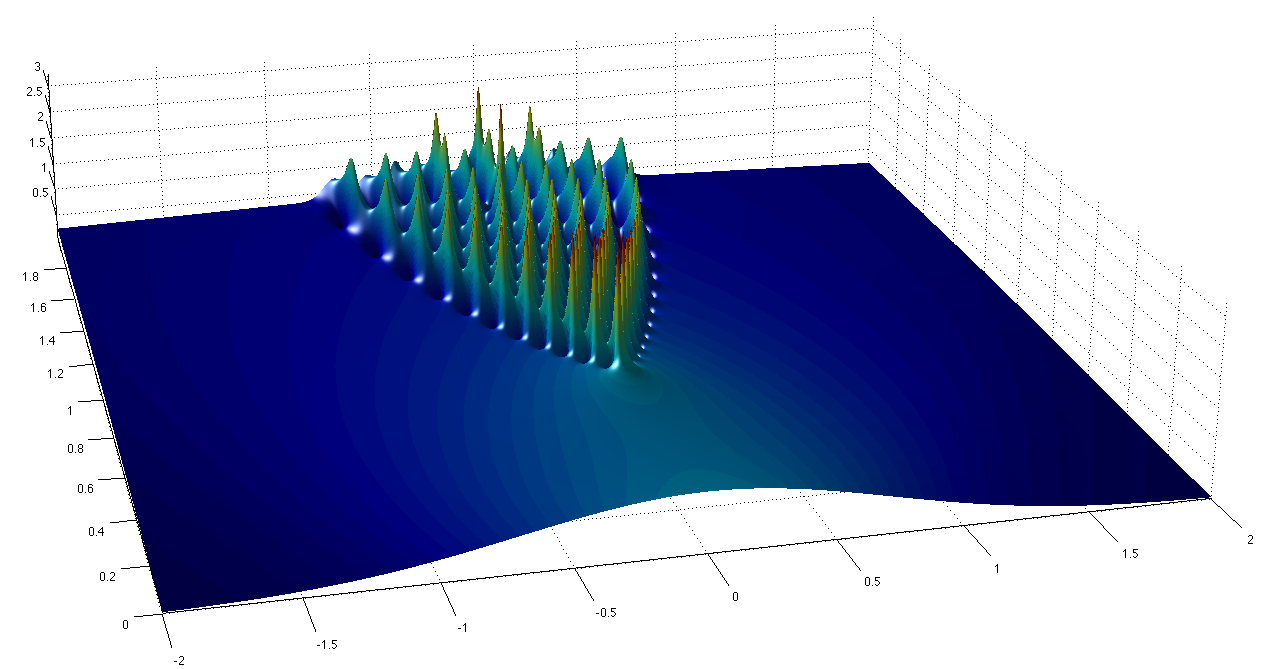}
\caption{Absolute value $|q(x,t,\varepsilon)|$ of a solution  $q(x,t,\varepsilon)$ to the focusing NLS (\ref{FNLS})
versus $x,t$ coordinates. Here $q(x,0,\e)=A(x)e^{\frac i\e \Phi(x)}$ with  $A(x)=e^{-{x^2}}$, $\Phi'(x)=\tanh x$ and $\varepsilon=0.03$.}
\label{isometry}
\end{center}
\end{figure}

Figure \ref{Caiorig}, as well as our numerical simulations shown on Figures \ref{Cai}, \ref{isometry},
clearly identify several spatio-temporal regions of distinct asymptotic regimes of {the}  $q(x,t,\varepsilon)$
in the semiclassical limit $\e\ra 0$. These regions (called {\it asymptotic regions}) are separated by some  curves in the $x,t$ plane 
that are asymptotically  independent of $\varepsilon$. They are called {\it breaking curves} or 
{\it nonlinear caustics}. 
In the  very first asymptotic region (containing the axis $t=0$) the solution $q(x,t,\e)$ can be approximated
by a slowly modulated plane wave $q_0(x,t,\e)=A(x,t)e^{\frac i\e \Phi(x,t)}$. 
Note that this approximation  fails  near (the first) breaking curve. 
A more complicated Ansatz that can be expressed in terms of Riemann
Theta-functions is required to approximate modulated nonlinear $2n$-phase waves in the asymptotic regions
beyond the first breaking curve, where $n$ can be $1,2,3,\cdots$.

A significant progress in the semiclassical 
asymptotics of the NLS (\ref{FNLS}) was achieved in  \cite{KMM} (pure soliton case) and \cite{TVZ1}
(pure radiation and radiation with solitons), where the order $O(\e)$ approximation $q_0(x,t,\e)$ 
 of $q(x,t,\e)$ was obtained
in the first two asymptotic regions. The $O(\e)$ error estimate is valid uniformly on compact subsets within 
the corresponding region. In both papers, the inverse scattering problem for the NLS (\ref{FNLS})
was cast as a two by two matrix  Riemann-Hilbert Problem (RHP), whose semiclassical asymptotics was obtained
through the combination of the nonlinear steepest descent method of P. Deift and X. Zhou (\cite{DeiftZhou}) and 
the $g$-function mechanism (\cite{DKMVZ}). The approximation $q_0$, obtained in   \cite{KMM}, \cite{TVZ1},
can be described in terms of some hyperelliptic Riemann surface $\Rscr=\Rscr(x,t)$, which depends on $x,t$ but does not depend on $\e$.
The Schwarz symmetry of the focusing NLS implies that the branchpoints and the branchcuts of $\Rscr$ are Schwarz symmetrical.
In this context, regions of different asymptotic behavior of 
$q(x,t,\varepsilon )$ corresponds to the different genera of  $\Rscr(x,t)$, and the approximation $q_0(x,t,\e)$
 is expressed in terms of the Riemann Theta-functions of $\Rscr$.
%enter in the asymptotic description. Note that $\Rscr(x,t)$ does not depend on $\e$. 
In the very first (genus zero) region 
(that contains the line $t=0$), 
the approximation $q_0(x,t,\ve) $
of $q(x,t,\varepsilon)$ is expressed through the branch-point $\a(x,t)$ of $\Rscr(x,t)$ %(the branch-points are Schwarz-symmetrical) 
as (see \cite{TVZ1})
\be\label{lotg0a}
q_0(x,t,\ve)=A(x,t)e^{\frac i\e \Phi(x,t)}=\Im \a(x,t)e^{- \frac{i}{\ve} \int^{(x,t)}_{(0,0)}\{2\Re\a(\x,t)d\x+[4(\Re\a(x,\t))^2-2 (\Im\a(x,\t))^2]d\t\}}+O(\e), 
\footnote{This formula was proven in \cite{TVZ1}, but, as stated in Theorem 1.1 there (with only $d\x$ part of the differential form), is correct only for $t=0$.} 
\ee
%where  the error estimate is uniform on compact subsets of the genus zero region (away from the breaking curve). 
We will often refer
to this $q_0$ as a modulated plane wave (or as genus zero)  approximation of the solution $q$. 
The next asymptotic region (behind the first breaking curve), studied in 
 \cite{KMM}  and \cite{TVZ1}, corresponds to  $\Rscr(x,t)$ of genus two; the corresponding approximation $q_0$ in this region
has the form of a modulated nonlinear $2$-phase wave.

The approximation formulae in the higher genera regions (genus $4,6,$ etc.) are, in a certain sense, similar to
%does not differ significantly from 
that in the genus
two region ({the} existence of such regions, though, remains  a challenging question, see \cite{LyngMiller}
for recent progress in this direction). However,  approximation near  breaking curves, to our best knowledge, remained to be   studied.
The first breaking curve consists generically of two smooth branches that form a wedge (tip) when joining  together, see Figures \ref{Cai}, \ref{isometry}. 
In the recent paper \cite{BertolaTovbis1}, we constructed the approximation
near the smooth parts (branches) of the first breaking curve: it consists of the modulated plane wave approximation $q_0$ from the genus zero region
plus correction terms. The shape of the corrections is depicted on Figure \ref{3dpic}. They form ranges of peaks and depressions aligned along
the breaking curves, see Subsection \ref{limalbreakc}. However, the origin of these ranges of  peaks and depressions (spikes and anti-spikes)
%\removelong{, which bear some resemblance with the formation of the rogue (freak), }
that is,  description of the approximation at the tip of the breaking curve, 
remains somewhat of a mystery. {\it The {\bf main goal} of the present paper is to describe the mechanism of formation of the spikes, to derive the formula for
approximation  around the tip of the breaking curve and to prove the corresponding error estimates.}  

The summary of {\bf our findings} can be stated as: i) the approximation near the tip consists of the {\it modulated plane wave approximation 
$q_0$ from the genus zero region
with $O(\e^\frac 25)$ error term} that is expressed in terms of the {\em tritronqu\'ee} solution to the first Painlev\'e\  equation (P1);
ii) evaluated near a pole of the {\em tritronqu\'ee} solution, the {\it error term becomes commensurable with the leading order term}  and contribute 
an order $O(1)$ correction to $q_0$; iii) these {\it corrections have the universal shape of a rational (Peregrine) breather}, see Figure \ref{NLSpeak} 
and 
%\replace{they are localized in the triangular-shaped lattice (see Figures   \ref{Cai}, \ref{isometry}), defined by the poles}
{their location is determined by the location of poles} of the 
{\em tritronqu\'ee} solution, see Figure \ref{Painplane}.

These facts emphasize  the universal role of the {\em tritronqu\'ee} solution to P1 in the modeling of the transition from a steepening modulated plane
wave (gradient catastrophe) to a nonlinear 2-phase wave behavior near the tip of the breaking curve (this transition resembles  formation of
rogue waves). In fact, the special role of P1 solutions in transitional regimes near critical points was first observed in the context of the 
orthogonal polynomials with varying exponential weights, see \cite{FIK} and \cite{ItsKapaevFokasBook} with some prior physical literature references there, 
see also \cite{ArnoDu}, where  the nonlinear steepest descent method was used for error estimates. In the context of the focusing
NLS, the form of the error term  containing the  {\em tritronqu\'ee} solution to P1, as well as the localization of all the poles of the  {\em tritronqu\'ee} solution
in a certain sector of the complex plane,  were conjectured in \cite{DubrovinGravaKlein}. In all these results and conjectures, the error terms
expressed through solutions of P1 were considered {\it only away from the poles of} these solutions. The main contribution of this paper is that we: 
{\it analyze the error terms {\bf both near the poles and away from poles}; linked the {\bf spikes of an NLS solution with the poles} of the 
{\em tritronqu\'ee} solution to P1, and;  calculated the universal shape of the spikes.} The exact localization of the spikes, linked with the 
 localization of the finite poles of the 
{\em tritronqu\'ee}, is  yet to be established. There are some recent analytical (\cite{JoshiKitaev01}) and numerical (\cite{Novokshenov}) results on 
this issue indicating the triangular shape of the lattice of the poles.

What is the class of solutions to the NLS (\ref{FNLS}) for which our results are valid?
 As it was mentioned above, our results are based on Deift-Zhou nonlinear steepest descent analysis (\cite{DeiftZhou}) of the semiclassical
($\e\ra 0$) limit of a matrix RHP that represents 
the inverse scattering problem for the  NLS (\ref{FNLS}). Therefore, in a broad sense, the  results of this paper should be applicable to any 
``generic'' solution $q(x,t,\e)$ for the  NLS (\ref{FNLS}) that: a) undergoes a transition from a modulated plane wave to a nonlinear 
2-phase wave behavior at a gradient catastrophe
point $(x_0,t_0)$, and; b) the nonlinear steepest descent method is applicable to the RHP, representing the 
inverse scattering problem for $q(x,t,\e)$, in a vicinity of  the point $(x_0,t_0)$ in the $x,t$ plane. This description, however, cannot be considered 
satisfactory because of its vagueness. Although, in principle, it is possible to clarify the technical issues involved in the above description,
%(see Remarks .... below), 
the authors have not found a sufficiently brief and rigorous way of doing so.  Therefore, we decided to formulate
our results for a much narrower class of solutions $\Uscr$ to the  NLS (\ref{FNLS}), defined in Section \ref{sectreview}, with some follow up
comments regarding the general situation (see Remark \ref{rem2}). It was shown in \cite{TVZ3} that  each $q\in\Uscr$ 
possesses the following properties: 
there exists a point of gradient catastrophe $(x_0,t_0)$ as required by the condition a) above, and;
 the nonlinear steepest descent method is applicable not only in a vicinity
of  $(x_0,t_0)$ but also for all $(x,t)$ with $x\in\R$ and $t\in[0,t_0]$.

A solution $q=q(x,t,\e)$ of an integrable equations, such as the NLS (\ref{FNLS}),
can be defined by its initial (Cauchy) data as well as by its scattering data. 
Since  the input data into the inverse scattering problem (including its RHP formulation) contains the scattering data,
{\it it is much more convenient for us to define solutions $q(x,t,\e)$ through  their scattering data}. 
So, to simplify our analysis, the class of solutions $\Uscr$ that we consider consists of
 solitonless solutions $q(x,t,\e)$, whose initial datum are defined through their reflection coefficients $r_0=r_0(z,\e)$.
The choice of $r_0$ is such that 
for every 
fixed $\e>0$, $r_0$
is continuous on $z\in\R$ and has an  exponential decay as $z\ra\pm\infty$ (see details in Section \ref{sectreview}).
Then the corresponding initial data $q=q(x,0,\e)$ belongs to the weighted $ L^2(\R)$ with the weight $1+x^2$, 
%\red{[Check that!]} 
so the (direct)
scattering transform $\Sscr_\e$ between $q$ and $r_0$ is well defined, see \cite{Zh1}, \cite{Zh2}.

Getting into a little bit more details, we can say that, roughly speaking, we consider initial datum $q$ of the form 
\be\label{qthroughr}
q(x,0,\e)=\Sscr_\e^{-1}r_0(z,\e)~~~~~~{\rm with}~~~ r_0(z,\e)=e^{\frac{2i}\e f_0(z)},
\ee
%where $\Scrp_\e^{-1}$ denotes the inverse scattering transform corresponding to (\ref{FNLS}) and  
where functions $f_0$, defined by (\ref{f'}), (\ref{jump}), are called admissible functions. (See Definition \ref{adr}
for the corresponding admissible reflection coefficients  $r_0$.) Notice that,
according to (\ref{f'}), (\ref{jump}),  we consider a special but nevertheless wide class of admissible reflection coefficients. 
Given an admissible reflection coefficient $r_0=r_0(z,\e)$, what do we know about the semiclassical limit of the corresponding initial data $q$?
%In other words, if $q=\Sscr^{-1}_\e r_0$, what can be stated about $\lim_{\e\ra 0} q(x,0,\e)$?   
According to 
%\cite{TVZ1}, 
\cite{TVZ3}, for any admissible $r_0$ there exist a pair of smooth functions $A(x),\Phi(x)$ with
$\lim_{|x|\ra\infty}A(x)=0$ and $\Phi'(x)$ having finite limits at $\pm\infty$, such that
\be\label{compappr} 
q(x,0,\e)=A(x)e^{\frac i \e \Phi(x)} +O(\e)~~~~~{\rm as}~~\e\ra 0
\ee
uniformly on compact subsets of $\R$. It would be certainly nice to remove the ``compact subsets'' condition
from (\ref{compappr}), but this is the subject of a different project. 
It could be added here that in numerical/experimental applications (like \cite{TH}), the data is always truncated to a finite interval 
and hence the control of the approximation over compact sets in these cases is sufficient.
%}\red{[I am not sure about that phrase.]

\begin{example}\label{exourfamily}
To illustrate the above discussion, consider the solitonless initial data
%For example for the potential 
\be\label{ourfamily}
\hat q(x,0;\ve):=
{\frac 1 {\cosh (x)}}e^{-\frac{\mu i}{\ve}
{\ln\cosh x}+i\pi},
\ee 
$\mu\geq 2$, whose reflection coefficient $\hat r_0(z,\e)$ is explicitly known (\cite{TVZ1}). For example,  
\begin{equation}
\label{eq:5in}
\hat r_{0}(z)=  -i\e 2^{-\frac {2i}\e}\frac{\G(\hf+\frac i\e(z-1)) \G^2(\hf-\frac{iz}{\e})}{ \G^2(-\frac i\e)\G(\hf+\frac i\e(z-1))},~~~\Im z \geq 0,
\end{equation}
when $\mu=2$.
%where $w_0=-\frac i\e$ and $w=-(z+1)\frac i\e+\hf$. 
Notice that $\hat r_{0}(z)$ is {\em not an admissible
reflection coefficient} since, for example, $\hat f_0(z,\e)=-\frac{i\e}2\ln \hat r_0(z,\e)$ has logarithmic singularities 
at the poles $z_n=1+(n+\hf)ni$, $n\in\N$, of $\hat r_{0}(z)$.
On the other hand, retaining the first two terms in the small $\e$ expansion of $\hat f_0(z,\e)$ 
(calculated by the Stirling formula), 
we obtain
\be\label{f0m=2,w}
f_0(z):=\lim_{\e\ra 0}\hat f_0(z,\e)+\frac{\pi\e}2=(1-z)\left[i\frac \pi 2 +\ln(1-z)\right]
+z\ln z+\ln 2+\frac{\pi\e}2,
\ee
where $\Im z\geq 0$ and $z\neq 1$. With a proper choice of logarithmic branches  (\cite{TVZ1}),
one can  check that $w(z)=\Im f_0(z)\sign(1-z)= \frac \pi 2 |1-z|$, $z\in\R$,
is an admissible function (Definition \ref{adw}).
Then a corresponding reflection coefficient $r_0$, defined on $z\in[-1,1]$ as $r_0(z,\e)=e^{\frac{2i}\e f_0(z)}$ 
(see Definition \ref{adr} for full details), is an admissible reflection coefficient. It was proven in \cite{TVZ1}
that $q(x,0,\e)=\Sscr^{-1}_\e r_0(z,\e)$ is an order $O(\e)$ approximation of $\hat q(x,0,\e)$ as $\e\ra 0$ on compact subsets of $\R$.
\end{example}

Generalizing on the above example, one can consider the class of solutions  $q\in\Uscr$ as obtained by replacing  actual reflection coefficients 
$\hat r_0=\Sscr_\e \hat q$ of
solitonless initial datum of the form $\hat q(x,0,\e)=A(x)e^{\frac i \e \Phi(x)}$ by their small $\e$ admissible approximations  $r_0=r_0(z,\e)$.
Although there is {no} proof that the solution $q(x,t,\e)$  defined by the initial data $q(x,0,\e)= \Sscr^{-1}_\e r_0(z,\e)$ will stay close
to  $\hat q(x,t,\e)$ on some time interval $t\in[0,t_*]$, $t_*>0$,  the idea of replacing an actual scattering data with its convenient 
small $\e$ approximation was widely used in semiclassical asymptotics
 of integrable systems starting with the pioneering papers \cite{LaxL} 
for the Korteweg - de Vries equation and through all  analytical
studies for the NLS, sine-Gordon, modified NLS that the authors aware of (\cite{KMM},\cite{TVZ1},\cite{BM}, \cite{DMM}). The only notable exception is
\cite{JM}, where a very simple form of the initial data allows for direct estimates.


\subsection{Semiclassical limit along the breaking curve}\label{limalbreakc}

\begin{figure}[t]
\begin{center}
\includegraphics[width=0.5\textwidth]{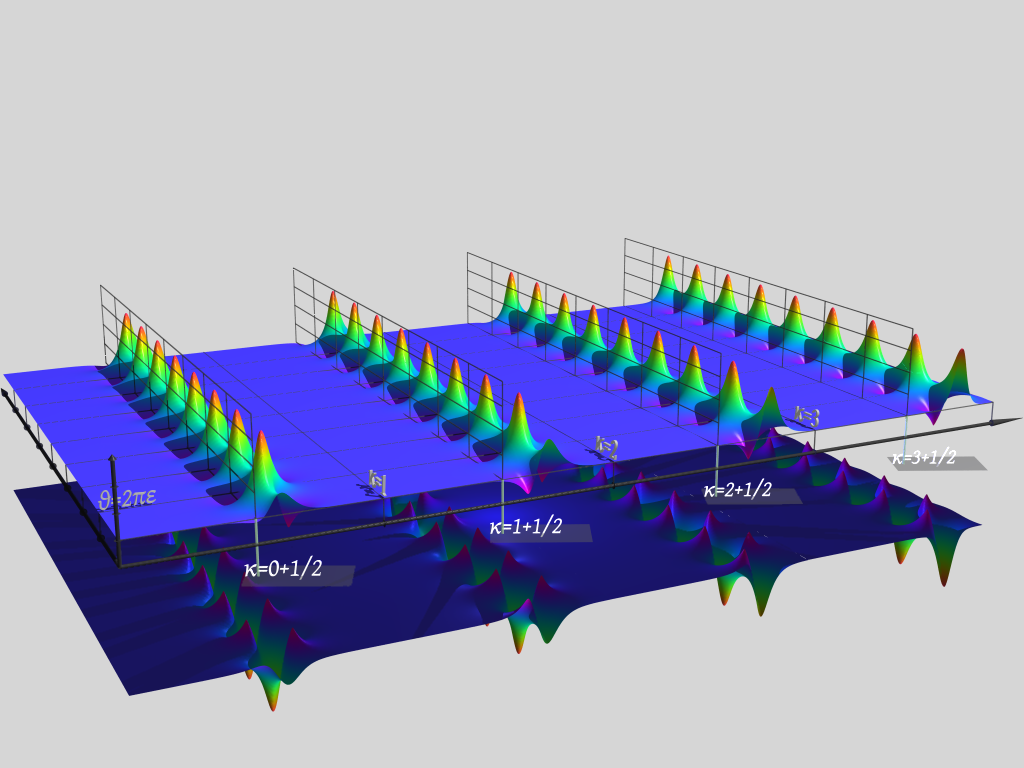}\includegraphics[width=0.5\textwidth]{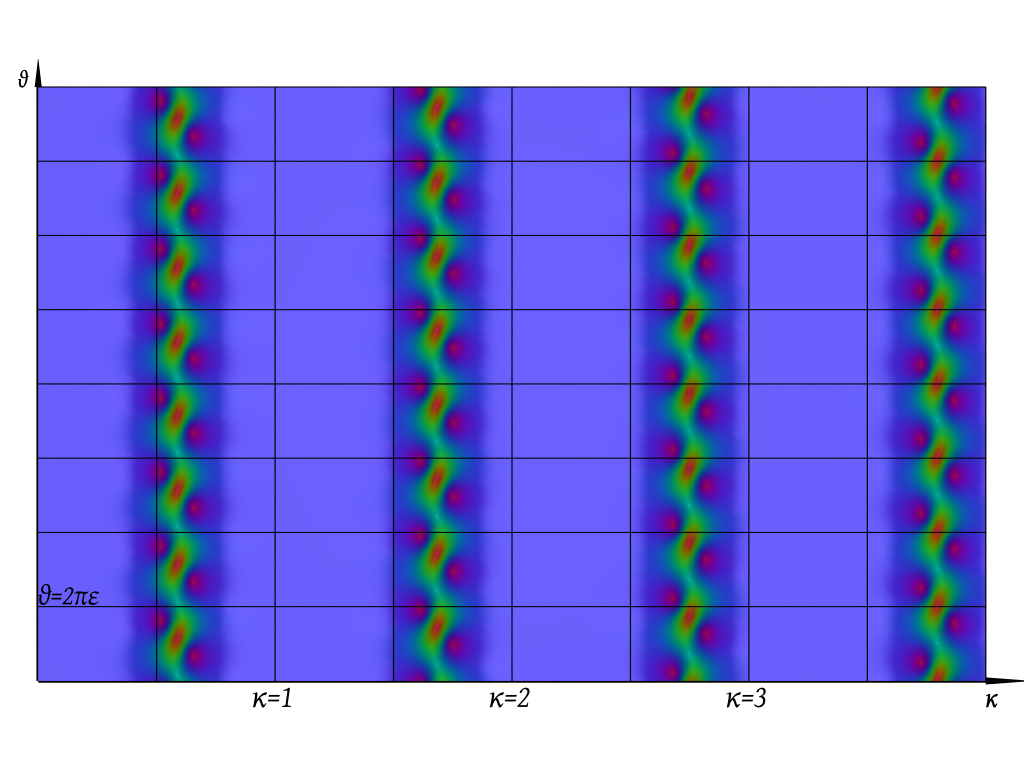}
\end{center}
\caption{The graphs depict a typical ratio $ \frac{\tilde q(x,t,\e)}{q_0(x,t,\e)}$ in  a vicinity of a regular point on the breaking curve, where 
%point $x = 0.1\ ,\ \ t = 0.1570482549$ with $\varepsilon = 10^{-10}$.
$q_0$ is the genus zero approximate solution and
$\tilde q$  is the leading order approximation near the breaking curve, constructed in \cite{BertolaTovbis1}. The graph is shown
in the variables $\theta,\varkappa$ and the scale is not uniform: $\varkappa$ measures distances in the scale $\varepsilon |\ln \varepsilon|$ whereas 
$\theta$ in the scale $\varepsilon$. 
The distance between consecutive ``ranges", should be much longer than the separation of the neighboring  peaks within the same range. 
The typical {\em size} of the hills is $\varepsilon$ in all directions while the {\em separation} in the longitudinal 
direction is $\varepsilon|\ln \varepsilon|$. This picture is consistent with a full--blown genus-two regime, where the solution is 
quasi-periodic at the scale $\varepsilon$: as we progress into the genus-2 region the separation reduces to the natural scale $\varepsilon$. 
If we were to plot this in the $(x,t)$ plane the only difference would be  a linear change of coordinates and the ``ranges'' would 
be parallel to the breaking curve.}
\label{3dpic}
\end{figure}

A detailed study of asymptotic behavior of $q(x,t,\e)$ along the first breaking curve  (a neighborhood of the tip of this curve was excepted)
together with error estimates were conducted in our previous work \cite{BertolaTovbis1}. Considering one of the pieces of the breaking curve
(to the left, $x<0$, or to the right, $x>0$, from the tip $x_0=0,t_0\approx \hf$, see Fig. \ref{isometry}, where the $x,t$-plane is shown upside-down), 
%we introduced scaled coordinates $S(x,t)=\hf\vartheta(x,t)+\frac{i}{2}\kappa(x,t)\e\ln\e$,
%that map a neighborhood of the left (right) breaking curve onto a horizontal strip in $\C$, containing negative (positive) real semi-axis.
we introduced two scaled coordinates $\vartheta$ measuring lengths of order $\mathcal O(\varepsilon)$ in the tangent direction to the 
breaking curve, and $\kappa$ measuring lengths of order $\mathcal O(\varepsilon\ln \varepsilon)$ in the transversal direction. 
In these coordinates  $\kappa>0$ was the interior of the oscillatory region (see Fig. \ref{Cai}, \ref{3dpic}). The shape 
%In the coordinates $\vartheta,\kappa$, where $\vartheta,\kappa$
%the oscillatory region of the left breaking curve is in the upper $S$-half-plane
%\red{[is that also true for the right breaking curve?]}, and the shape 
of the oscillations in the $\kappa,\vartheta$-plane is depicted on Fig. \ref{3dpic}.

The tip-point $x_0,t_0$ of the breaking curve is called a point of  {\bf gradient catastrophe}, or {\bf elliptic umbilical
singularity} (\cite{DubrovinGravaKlein}).)  It is evident from the numerical simulations shown on Fig.  \ref{Cai} and Fig. \ref{isometry},
as well as from the analysis of the spectral plane (presented below),  that the behavior of the solution
$q(x,t,\e)$ at the tip of the breaking curve is very different from the behavior elsewhere on the breaking curve.

The main goal of this paper is to analyze the leading order asymptotic behavior of
the solution $q(x,t,\varepsilon )$ on and around this special point of transition.   More precisely 
we will examine a 
neighborhood $D$ of $(x_0,t_0)$ that is  shrinking at the rate { $\mathcal O(\varepsilon^\frac 45)$} as $\e\ra 0$.
As the first step, we construct a map $v=v(x,t,\e)$, that maps $D$ onto a bounded disk $V\in\C$, where
$V$ is independent of $\e$ and $v(x_0,t_0,\e)=0$. It turns out that the leading order behavior of $q(x,t,\e)$ in $D$
can be conveniently described through a specific   {\em tritronqu\'ee}  solution (see Section \ref{tritron}) to the Painlev\'e\  I (P1) equation
\be
y''(v) = 6 y(v)^ 2 - v.\label{P1ODEintro}
\ee
That is why  the map $v(x,t,\e)$  and the
complex $v$-plane will be referred to as  the {\em Painlev\'e\  coordinatization} of $D$ and  the Painlev\'e\  plane respectively. 
Note that the map $v(x,t,\e)$ near the point of gradient catastrophe 
plays a similar  role to the map $S(x,t,\e)$ for the rest of the breaking curve.

The point of gradient catastrophe $(x_0,t_0)$ for a solution $q\in\Uscr$ of the NLS (\ref{FNLS}) can be defined 
in terms of the space-time (physical) 
variables $(x,t)$ as the point, where the genus zero approximation (\ref{lotg0a})
of the solution $q(x,t,\e)$   develops an infinite $x$-derivative in  $A(x,t)$
and/or in  $\Phi_x(x,t)$ (while $A(x,t)$ and $\Phi_x(x,t)$ stay finite). In terms of the time evolution of the spectral data (see 
Fig. \ref{phasediagram} below),
the point of gradient catastrophe is the point where the birth/collapse of the new main arc (band) happens exactly at 
the end of an existing main arc (see Fig. \ref{phasediagram}).

As it was mentioned above, the asymptotics of $q(x,t,\e)$ within the region of a given genus 
is given explicitly in terms
of the Riemann Theta-functions, associated with $\Rscr(x,t)$, with the accuracy $\mathcal O(\e)$, 
see \cite{TVZ1}, \cite{KMM}.
(In the case of genus zero, see (\ref{lotg0a}).)
The leading order approximation $\tilde q(x,t,\e)$ of   $q(x,t,\varepsilon )$ in order $O(\e\ln\e)$ strips
around left and right branches of the breaking curve, found in \cite{BertolaTovbis1}, has the accuracy $O(\sqrt{\e})$.
%\removelong{It was conjectured by B. Dubrovin, T. Grava and C. Klein in \cite{DubrovinGravaKlein}
%that the error estimate near the gradient catastrophe is of order $O(\e^{2/5})$ and is expressible in terms of the  {\em tritronqu\'ee} solution of P1
%(see  Remark \ref{remarkConjecture} for more details on their conjecture).}

\subsection{Description of results}\label{descreslts}

%\removelong{The results of this paper not only confirm the conjecture of Dubrovin, but in fact go  beyond it (see also Remarks \ref{classU}, \ref{initscatt}, \ref{conjDub}), 
%as w}
We {provide}  the leading   
order behavior together with the accuracy estimate {\bf in the whole domain $D$ around the point of gradient catastrophe} $x_0,t_0$,
that include the oscillatory part of $D$.  The following notations are useful in describing our results.

Let:  $V$ denote a compact neighborhood of the origin $v=0$ of the independent variable \yellow{$v$} of the Painlev\'e\ transcendent 
(for example a bounded disk of arbitrary large but fixed radius);
 $V_p=\{v_{p,1},\cdots,v_{p,N}\}\subset V$ denote the set of poles of the { tritronqu\'ee} solution $y(v)$ in $V$;
$B_{\d,j}$ denote the disk of radius $\d>0$ centered at $v_{p,j}$, $j=1,\cdots, N$, $B_\d=\cup_{1}^N B_{\d,j}$ and $K_\d=V\setminus B_\d$.
Denoting   
\be\label{aib}
a=-\hf \Phi_x(x_0,t_0),~~~~~~b=A(x_0,t_0),
\ee
(so that, according to (\ref{lotg0a}),  
$a+ib=\a(x_0,t_0)$),  we  prove:
%\removelong{ the following statements  about a solution $q(x,t,\e)$ that undergoes the gradient catastrophe at $x_0,t_0$}

\begin{enumerate}
\item (Thm. \ref{thmpeakshape}) There is a {\bf one to one correspondence between the poles of the tritronqu\'ee solution} $y(v)$ within $V$ and the
{\em spikes of the NLS solution $q$} within $D$.  Each spike  is centered at the corresponding $(x_{p,j},t_{p,j})=v^{-1}(v_{p,j})$,
where 
\bea\label{vxte}
v(x,t,\e)=\frac{e^{-i\pi/4}}{\e^\frac 45}\sqrt{\frac{2b}{C}}\le[x-x_0+2(2a+ib)(t-t_0)\ri]\le(1+ \mathcal O(\varepsilon^{\frac 25})\ri)
\eea
uniformly in $D$, with the  nonzero constant  $C$ explicitly defined by (\ref{expressC}) {in terms of the scattering data};

\item Each spike has the fixed height of $3|q_0(x_0,t_0,\e)|+O(\e^{1/5})$, 
%\green{[we have $3|q|$ in the abstract] [I have changed it there too!]}
where $q_0$ is the genus zero approximation of $q$,
 i.e., {\bf the height of each spike is three times} the amplitude {at the gradient catastrophe},
see Theorems \ref{ThmAmplitude}, \ref{thmpeakshape}; 

\item Each spike has the {\bf universal shape} of the 
(scaled) {\bf rational breather solution to the NLS eq.} (\ref{FNLS}), see Fig. \ref{NLSpeak}, i.e,
\be\label{spikeappr}
q(x,t,\e)  ={\rm e}^{\frac i \ve \Phi(x_p,t_p)} Q_{br}\le(\frac {x-x_{p,j}}\ve,\frac {t-t_{p,j}}\ve\ri)
(1 + \mathcal O(\ve^\frac 15)),
\ee
where the rational breather 
\be\label{ratbreather}
 Q_{br}(\xi,\eta) = {\rm e}^{-2i \le(a \xi + (2a^2 - b^2) \eta\ri)} 
b\, \le( 1 - 4\frac {1+ 4ib^2 \eta }{1 +  4b^2 (\xi  +4 a \eta)^2 + 16 b^4 \eta^2} \ri)
\ee
satisfies the NLS eq. (\ref{FNLS}) with space-time variables $\xi,\eta$. 
This breather approximation of the  spike
is valid in the domain $\tilde B_j=v^{-1}(B_{\d,j})$ of $(x_{p,j},t_{p,j})$, where  $\d=O(\e^{1/5})$,
see Theorems \ref{ThmAmplitude}) and \cite{BertolaTovbis1}. 
The size of each spike in the physical plane (the size of $\tilde B_j$) is thus $O(\e)$, which is consistent
with the size  of spikes along the breaking curve (away from $(x_0,t_0)$) { and within the bulk of the genus two region}, see above.
The two zeroes (``roots'') and the maximum of each breather, shown on  Fig. \ref{NLSpeak}, occur at the same time
 (within the accuracy of our approximation).
{\em  We note here that this universal shape is a completely new result, which, to our best knowledge,
 was never even conjectured or observed numerically}.

\item  In Thm. \ref{mainthmgen0} we show that if $\d>0$ is a small fixed number then  
\bea\label{corrawayspike}
&q(x,t,\e)=\left(b-2 \ve^{\frac 25} \Im \le(\frac {y(v)}{C }\ri)+\mathcal O(\varepsilon^{\frac 35})\ri) \times\cr
&\exp \frac{2i}{\varepsilon} \le[\hf\Phi(x_0,t_0)- \le(a( x-x_0) - (2a^2-b^2)(t-t_0)\ri)  
+  \ve^\frac 65\Re\le(\sqrt{\frac{2i}{C b }}H_I(v)\ri) \ri]
\eea
uniformly in $\hat K_\d=v^{-1}(K_\d)$ (i.e. uniformly over compact sets of the $v$ plane that do not contain any pole), where $C$ is a nonzero constant explicitly given by (\ref{expressC}), $y(v)=y(v(x,t,\e))$ is the 
{ tritronqu\'ee} solution and  $H_I = \frac 1 2 (y'(v))^2 + v y(v) - 2y^3(v)$. 
Equation (\ref{corrawayspike}) is consistent with the conjecture of \cite{DubrovinGravaKlein} 
(see Remark \ref{remarkConjecture}), although that conjecture is formulated for a different class of solutions (defined 
through their initial data).

\item If $\d=O(
{\e^{\nu}})$, where $\nu \in(0,\frac 15)$, and  $\hat K_\d=v^{-1}(K_\d)$, then equation (\ref{corrawayspike}) will 
be uniformly valid in $\hat K_\d$ provided that $\e^{\frac 15}$ in the error term will be replaced by  
$\e^{\frac 15-\nu}$, i.e.
\bea\label{corrawayspikeext}
&q(x,t,\e)=\left(b-2 
{\ve^{\frac 25}} \Im \le(\frac {y(v)}{C }\ri)+\mathcal O(\varepsilon^{\frac 35 -3\nu})\ri) \times\cr
&\exp \frac{2i}{\varepsilon} \le[\hf\Phi(x_0,t_0) - \le(a( x-x_0) - (2a^2-b^2)(t-t_0)\ri)  
+  
{\ve^{\frac 65}}\Re\le(\sqrt{\frac{2i}{C b }}H_I(v)\ri) \ri].
\eea
{Note that -- since $y(v)$ has a double pole and $H_I(v)$ a simple pole -- the term $y(v)$ is actually of order $\ve^{-2\nu}$ and $H_I(v)$ of order $\ve^{-\nu}$; 
clearly the description in terms of the {\em tritronqu\'ee} cannot be pushed ``too close'' to the pole/spike.}
\end{enumerate}

Note that the results above hold uniformly within the specified regions $\cup_j\tilde B_j$ and $\tilde K_\d$
and hence are not sensitive to the actual location of the poles of the tritronqu\'ee solution $y(v)$.
That is to say that the {\em actual} behavior of a solution $q$ near the point of gradient catastrophe 
{depends on the location of the spikes and hence of the poles of $y(v)$, but does not affect the description of the individual spike.}
%\removelong{(the location of the spikes), of course,
%depends on the location of the poles of $y(v)$, but that does not affect our description of an individual spike}.
%\footnote{\yellow{Localization of all the poles  of the tritronqu\'ee solution $y(v)$ in a certain sector of the complex $v$-plane was
%also conjectured in \cite{DubrovinGravaKlein}. Our numerics in Fig. \ref{phasediagram} is consistent with this conjecture.}}

We also make the  Conjecture \ref{P1conjecture} that the amplitudes of the spikes near any (degenerate)
gradient catastrophe point in the genus zero phase are odd multiples of the amplitude at the point itself. 
In addition we can speculate that the shape of the spikes in the higher-degeneracy cases should be related to the higher rational breathers recently investigated in \cite{Ankiewicz010}.

Among other results obtained in this paper we mention the  proof that the two branches of the breaking curve form a corner  (wedge) at the point of gradient catastrophe $(x_0,t_0)$
and give explicit expression, see (\ref{wedgeangle}), of the angle between the breaking curve in terms of $C$ and $\a(x_0,t_0)$.
We further prove that the map $v(x,t,\e)$ maps this corner {into} the sector  $\frac{2\pi}{5}<\arg v< \frac{4\pi}{5}$
of the complex $v$-plane, see Fig. \ref{Painplane}.  This is consistent with another conjecture, stated
in \cite{DubrovinGravaKlein}: all the 
poles of the { tritronqu\'ee} solution $y(v)$ are contained within the sector $\frac{2\pi}{5}<\arg v< \frac{4\pi}{5}$. This 
 is a longstanding 
question in the theory of Painlev\'e\  equations.  
According to our results, {\em the set of spikes near the point of gradient
catastrophe is, in fact, the visualization of the poles of the { tritronqu\'ee} solution $y(v)$ to P1}. 
In this sense numerical simulations, shown on Fig. \ref{Cai} and \ref{isometry} (as well as similar 
computations in other papers), are consistent with the conjecture from \cite{DubrovinGravaKlein}, although they do not amount to a proof. 
%\green {[Do we need the following sentence?] We want to emphasize that 
% the above-mentioned statements 1 - 5 do not depend on the localization of the poles $v_p$ of the { tritronqu\'ee} solution $y(v)$,
%i.e., {\bf statements 1 - 5 do not rely on the conjecture of Dubrovin} on the localization of the poles.}
The very first real pole of the real-analytic { tritronqu\'ee} solution
 was numerically calculated
in \cite{JoshiKitaev01}. Applied to our case, the result of \cite{JoshiKitaev01} implies that
the very first pole $v_{p,1}$ of
$y(v)$ on the ray $\arg v=\frac{3\pi}{5}$ has $|v_{p,1}|\approx 2.38$. Using (\ref{vxte}), we calculate $(x_{p,1},t_{p,1})=(0,2.79126)$
for the NLS evolution of the initial data $q(x,0,\e)={\rm sech} x$. Numerical simulation of this evolution with $\e=\frac{1}{33}$ shows
the first spike at $t_s\approx 2.8$, which is in a very good agreement with $t_{p,1}$, see 
Example \ref{firstspike} and Fig. \ref{numericpeak}

\br\label{conjDub}
Statement 4 from the above list  is consistent with Dubrovin's conjecture \cite{DubrovinGravaKlein} for solutions $q\in\Uscr$.
%(so, if you wish, a modified version of the conjecture). 
\er

While reducing the (original)  matrix RHP, associated with  the inverse scattering transform, to the model RHP, 
the error is controlled through the so-called  local parametrices.
 At  the regular points $(x,t)$ in the physical plane, these parametrices can be constructed through the Airy functions. As we show in Section \ref{sectPRHP}, 
at the point of gradient catastrophe $(x_0,t_0)$, the parametrix $\mathcal P$ is constructed  through the  { tritronqu\'ee} solution $y(v)$ of the P1.
It is well known that Painlev\'e\ equations can be expressed as conditions of isomonodromic deformations for certain rational 
$2\times 2$  systems of ODEs with rational coefficients  \cite{JMU1,JMU2}.  
The  parametrix $\mathcal P$ at the point of gradient catastrophe is built through the fundamental solution  $\Psi(\x;v)$
to the system of ODEs associated with the P1. 
The occurrence of 
parametrices built out of Painlev\'e\ associated linear systems is 
not unexpected here, as they often  appear in various 
RHPs related to  random matrices/orthogonal polynomials, for example:
\begin{itemize}
\item The case of random matrices with a soft-edge where the density vanishes to order $(z-\a)^{2 k+1/2}$, corresponding to the even P1 hierarchy \cite{ClaeysVanlessen2};
\item The case of random matrices where a spectral band splits into two (P2 equation \cite{BleherIts} and hierarchy);
\item The trailing edge of the region of oscillations in the small--dispersion limit of KdV \cite{ClaeysGrava1} (PII equation).
\end{itemize}

The novelty  of our work lies in the fact 
that the matrix  $\Psi(\x;v)$, and thus,  the parametrix $\mathcal P$, {\bf is not defined }(has poles)
at the poles $v_p\in V$  of the { tritronqu\'ee} solution $y(v)$. To our best knowledge, 
this paper contains {\it the first example of parametrices with singularities,
that were successfully  used to control the errors} at and around the singularities.

Solving the RHP at or near the pole $v_p$ of $y(v)$, i.e., studying the shape of the spike,  require several additional steps, which can be briefly listed as:
\begin{itemize}

\item Factorization $\Psi(\x;v)=G(\xi,v)\hat\Psi(\xi,v)$, where $G(\xi,v)$ is a ``simple'' matrix with singularity at $v=v_p$, and $\hat\Psi(\xi,v)$ is regular at
$v=v_p$. This factorization was introduced by D. Masoero in \cite{Masoero}. The existence of the limit
of $\hat\Psi(\xi,v)$ as $v\ra v_p$ that is uniform in a certain region of the spectral $\xi$-plane
(see Appendix \ref{towards}), is an important part in establishing the shape of spikes;

\item  Construction of $\Psi_1$, solution of a {\em modified} model RHP,  that is needed at the poles $v=v_p$ of 
the { tritronqu\'ee} solution $y(v)$. $\Psi_1$ was obtained by means of the
{\bf discrete Schlesinger isomonodromic deformations} of $\Psi_0$, {solution to the ``standard'' model RHP}.
This type of deformations were used, for example, in \cite{Bertola:Lee, BertolaTovbis1};

\item Construction of the new parametrix $\mathcal P_1$ for the modified model RHP, using $\hat\Psi(\xi,v)$. The formula $3b+O(\e^{1/5})$ for the height of 
spikes follows  immediately from $\mathcal P_1$; 

\item Additional transformation of $\Psi_1$, called  partial Schlesinger transformation, is used to obtain the shape of the spikes, see (\ref{spikeappr}).

\end{itemize}

Finally, it is clear that our method can be  modified to handle higher order (degenerate) gradient catastrophes, where $k$ main arcs, $k>1$,
simultaneously emerge from the endpoints of an existing main arc. The parametrices in these cases can be written in terms of the higher members of the P1 hierarchy,
and  one should  expect the height of the spikes to be $2k+1$ times the amplitude at the point of gradient catastrophe.

To summarize our results about the 
typical behavior of a solution  $q=q(x,t,\e){\in\Uscr}$ in a full $O(\e^\frac 45)$-scaled
neighborhood $D$ around the point of its gradient catastrophe $(x_0,t_0)$:
\begin{itemize}
\item  The poles of the { tritronqu\'ee}  solution $y(v)$ that belong to $V$
are mapped into $D$ by the map $(x,t)=v^{-1}(v)$; 
\item Every pole $v_p$ from $V$, wherever in $V$ it is located, generates
a spike of the size $O(\e)$ and of the universal height $3|q_0(x_0,t_0,\e)|$ centered around $(x_p,t_p)=v^{-1}(v_p)$. All the spikes
have the universal shape of the scaled rational breather; 
\item In between the spikes, the solution $q$ is approximated by the constant
$q(x_0,t_0,\e)$ term with accuracy $O(\e^\frac 25)$, where the correction term is explicitly given in terms of $y(v)$; 
\item While the actual behavior of $q$ in $D$ depends on where the poles of the { tritronqu\'ee}  solution $y(v)$ are located,
our universal description of the spikes of $q$  is valid regardless  of the location of the poles of $y(v)$.
\end {itemize}
%where the corresponding poles of $y(v)$ are located.

%
\section{A short review of the  {zero dispersion limit of the} inverse scattering transform}
\label{sectreview}
%
%\removelong{Given an initial data (potential)  $q(x,0,\e)$ for the (\ref{FNLS}) that \replace{is  decaying}{has a sufficient decay} as $x\ra\pm\infty$, 
%the direct scattering transform for the NLS  (\cite{ZS})
%produces {the scattering data, namely: the reflection coefficient $r_0(z,\ve)$ and the  points  of discrete
%spectrum together with their norming constants (solitons). The time evolution of the scattering data is simple
%and well-known. Thus, to find the evolution of a given potential at a time $t$, 
%one needs to solve the inverse scattering 
%problem at this time $t$. Equivalently, we can stipulate that the initial data is assigned directly 
%through the  scattering data;
%thus,  one can produce a solution to the NLS (\ref{FNLS}) by choosing  some 
%scattering data and solving the inverse scattering problem for $t=0$ (initial data) and for $t>0$ (evolution
%of the initial data). The latter approach allows one to avoid solving the direct scattering problem in the semiclassical limit and addressing
%many delicate issues associated with it (see \cite{Abel}  for more details). 
%Since we are interested in studying the generic structure of solutions of the NLS near the point of gradient catastrophe, 
%it will be convenient for us to define
%a solution to the NLS (\ref{FNLS}) by its scattering, not by its initial data.}} 

At any time $t$, the inverse scattering problem for a solitonless solution $q=q(x,t,\e)$ of (\ref{FNLS})  with
a fixed (not infinitesimal) $\ve$ is reducible to the following matrix RHP.

\begin{problem}\label{RHPG}
Find a  matrix $\GGamma(z)$ analytic in $\C\setminus \R$ such that 
\bea\label{rhpgam}
\GGamma_+(z) = \GGamma_-(z) \begin{bmatrix} |r_0(z,\ve)|^2 +1  & \ov r_0(z,\ve){\rm e}^{-\frac {2i} 
\varepsilon \le(2 t z^2 + x z\ri)}\\ r_0(z,\ve) {\rm e}^{\frac {2i}\varepsilon \le(2 t z^2 + x z\ri)} & 1\end{bmatrix} \ ,\ \ z\in \R,\\
\GGamma(z) = \1 +\frac 1  z  \GGamma_1  +\mathcal O(z^{-2})\ ,\ \ \ \ z\to \infty,
\label{gaminf}
\eea
where $r_0(z,\e)$ is the reflection coefficient of $q$.
(In the case  with solitons, there are additional jumps across small circles surrounding
the points of discrete spectrum, see \cite{ZS}.)  Then 
\be\label{nlssolfull}
q(x,t,\ve):= -2\le(\GGamma_1\ri)_{12}.
\ee
%is the solution of the initial value problem (\ref{FNLS}) for the NLS equation. 
\end{problem}

The jump matrix for the RHP admits the factorization
\bea\label{factoriz}
\begin{bmatrix} |r|^2 +1  & \ov r\\ r& 1\end{bmatrix} = \begin{bmatrix}  1 & \ov r \\ 0 & 1\end {bmatrix} 
\begin{bmatrix} 1 &0 \\
r&1 \end {bmatrix},
\eea
where $r = r(z;x,t,\e)=r_0(z,\e){\rm e}^{\frac {2i}\varepsilon \le(2 t z^2 + x z\ri)}$.

Inspection of the RHP shows that the matrix $\GGamma(\ov z)^*$ (where $^\star$ stands for the complex--conjugated, transposed matrix)
solves the same RHP with the jump matrix $M(z)$ replaced by $M^{-1}(z)$ and hence: 

\bp
\label{propsymmetry}
The solution $\GGamma(z)$ of the RHP for NLS has the symmetry
\be
\GGamma(z)( \GGamma(\ov z))^* \equiv \1~.
\ee
\ep

%We will consider solutions $q$ to the NLS (\ref{FNLS}) for which $r_0(z,\ve)$ has the form 
%\be r_0(z,\ve) = {\rm e}^{\frac i {2\ve} f_0(z)}\label{r0fake} \ee
%and $f_0(z)$, described below, is {\em independent} of $\ve$. 
%The motivation behind this choice is that the {\bf formal} WKB\yellow{-type of} analysis for 
%initial data (\ref{q0pure}) does 
%lead to scattering data $r_0(z,\ve)$ which have the above form (\ref{r0fake}).
%A smooth dependence of $f_0(z)$ on small positive $\ve$ would not change significantly the result, 
%but complicate unnecessarily the discussion.
%\remove{(and offer no practical benefits since the scattering data for the initial data (\ref{q0pure}) do not have this form)} 

%\removelong{In  considering the semiclassical  limit of (\ref{FNLS}), one has to consider the semiclassical limit of the 
%corresponding scattering transform.
% For the case of decaying potentials of type (\ref{IDe}), this limit was discussed in \cite{TVZ3} (inverse scattering) and \cite{Abel} 
%(direct scattering), where it was shown to be
%a correspondence between 
%\be\label{ptlim}
%\a(x,t)= -\hf \Phi_x(x,t)+iA(x,t)
%\ee
%on the potential (physical) side and 
%\be\label{sctlim}
%f(z)=f_0(z)-xz-2tz^2
%\ee
%on the scattering side.
%Here $t\geq 0$ is fixed and $f_0(z)$ has the meaning of the ``scaled'' logarithm
%of the reflection coefficient $r_0(z,\ve)$ that corresponds to the initial data (\ref{IDe}), that is, 
%\be \label{f0}
%f_0(z)=\frac{i}{2}\lim_{\ve\ra 0}\ve r_0(z,\ve)~.
%\ee}

We shall now specify the set $\Uscr$ of  solution to the NLS (\ref{FNLS});  
it consists of solitonless solutions with admissible reflection coefficients  
(\cite{TVZ3}), as defined below.

\begin{figure}[th]
\begin{center}
\resizebox{0.4\textwidth}{!}{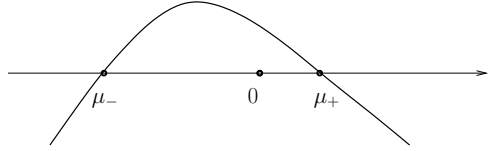}
\end{center}
\caption{ Typical graph of $w(z)$.}
\label{wgraph}
\end{figure}

\begin{definition} \label{adw}
An absolutely continuous, piecewise $C^1$, real function $w(\z)$ 
with a locally square-integrable derivative is {\bf admissible} if it 
satisfies the  following additional conditions:
\begin{enumerate}
\item there exists $\m_-<0<\m_+$ such that  $w(\z)$ is positive on $(\m_-,\m_+)$ 
and negative on  the union $(-\infty,\m_-)\cup(\m_+,\infty)$;
\item $w'(\m_-\pm 0)>0
%=\lim_{\z\ra(\m_-)^\pm} w'(\z)>0
$ and $w'(\m_+\pm 0)
%=\lim_{\z\ra(\m_+)^\pm}-w'(\z)
<0$;
\item $\exists k\ge 0$ such that $w'(\z)=\mp k+o(1)$ as $\z\ra\pm\infty$;
\item $\sign(\m_+-\z)w'(\z)-k\in L^1(\R)$.
\end{enumerate}\end{definition}

%\noindent 
%\blue{
Given an admissible function $w(\z),~\z\in\R$, we  construct the  function $f_{0}(z)$ (unique up to a real constant), 
that is analytic and Schwarz reflection invariant in $\C\setminus\R$ in the following way: first, using the Cauchy transform, we construct  $f'_{0}(z)$ by
\begin{equation}
\label{f'}
f_{0}'(z)=
 ik\sign(\Im z)
+ \frac{1}{\pi}\int_{-\infty}^\infty
\frac{\sign(\m_+-\z)w'(\z)-k}{\z-z}d\z;
\end{equation}
then $f_{0}(z)$ is an antiderivative of $f'_{0}(z)$ satisfying
\be \label{jump}
\Im f_{0}(\z)_+=w(\z)\sign(\m_+-\z),~~~~~~~\z\in\R,
\ee
where the subscripts $\pm$ indicate limiting values on the real axis from the upper/lower 
complex half plane respectively. 
Notice also that: 
$\Re f'_{init}(\z)_\pm=\Hscr [\sign(\m_+-\z)w'(\z)-k]$ 
for almost every  $\z\in[\m_-,\m_+]$ (here $\Hscr$ denotes the Hilbert transform); 
%\be\label{Ref'pm} \Re f'_{init}(\z)_\pm=\frac{1}{\pi}\int_{-\infty}^\infty
%\dfrac{\sign(\m_+-s)w'(s)-k}{s-\z}ds=\Hscr [\sign(\m_+-\z)w'(\z)-k] 
%~~~~~\z\in[\m_-,\m_+];\ee}
as $z\ra\infty$, 
\begin{equation}\label{asympt}
f'_{0}(z)= ik\sign(\Im z) -\frac{1}{z\pi}\int_{-\infty}^\infty [\sign(\m_+-s)w'(s)-k]ds+o(z^{-1})~,
\end{equation}
and;  $\Im f_0(z)$ has a jump across the real axis 
given by
\begin{equation}\label{rhpf'}
f_{0}(\z)_+-f_{0}(\z)_-=2i\sign(\m_+-\z)w(\z), \ \ \ \ \z\in\R.
\end{equation}

%\blue{
\begin{definition}\label{adr}
A  reflection coefficient $r_{0}$ continuous on $\R$ is called admissible if there exists an 
admissible function $w(\z), ~\z\in\R$, such that
\be \label{r}
\le\{
\begin{array}{cc} 
r_{0}(\z,\e)=e^{-\frac{2i}{\e}f_{0}(\z)_+}, \ \ \mbox{when $\z\in[\m_-,\m_+]$},\\
|r_{0}(\z,\e)|\leq ce^{-\frac 1\e {\min\{c_1|\z-\m_\pm|,c_2 \}}}, \ \ \mbox{when $\z\notin[\m_-,\m_+]$},
\end{array}\right.
\ee
where  $c, \ c_1, \ c_2$ are some
positive constants  {and $f_0$ is as described above}. 
%\marginpar{\red{What is $v$?}}
\end{definition}
%}

%\blue{
In other words, a reflection coefficient $r_{0}$ is admissible if: A) it 
is exponentially decaying (with respect to $\e$) outside some interval $[\m_-,\m_+]$
and exponentially growing inside this interval; 
B) for any fixed $\e>0$ it is exponentially decaying as $\z\ra\pm\infty$;
C) the values of $f_{0}(\z)=
\hf i\e\ln r_{0}(\z)$ on $[\m_-,\m_+]$ are boundary values of some analytic in 
the upper half-plane function $f_{0}(z)$, such that (\ref{jump}) and 
(\ref{asympt}) hold, where $w(\z)$ and $k$ satisfy all the requirements 
in Definition \ref{adw}.
%}

%Since we have taken the perspective that we are starting with 
\bd
\label{defUscr}
%\blue{
Let $\Sscr_\e$ denote the (direct) scattering transform for  solitonless initial data  of the NLS (\ref{FNLS}). 
The set  $\Uscr$ of solutions $q(x,t,\e)$ to (\ref{FNLS}) is the set of solitonless solutions with 
initial data of the form $q(x,0,\e)=S^{-1}_\e r_0(z,\e)$, where $r_0(z,\e)$ are
%that we consider is the set of all solitonless solutions $q$ to the NLS (\ref{FNLS}) that have (at $t=0$) 
admissible reflection coefficients.
\ed

The most advanced results to our knowledge  about the correspondence between the initial and the scattering data for the focusing NLS (with a fixed $\e$) can be found in
\cite{Zh1}, \cite{Zh2}. According to them, 
since an admissible reflection coefficient $r_0$ (with a fixed $\e>0$) is continuous, piece-wise differentiable on $\R$ and is
exponentially decaying at $\z\ra\pm\infty$, the corresponding initial data $q(x,0,\e)$ belongs to the weighted  $L^2(\R)$
with the weight $1+x^2$. Moreover, the additional assumption $r_0\in C^\infty(\R)$ would imply that  $q(x,0,\e)$ is in the Schwarz class.  
In the semiclassical limit, one can take advantage of the
 steepest descent method for RHP (\ref{RHPG})-(\ref{gaminf}) 
to prove  (see \cite{TVZ3}) that a solution $q\in\Uscr$  %(Def. \ref{defUscr})
has the modulated plane wave (genus zero) approximation $q(x,t,\e)= q_0(x,t,\e)+O(\e)$ that is valid uniformly on the compact subsets of the genus zero region. 
%strip $x\in\R$, $0\leq t <t_0$,
%where $(x_0,t_0)$ is the point of gradient catastrophe for $q$. 
Here 
$q_0(x,t,\e)= A(x,t)\exp\{\frac{i}{\e}\Phi(x,t)\} $, where
$A(x,t),\Phi(x,t)$ are defined implicitly through (\ref{aib}) and  the modulation equation (\ref{modeqint}). These functions
are real-analytic in the genus zero region of the $(x,t)$ - plane 
%$\in\R\times[0,t_0)$ 
with $A(x,t)\ra 0$ and $\Phi_x(x,t)\ra -2\m_\pm$ exponentially fast
as $x\ra \pm\infty$.
%
%
%

%\begin{remark}\label{classofq}
%The class of solutions $\Uscr$ is not overly restrictive: for example, if $r_0$ is the reflection coefficient for the family
%of initial data  (\ref{ourfamily}) studied in \cite{TVZ1} with $\m\geq 2$ (solitonless case), the corresponding 
%$w(\z)=\hf \lim_{\e\ra 0} (\e\ln |r(\z,\e)|)$ is an admissible function. In particular, in the case $\m=2$ we have
%$w(\z)=\Im f_0(\z)=\frac{\pi}{2}(1-|x|)$, where $f_0$ is given by (\ref{f0m=2}), provided that the proper branches of 
%logarithms were chosen. 
%\end{remark}

\begin{proposition}\label{propclassU}
 Any solution $q=q(x,t,\e)$ from $\Uscr$ has the following properties (\cite{TVZ3}): 
\bi
\item the genus of $q(x,t,\e)$ for all $x\in\R$ and $t\geq 0$ cannot exceed two; 
\item the genus zero region contains a strip $x\in\R,~0\leq t<t_*$ with some $t_*>0$; It has asymptotes with the slopes
$-\frac{1}{4\m_\mp}$ as $x\ra\pm\infty$ in the $(x,t)$-plane;
\item there is a point of gradient catastrophe $x_0,t_0$ on the boundary of the genus zero region; 
\item the modulated plane wave $q_0(x,t,\e)$ is an order $\mathcal O(\e)$ approximation of the solution $q(x,t,\e)$ in the genus zero region,
which is uniform on compact subsets.
\ei
\end{proposition}
Proposition \ref{propclassU} shows that $q_0(x,0,\e)=A(x,0)e^{\frac i\e \Phi(x,0)}$
%the initial condition $A(x,0,\e)e^{\frac i\e \Phi(x,0,\e)}$ of $q(x,t,\e)$ 
%value $A(x,0)e^{\frac i\e \Phi(x,0)}$ of $q_0(x,0,\e)$ as $\e\ra0$, where $A(x,0),\Phi(x,0)$ are real analytic functions.
approaches the initial data $q(x,0,\e)$ of a solution $q\in\Uscr$  as $\e\ra 0$ uniformly on compact subsets of $\R$.

\br\label{rem2}
The statements 1-5 from Section \ref{descreslts} are formulated  for solutions of the class $\Uscr$.
However, they could be  extended for the cases when $f_0(z)$ may have singularities (including logarithmic branch-cuts) in $\C\setminus \R$,
provided that in some vicinity of the gradient catastrophe $x_0,t_0$  the contour $\g_m$ of the RHP for the $g$-function 
(which will be introduced in the next section) 
lies within the domain of analyticity of $f_0(z)$ in $\C\setminus \R$. This is the case, for example, when, similarly to 
 (\ref{f0m=2,w}), we define  $f_0(z)=-\lim_{\e\ra 0}\frac{i\e}2\ln S_\e\hat q_0(x,0,\e)+\frac{\pi\e}2$, where $q_0$ is given by
(\ref{ourfamily}) with
 $\m\in (0,2)$ (radiation with solitons), see \cite{TVZ1}.
Our results, apparently, should also be extendable to the case  $f_0=f_0(z,\e)$, provided that for all $z$ on and around $\g_m$
the dependence of $f_0(z,\e)$ on $\e$ is smooth. 
\er

In order to study the dispersionless limit $\varepsilon\to 0$, the RHP (\ref{rhpgam})-(\ref{gaminf}) undergoes a 
sequence of transformations (that are briefly recalled in Section \ref{modelpr}) along 
the lines of the nonlinear steepest descent method \cite{DKMVZ, TVZ1}, which reduce it to an RHP that allows for an approximation by the so-called model RHP.
The latter RHP has  piece-wise
constant jump matrices (parametrically dependent on $x,t,\e$)  and, in general,  can be solved explicitly
in terms of the Riemann Theta functions,  
or, in simple cases, in terms of algebraic functions. The $g$-function, defined below,
is the key element of  such a reduction.
\subsection{The $g$-function}
\label{sectg}
The $O(\e)$ order approximation $q_0$ of a solution $q\in\Uscr$ is determined by $f_0(z)$.
Given $f_0(z)$, we introduce  the $g$-function $g(z)=g(z;x,t)$ as the solution to
the following scalar RHP:
\begin{enumerate}
\item  $g(z)$ is
analytic (in $z$) in $\bar\C\setminus 
\g_m$ (including analyticity at $\infty$); 
\item $g(z)$ satisfies the jump condition
\be\label{rhpg}
g_+ + g_-=f_0-xz-2tz^2~~~~ {\rm on}~~\g_{m},
\ee
for  $x\in\R$ and $t\ge 0$, and; 
\item
$g(z)$ has the endpoint behavior
\be\label{modeq}
g(z)=O(z-\a)^{3\over 2}~ + ~{\rm an~analytic~ function~ in~ a~ vicinity~ of~} \a. 
\ee
\end{enumerate}

\begin{figure}[t]
\resizebox{0.99\textwidth}{!}{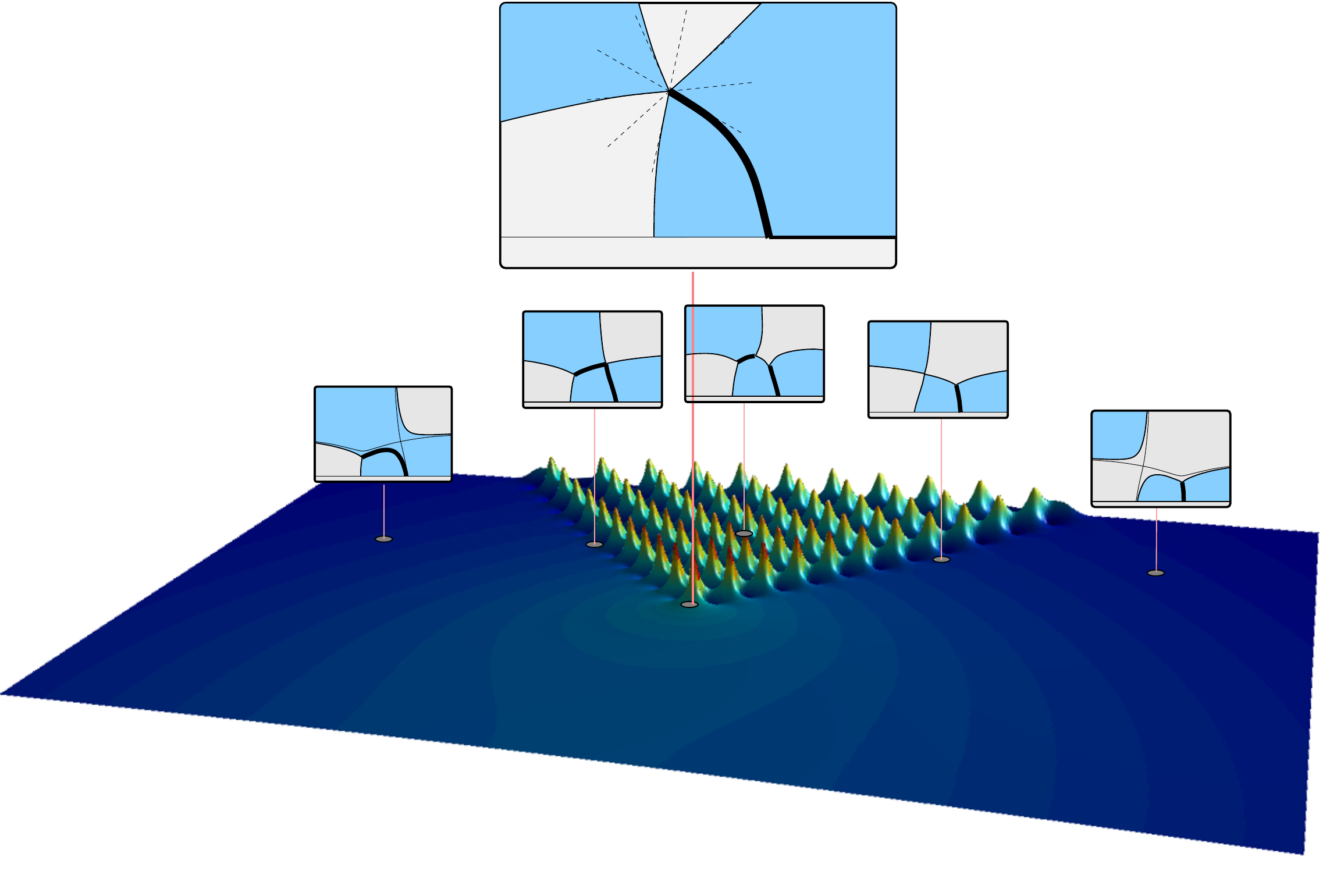}
\caption{The typical zero-dispersion phase diagram for a one-hump initial data. Representative at different points in the $(x,t)$--plane are the level-curves of $\Im (h)$. 
Only the upper-half spectral plane is depicted. The shape of the level 
curves was obtained numerically on a simple example. The plot of the amplitude of $q(x,t,\e)$ corresponds to the initial data $q(x,0,\e)={\rm sech}(x)$, $\ve = \frac 1 {33}$.
This is a pure-soliton case, which was used  here only for the purpose of an effective illustration, 
 as our results are valid for any generic (non-degenerate) gradient catastrophe.} 
\label{phasediagram}

\end{figure}

Here:
\begin{itemize}
\item  $\g_m$ is a bounded Schwarz-symmetrical contour
(called the main arc) with the endpoints $\bar\a, \a$, oriented from $\bar\a$ to $ \a$ and
intersecting $\R$ only at  $\m_+$;
\item  $g_\pm$ denote the values of $g$ on the positive
(left) and negative (right) sides of $\g_m$;
\item the 
function $f_0=f_0(z)$, representing the initial scattering
data, is   Schwarz-symmetrical and H\"{o}lder-continuous on $\g_m$.
\end{itemize}
Taking into the account Schwarz symmetry, it is clear
that  behavior of $g(z)$ at both endpoints $\a$ and  $\bar \a$ should be the same.

Assuming $f_0$ and $\g_m$
are known, {the} solution $g$ to the {scalar} RHP (\ref{rhpg}) without the endpoint condition (\ref{modeq}) 
can be obtained by the Plemelji formula
\be\label{gform}
g(z)={{R(z)}\over{2\pi i}} \int_{\g_m}{{f(\z)}\over{(\z-z)R(\z)_+}}d\z~,
\ee
where $R(z)=\sqrt{(z-\a)(z-\bar\a)}$. 
{ We fix the branch of $R$ by requiring that $\lim_{z\ra\infty}\frac{R(z)}{z}=1$.} If $f_0(z)$ is analytic in some region $\Sscr$ that 
contains $\g_m\setminus\{\m_+\}$, the formula for $g(z)$   can be rewritten as
\be\label{gforman}
g(z)={{R(z)}\over{4\pi i}} \int_{\gt_m}{{f(\z)}\over{(\z-z)R(\z)_+}}d\z~,
\ee
where $\gt_m\subset\Sscr$ is a negatively oriented loop
around $\g_m$ (which is ``pinched'' to $\g_m$ in $\m_+$, where $f$ is not analytic)
that does not contain $z$. 
Introducing function 
$h=2g-f$,
we obtain 
\be\label{hform}
h(z)={{R(z)}\over{2\pi i}} \int_{\gt_m}{{f(\z)}\over{(\z-z)R(\z)_+}}d\z~,
\ee
where $z$ is inside the loop $\gt_m$. 
The endpoint condition
(\ref{modeq})  can now be written as
\be\label{modeqh}
h(z)=O(z-\a)^{3\over 2}~~{\rm as}~~z\ra \a,
\ee 
or, equivalently,
\be\label{modeqint}
\int_{\gt_m}{{f(\z)}\over{(\z-\a)R(\z)_+}}d\z=0~.
\ee
The latter equation is known as a {\it modulation equation}. The function $h$ plays a prominent role 
in this paper. Using the fact that the Cauchy operator for the RHP (\ref{rhpg}), (\ref{modeq}) commutes with
differentiation, we have 
\be\label{h'form}
h'(z)={{R(z)}\over{2\pi i}} \int_{\gt_m}{{f'(\z)}\over{(\z-z)R(\z)_+}}d\z~,
\ee
where $z$ is inside the loop $\gt_m$.  Note that (\ref{modeqh}) implies that there are exactly three zero level curves of
$\Im h(z)$ emanating from $z=\a$.

In order to reduce the RHP (\ref{rhpgam})-(\ref{gaminf}) to the RHP with 
piece-wise jump matrices, called the {\em model RHP}, the signs of $\Im h(z)$ in the upper half-plane should satisfy the following conditions:
\bi \label{ineqh}
\item $\Im h(z)$ is negative on both sides of the contour (main arc) $\g_m$;
\item there exists a continuous contour $\g_c$ (complementary arc) in $\C_+$ that connects $\a$ and $\m_-$, so that 
$\Im h(z)$ is positive along $\g_c$. Since $\Im h(z)>0$ on the interval $(-\infty,\m_-)$,  the point $\m_-$ in
$\g_c$ can be replaced  by any other point of this interval, or by $-\infty$.
\ei

{ Note that the first sign requirement, together  with (\ref{rhpg}), imply that $\Im h(z)=0$
along $\g_m$. Since the signs of $\Im h(z)$ play an important role in the following discussion,
we call by ``sea'' and ``land'' the regions in $\C_+$, where $\Im h(z)$ is negative and positive
respectively. In this language, the complementary arc $\g_c$ goes on ``land'', whereas the main
arc $\g_m$ is a ``bridge'' or a ``dam'', surrounded by the sea, see Fig. \ref{YRHP}.}    
%\blue{
\begin{remark}
A point $(x_0,t_0)$ is a point of gradient catastrophe if the number of zero level curves of $\Im h(z;x,t)$
emanating from $z=\a(x,t)$ changes from $3$ at ordinary points to $5$ (or more) at $(x,t)=(x_0,t_0)$.
\end{remark}
\subsection{Reduction to the  model RHP }
\label{modelpr}

{ We start the transformation of the RHP (\ref{rhpgam})-(\ref{gaminf}) by deforming (preserving the orientation) 
the interval $(-\infty,\m_+)$ , which is a part of its jump contour, into some contour $\g^+$ in 
the upper half-plane $\C_+$, such  that $\m_+\in\g^+$. Let $\g_-$ be the Schwarz symmetrical image of $\g_+$. {Using the factorization (\ref{factoriz})} the   
RHP (\ref{rhpgam})-(\ref{gaminf}) {can be reduced to an equivalent one where}:
\bi
\item {the} right factor of (\ref{factoriz}) is the jump matrix on $\g_+$;
\item {the} left factor of (\ref{factoriz}) is the jump matrix on $\g_-$;
\item the jump matrix on the remaining part of $\R$ is unchanged.
\ei
It will be convenient for us to change the orientation of $\g_+$, which causes the change of sign in the off-diagonal entry of the corresponding jump matrix.
{On the interval  $(\m_+,\infty)$ we have  $\Im f_0(z)<0$ and it appears that the jump is exponentially close to the identity jump and hence it 
is possible to prove that it has no bearing on the leading order term 
of the solution (\ref{nlssolfull}) (as $\ve \ra 0$: see   \cite{TVZ1} for the case when $f_0$ is a one-parameter family that contains (\ref{f0m=2,w})
and  \cite{TVZ3} for the general case)}. {Therefore}  the leading order contribution in  (\ref{nlssolfull}) comes from the contour $\g=\g_+\cup\g_-$. 
In the genus zero case, the contour $\g$ contains points $\a,\bar\a$, which divide it into the main arc $\g_m$ (contained between $\bar\a$ and $\a$, and the complementary
arc $\g_c=\g\setminus \g_m$. 
According to the sign requirements (\ref{ineqh}), the contour $\g_m$ is uniquely determined as an arc of the level curve $\Im h(z)=0$ (bridge) that connects
$\m_+$ and $\a$, whereas $\g_c$ can be deformed arbitrarily ``on the land''.
Because of the  Schwarz symmetry
\ref{propsymmetry}, it is sufficient to consider $\g$ only in the upper half-plane, i.e., it is sufficient to consider $\g_+$. 
}

Having found
{the branch-point  $\alpha$, the $g$-function $g(z)$ and
the contour $\gamma_m$,
we introduce} additional contours customarily called ``lenses'' that join $\alpha$ to $\mu_+$ on both sides of $\gamma_m$ 
(and symmetrically down under). These lenses are to be chosen rather freely with the only condition that $\Im h$ must be negative along them (positive in $\C_-$). 
This condition is guaranteed by (\ref{ineqh}).

\begin{figure}[ht]
\resizebox{0.94\textwidth}{!}{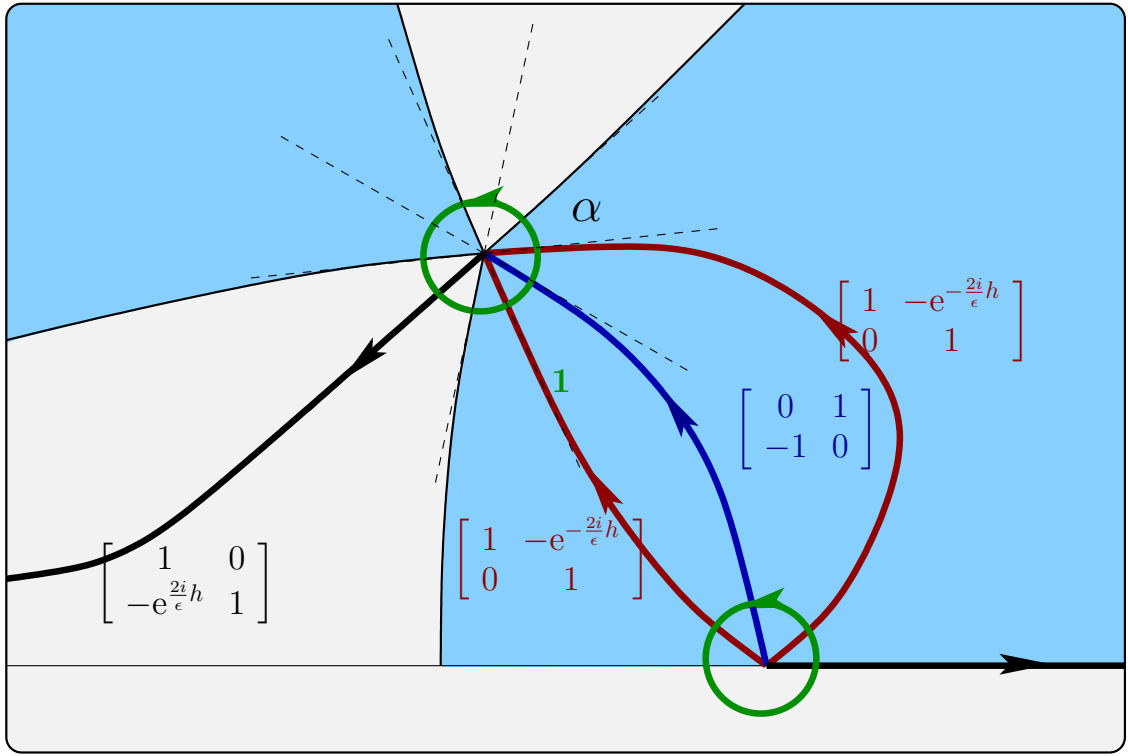}
\caption{The jumps for the  RHP for $Y$. The shaded region is where $\Im h<0$ (the ``sea''). The blue contour is the main arc $\g_m$, the black contour
in $\C_+$ is the complementary arc $\g_c$ and the red contours are the lenses.  The green circles show the boundaries of $\D_\a$ and $\D_+$.}
\label{YRHP}
\end{figure}

The two spindle-shaped regions between $\gamma_m$ and the lenses are usually called upper/lower {\bf lips} (relative to the orientation of $\gamma_m$. 
At this point one introduces the auxiliary matrix-valued function $Y(z)$ as follows
\be
Y(z) ={\rm e}^{-\frac{2i}\varepsilon g(\infty) \sigma_3}\GGamma(z)\le\{
\begin{array}{cc} 
\ds {\rm e}^{\frac {2i}\varepsilon g(z) \sigma_3} & \mbox{ outside the lips,}\\
\ds{\rm e}^{\frac {2i}\varepsilon g(z) \sigma_3} \left[
\begin{array}{cc}
1 & -{\rm e}^{-\frac {2i}\ve h(z)}\\
0& 1
\end{array}
\right] & \mbox{ in the upper lip in $\C_+$,}\\
\ds {\rm e}^{\frac {2i}\varepsilon g(z) \sigma_3}\left[
\begin{array}{cc}
1 & {\rm e}^{-\frac {2i}\ve h(z)}\\
0& 1
\end{array}
\right] & \mbox{ in the lower lip in $\C_+$.}
\end{array}
\ri.
\label{GtoY}
\ee
The definition of $Y(z)$ in $\C_-$ is done respecting the symmetry in Prop. \ref{propsymmetry}, namely
\be
Y(z) = (Y(\ov z)^\star)^{-1},\ \ z\in \C_-.
\ee
The jumps for the matrix $Y(z)$ are reported in Fig. \ref{YRHP}.

\paragraph{The model RHP.}
 In the limit $\ve\ra 0$, according to the signs (\ref{ineqh}), the jump matrices on the complementary arc $\g_c$ and on the lenses are approaching the identity matrix $\1$
exponentially fast. Removing these contours from the RHP for $Y(z)$, we  will have only one remaining contour $\g_m$ with the constant jump matrix 
$\left[
\begin{array}{cc}
0 & 1\\
-1& 0
\end{array}
\right]  $ on it. This is the model RHP. Calculating the (1,2) entry of the residue at infinity (see (\ref{nlssolfull})) of the solution to the model RHP, one obtains the leading 
order term of the genus zero solution as follows (\cite{TVZ1})
\be\label{lotg0}
q_0(x,t,\ve)=\Im \a(x,t)e^{ \frac{i}{\ve} \Phi(x,t)  }; 
\ee
where 
\be
\Phi(x,t) = 4 g(\infty;x,t) \label{phase}.
\ee
A direct calculation of $g(\infty;x,t)$ that uses (\ref{dgxt}) yields (\ref{lotg0a}).

To justify removing contours with exponentially small jump matrices, 
one has to calculate the error estimates coming from neighborhoods of points $\a$, $\bar \a$ and $\m_+$ (for$\a$ and $\m_+$, these neighborhoods are shown as green
circles in Fig. \ref{YRHP}). This is accomplished through the use of  {\em local parametrices}.
We shall consider the construction of the parametrices near the point $\mu_+$  as already done and known to the reader, see \cite{TVZ1}. 
The only information that we need is that these parametrices allow to approximate the exact solution to within an error term $\mathcal E(z) = \1 + \mathcal O(\epsilon)$
uniformly on compact subsets of the genus zero region.
\section{Analysis near the gradient catastrophe point}\label{sectanal}

Let $\a=\a(x,t)\in\C_+$ be the branch-point in the genus zero region, where $(x,t)$ is close to the 
point of gradient catastrophe $(x_0,t_0)$. 

For generic values of $(x,t)$ the function $h(z;x,t)$, according to (\ref{modeqh}), has the behavior
$\frac{i}{\varepsilon}h(z; x,t)\sim \mathcal O(z-\a)^{\frac{3}{2}}$; 
at the point $(x_0,t_0)$  of gradient catastrophe the behavior is instead
$\frac{i}{\varepsilon}h(z:x_0,t_0)\sim \mathcal O(z-\a)^{\frac{5}{2}}$. Thus for $(x,t)$ in the vicinity of this point we obtain
\be
\label{hser2}
\frac{i}{\varepsilon}h(z;x,t)=\frac{1}{\varepsilon}(z-\a)^{3/2}\left( C_0+C_1(z-\a)+ \mathcal O(z-\a)^2\right),
\ee
where $\a=\a(x,t)$ is the branch-point and $C_0, C_1$ are some functions of $x,t$.
The gradient (umbilic) catastrophe point is the one for which $C_0(x_0,t_0)=0$ but
$C_1(0,0)\neq 0$, this latter inequality being our standing genericity assumption.

According to (\ref{h'form}) and (\ref{hser2}), 
\be \label{C00}
C_0=C_0(x,t)=\lim_{z\ra\a}\frac{2 ih'(z)\sqrt{z-\bar \a}}{
3 R(z)}=\frac{\sqrt{2ib}}{3\pi}\int_{\gt_m}  {\frac{f'(\z)}{(\z-\a)R(\z)_+}}d\z~.
\ee
\bl 
\label{lemmaC1}
The value of $C_1$ at the point of gradient catastrophe is given by 
\be \label{C1}
C_1=\frac{2\sqrt{2ib}}{15\pi}\int_{\gt_m} {{f''(\z)}\over{(\z-\a)R(\z)_+}}d\z~.
\ee
\el
{\bf Proof}.
To obtain $C_1$, we notice that at the point of gradient catastrophe $(x_0,t_0)$
\be\label{h''form}
h''(z)={{R(z)}\over{2\pi i}} \int_{\gt_m}{{f''(\z)}\over{(\z-z)R(\z)_+}}d\z~,
\ee
where $z$ is inside the loop $\gt_m$. (This formula is not correct when $(x,t)\not=(x_0,t_0)$.) Then, similarly to (\ref{C00}),
\be
C_1=C_1(x,t)=\lim_{z\ra\a}\frac{4ih''(z)\sqrt{z-\bar \a}}{15 R(z)}=\frac{2\sqrt{2ib}}{15\pi}\int_{\gt_m} {{f''(\z)}\over{(\z-\a)R(\z)_+}}d\z~.
\ee
\QED

The goal of this section is that of introducing a suitable conformal coordinate $\zeta$ near $z=\a$ as in the definition below.

\bd[Scaling coordinate]
\label{defzetatau}
The {\bf scaling coordinate} $\zeta(z) =\zeta(z;x,t,\varepsilon)$ and the {\bf exploration parameter} $\tau =\t(x,t,\e)$ are  defined by 
\be
\frac {i}\varepsilon h(z;x,t) = \frac 45 \zeta^{\frac 5 2 } {(z;x,t,\ve)} + \tau {(x,t)}  \zeta^{\frac  3 2 } {(z;x,t,\ve)}\label{hzeta},
\ee
where $\zeta(\a;x,t,\ve)\equiv 0$ and $\zeta(z;x,t,\ve)$ is analytically invertible in $z$ in a fixed neighborhood of $z=\alpha$.
\ed
The expression (\ref{hzeta}) is the {\bf normal form} of the singularity defined by $h(z;x,t)$ (in the sense of singularity theory \cite{ArnoldGZV-1}).

The detailed analysis of   $\tau(x,t;\ve)$ on space-time will be accomplished in Sect. \ref{mapxt}; for the remainder of this section we dwell a bit on the details of the construction of $\tau,\zeta$ starting from the power-series expansion of $h(z;x,t)$.

Let us denote the expansion of  $h(z,x,t)$ as 
\be\label{hser3}
ih(z;x,t) = C_0(x,t) (z-\a)^\frac 3 2 + C_1(x,t) (z-\a)^\frac 5 2  + \mathcal O( z-\a)^\frac 72.
\ee

For $x=x_0,~ t  = t_0$ we have
\be
\frac i \varepsilon h(z; x_0,t_0 )  = C_1 (z-\a)^{\frac 5  2} \le( 1+ \mathcal O( z-\a)\ri)
\ee
and then the function $\zeta$ and the parameter $\tau$ are defined by the formula 
\be
\zeta(z) := \zeta(z;  x_0,t_0,\varepsilon) := \le(\frac {5i}{4 \varepsilon} h(z; x_0,t_0 )  \ri) ^{\frac 25}\ \ \ \Leftrightarrow\ \ \ 
\frac i \varepsilon h(z; x_0,t_0) = \frac 45 \zeta^{\frac 52},\ \ \ \tau =0.
\ee
Thus, the function $h(z;x_0,t_0)$ {has}  a {\bf singularity} (in the sense of {\em singularity theory}, i.e. the study of normal 
forms of degeneracies of critical values) at $z=\a$. For $x\neq x_0, \ t\neq t_0$ this function undergoes a (smooth) deformation by 
which the coefficient $C_0(x,t)$ acquires a nonzero value that, consequently, is inherited by $\tau$.

In the language of singularity theory this defines a (partial)  {\bf unfolding} of the singularity. 
It is a standard theorem \cite{ArnoldGZV-1} that for any such deformation there is a family of changes of coordinates $z\mapsto w $ so that 
\be
i h(z;x_0,t_0)  = w^{\frac 52} + T(x,t) w^\frac 3 2\ ,\qquad w =  w(z;x,t),~~ w(\a(x,t) ;x,t) \equiv  0, \label{wconf}
\ee 
where $T(x,t)$ and $w(z;x,t)$ have the same smoothness class as the family of the deformation. 
\br
To be more specific, changing variable from $(z-\a)$ to $q = \sqrt{z-\a}$ we then have a singularity for $ih$ of type $A_4$, with additional symmetry
\be
ih(q) = -ih(-q)\ .
\ee
Then the theorem guarantees the existence of a conformal change $Q=Q(q)$ such that any deformation can be recast into
\be
ih(q; x,t) = Q^5 + T_1 Q^3 + T_2 Q^2 + T_1 Q + T_0,
\ee
where $Q$ is a local bi-holomorphic equivalence depending analytically on the deformations. The oddness forces $T_0= T_2=0$ and the
 fact that our particular deformation for $ih$ starts with $q^3$ forces $T_1=0$. Since the theorem guarantees the existence of 
such {\em analytic} family of change of coordinates, a computation manipulating series allows to easily set up a recursive 
algorithmic procedure to find this function, see the next Remark \ref{arnoldser}. 
The most pertinent reference is Chapter 8 in \cite{ArnoldGZV-1}.
\er
\br
\label{arnoldser}
It is not hard to find -- recursively -- the expansion of $w$ and $T$ in terms of the coefficients of the series of $ih$.
If we set (for brevity we shift $\a$ to the origin, without loss of generality)
\bea
i h (z) = C_0 z^\frac 32 + C_1 z ^\frac 5 2 + \sum_{j=2}^\infty C_j z^{\frac {2j+3}2 } \ ,\qquad 
w = w_1  z + \sum_{j=2}^\infty w_j z^j
\eea
we can equate the series expansions of 
\be
ih(z) =  w^{\frac 52} + T w^\frac 32 \ .
\ee
From the coefficient of $z^\frac 32$ we have $T = \frac {C_0}{{w_1}^\frac 32}$ and all the remaining coefficients of the expansion of $w$ can be 
determined in terms of $w_1$  and the $C_j$'s. The first few are 
\bea
w_2 =  - \frac 23 \frac {(w_1^{1/2}-C_1^\frac 15) (w_1^2 + w_1^\frac 32 C_1^\frac 15 + w_1 C_1^\frac 2 5 + w_1^\frac 1 2  
C_1^\frac 3 5  + C_1^\frac 4 5) w_1}{C_0},\label{47}\\
w_3 = \frac {(9w_1^{5} - 8 c_1 w_1^\frac 5 2  - 6 C_2 C_0 - C_1^{2}) w_1}{9 C_0^2}.\label{48}
\eea
The requirement that each term in the  expansion should be {\bf analytic} at $C_0=0$  determines $w_1$ uniquely. 
For example, from (\ref{47}) we must have  $w_1 = C_1^\frac 25 + \mathcal O(C_0)$: plugging this into (\ref{48}) one sees that there 
can be at most a simple pole at $C_0=0$ and setting the residue to zero we determine the next coefficient in the expansion of $w_1$. For example we have 
\bea
w_1^\frac 1 2 = C_1^\frac 1 5 + \frac {3C_0 C_2}{25C_1^\frac 9 5} +  \frac {C_0^2(45 C_3 C_1 - 72 C_2^2)}{625 C_1^{\frac {19}{5}}} + \mathcal O(C_0^3),  \\
w_1 = C_1^\frac 25  -  \frac {6 C_0 C_2}{25C_1^\frac 8 5}  + \mathcal O(C_0^2),\\
T =     \frac {C_0}{C_1^\frac 35} + \mathcal O(C_0^2).
\eea
While it is clear that this recursive procedure determines a formal expansion whose coefficients are analytic at $C_0=0$, it is not clear 
whether the expansion should be convergent. However the above-mentioned theorem guarantees the existence of such analytic expansion and hence 
it must coincide with this formal manipulation. 
\er

In order to translate the normal form (\ref{wconf}) into the desired one (\ref{hzeta}) we need to perform a simple rescaling
\be
w = \varepsilon ^{\frac 25} \le(\frac 54 \ri)^{\frac 2 5 }\zeta \ ,\ \ T = \varepsilon^{\frac 2 5}\le(\frac 45\ri)^{\frac 35} \t.
\ee

The function $\zeta$ is locally univalent in a neighborhood of  $z=\a$ and $\zeta(\alpha)\equiv 0$. The function $\tau$ is analytic in $C_0$ at $C_0=0$. 
Their local behavior is 
\bea
\zeta =\ve^{\frac 2 5 } \le(\frac {5}4  C_1\ri)^\frac 25\le ( 1- \frac {6 C_0 C_2}{25 C_1^2} + \mathcal O(C_0^2)   \ri) (z-\a)(1+ \mathcal O(z-\a)),\\
\tau  = \varepsilon ^{ - \frac 2 5}C_0  \le(\frac 4 {5C_1} \ri)^{\frac 3 5 }  \le(1 + \mathcal O(C_0)\ri).\label{tau}
\eea
The determination of the root is fixed uniquely by the requirement that the image of the main arc (cut) where $\Im h \equiv 0$ be mapped to the {\bf negative real} $\zeta$--axis.

Repeating identical considerations for the behavior of $h$ near $\ov \a$ we define  $\wh \zeta(z; x,t) = \ov {\zeta(\ov z; x,t)}$.
Before we can proceed with the detailed asymptotic analysis of the Riemann--Hilbert Problem \ref{RHPG}, we need to establish more precisely the 
relation between the complex parameter $\t(x,t,\e)$ and the $(x,t)$-plane. 

We shall consider the {\bf scaling limit} in which $\t$ is {\bf uniformly bounded}; this means that $T(x,t)$ in (\ref{wconf})  must tend to zero as a  
$\mathcal O(\varepsilon^{\frac 25})$.  Therefore, we will be considering some shrinking neighborhood of the point of umbilic catastrophe $(x_0,t_0)$, 
which will be determined in more details in Sect. \ref{mapxt}.

\br
Incidentally, one could construct critical initial data for which there is a more degenerate gradient catastrophe 
$\frac i \varepsilon h(z;x_0,t_0) = \frac {i} \varepsilon (z-\a)^{\frac {2k+3}2}(C_k + \mathcal O(z-\a))$ with $C_k\neq 0$. The case $k=0$ correspond to a 
regular (non-gradient catastrophe) point
$(x_0,t_0)$, where  the local parametrix is written in terms of Airy functions. The case $k=1$ is the one under scrutiny now and corresponds to a 
parametrix written in terms of Painlev\'e\ I. For  $k\geq 2$ it is easy to speculate that the PI parametrix needs to be substituted by a member of 
the Painlev\'e\ I hierarchy. This will be investigated elsewhere.
\er

\subsection{The map $(x,t)\mapsto \t(x,t;\varepsilon)$.}
\label{mapxt}
The goal of this section is to determine the dependence of $C_0, \tau,$ etc. on the space-time variables $(x,t)$ 
%\blue
{near the point of graduate catastrophe $(x_0,t_0)$.}
Here and henceforth we use the notation $\a_0=a_0+ib_0=\a(x_0,t_0)$ and  
%\blue
{ $\D x=x-x_0,~\D t=t-t_0$}.

\bt
\label{thmalphaxt}
Near the point of gradient catastrophe $(x_0,t_0)$ the behavior of $\alpha$ is 
\bea
\Delta \a^2 =
{\KK^2}
\big (\Delta x + 2(\a_0+a_0)\Delta t\big ) + \mathcal O(\Delta t^2+ \Delta x^2),\label{Dalph^2}
\\
{{\rm where}~~~~ \KK^2 = 
%\blue
{\frac{8\sqrt{2ib_0}}{15iC_1}}= \le(
\frac i{4\pi } {\int_{\gt_m}{{f''(\z)}\over{(\z-\a_0)R(\z)_+}}d\z} 
\ri)^{-1}}.\label{KKdef}\eea
\et
{\bf Proof.}
The branch-point $\a(x,t)$ is determined implicitly by the modulation equations (\ref{modeqint})
that can be written (see \cite{TVZ1}) as
\bea
\vec F(\a,\ov \a,x,t) = \frac 1{2i\pi} \le[\oint_{\wh \gamma_m} \frac { f'(w)d w}{ R_+(w)} , \oint_{\wh \gamma_m} \frac {  w f'(w)d w}{ R_+(w)}\ri]^T = \vec 0, 
\label{modeqintpr}\\
f(z;x,t) = f_0(z) - z x - 2 t z^2,
\eea
where $\wh \gamma_m$ is a  closed contour around the main arc and $R(z)$ is chosen with the determination that behaves as $z$ for $z\to\infty$.
The Jacobian of $\vec F$ is (\cite{TVZ3}, Lemma 3.4)
\bea
\frac{\pa \vec F}{\pa(\a,\ov \a)} = \frac 1 2 \begin{bmatrix}
\frac {h'(z)}{R(z)} \big|_{z=\a} & \frac {h'(z)}{R(z)} \big|_{z=\ov \a}\\
\a \frac {h'(z)}{R(z)} \big|_{z=\a} & \ov\a \frac {h'(z)}{R(z)} \big|_{z=\ov \a} 
\end{bmatrix}.
\label{jacobmat}
\eea
On account of the Schwartz symmetry $h(z) = \ov {h(\ov z)}$, the determinant of this matrix is 
\be
\det \frac{\pa \vec F}{\pa(\a,\ov \a)} =\frac {ib}2 \le|\frac {h'(z)}{R(z)}\ri|^2_{z=\a}.
\label{jacob}
\ee
For $(x,t)$ away from the gradient catastrophe  we have $h(z) = C_0(z-\a)^{\frac 32}(1+ \mathcal O(z-\a))$ 
with $C_0=C_0(x,t)\neq 0$, so that the Jacobian (\ref{jacobmat}) is invertible  (as long as $b=\Im \a>0$) and the standard implicit function theorem yields 
$\a(x,t)$ as a smooth function of $(x,t)$.   At the point of gradient catastrophe we have $h(z) = C_1(x_0,t_0) (z-\a)^\frac 5 2(1 + \mathcal O(z-\a))$ 
with $C_1=C_1(x_0,t_0)\neq 0$; therefore the matrix (\ref{jacobmat}) is not invertible and -- in fact-- it is the zero matrix.
\begin{remark}\label{remdFdxt}
Note that the Jacobian matrix  $\frac {\pa \vec F}{\pa(x,t) }$ is
\be
\frac {\pa \vec F}{\pa(x,t) } = -\frac 1{2i\pi}\begin{bmatrix}
\oint \frac {d w}{R(w)}  & \oint \frac {4w d w}{R(w)}  \\
\oint \frac {w d w}{R(w)}  & \oint \frac {4w^2 d w}{R(w)}  
\end{bmatrix} = \begin{bmatrix}
1 & 4a \\
a & 4a^2 - 2b^2
\end{bmatrix},
\label{xtJacob}
\ee
where $\a=a+ib$.  {The contour of integration in the integrals above and below is $\wh \gamma_m$ defined below eq. (\ref{gforman}).}
For any $\a\in\C_+$, we have
\be
\det \frac {\pa \vec F}{\pa(x,t) } = -2b^2\neq 0\ .
\ee
\end{remark}

We expand each component $F_j$, $j=1,2$, of $\vec F$ around $\a_0=\a(x_0,t_0)$ and its
complex conjugate (we denote only the dependence on $\a$ with the understanding that $\vec F$ depends also on $\ov \a$) as
\be
F_j(\a;x,t) = F_j(\a_0;x,t ) + \pa_\a F_j\,\Delta \a + \pa_{\ov \a} F_j\,\Delta \ov \a + \frac 1 2 \le[\Delta \a,\Delta \ov \a \ri] H_j  \begin{bmatrix}\Delta \a\\ \Delta \ov \a\end{bmatrix} + \mathcal O(|\Delta\a|^3), 
\label{TaylorF}
\ee
where $\Delta \a=\a-\a_0$ and $H_j$ denotes the Hessian of $F_j$ evaluated at $(\a_0,x,t)$:
\bea
H_j = \frac 1{ {8}\pi i} \begin{bmatrix}
\ds 3\oint \frac { w^{j-1} f'(w) d w}{(w-\a)^2 R_+(w)} & \ds \oint \frac { w^{j-1}  f'(w) d w}{(w-\a)(w-\ov \a) R_+(w)}\\[8pt] 
\ds \oint \frac { w^{j-1}  f'(w) d w}{(w-\a)(w-\ov \a) R_+(w)} &\ds  3\oint \frac { w^{j-1}  f'(w)d w}{(w-\ov \a)^2 R_+(w)} 
\end{bmatrix}.
\label{Hess}
\eea

The off--diagonal entries of $H_j$  vanish because 
\bea
\frac 1{ {8}\pi i}\oint \frac {w^{j-1}  f'(w) \d w}{(w-\a)(w-\ov \a)R_+(w)} = -\frac 1{ {16}\pi b} \le[
\oint\frac {w^{j-1}  f'(w)\d w}{(w-\a)R_+(w)} - \oint\frac {w^{j-1}  f'(w)\d w}{(w-\ov \a)R_+(w)} 
\ri]  =\nonumber \\=
-\frac 1{ {16}\pi b}\le[  z^{j-1}  \frac {h'(z)}{R(z)} \bigg|_{z=\ov \a}^{z=\a}\ri]=0
\label{offdiag0}
\eea
Let us denote by $G_{j-1}$ the $(1,1)$ entry of $H_j$ ($j=1,2$), so that

\bea
G_m= \frac 3 { {8}i\pi} \oint \frac {w^m f'(w)\d w}{(w-\a)^2R_+(w)} = \frac 3{ {4}} \frac {d }{d z}\le(
z^m  \frac {h'(z)}{R(z)}\ri)\bigg|_{z=\a}\ ,\ \ \ m=0,1.
\label{Gjs}
\eea
Using the fact that $\frac {\pa \vec F}{\pa(\a,\ov \a)}=0$  {at $\a=\a_0,~\ov \a=\ov \a_0$},
we then have (we suppress the $(x,t)$ dependence for brevity)
\be
F_j(\a) = F_j(\a_0) +  \Re ( G_{j-1} \Delta\a^2)  + \mathcal O(|\Delta\a|^3).
\ee
Expanding also in $x,t$ near $x_0,t_0$, we obtain
\be
0= F_j(\a) = F_{j,x} \Delta x + F_{j,t} \Delta t +  \Re \le( G_{j-1}  \Delta\a^2\ri)  + \mathcal O(|\Delta\a|^3) + \mathcal O(\Delta t^2 + \Delta x^2).\label{qqwe}
\ee
Equation (\ref{qqwe}) shows that $\Delta \a$ is of order $\sqrt{|\Delta x|+|\Delta t|}$.
Solving equations (\ref{qqwe}) ($j=1,2$) for $\Delta\a$  using the expressions for the $x,t$-derivatives in (\ref{xtJacob}) yields
\be
(\Delta\a)^2 = -\frac 2{G_0 \ov G_1 - G_1 \ov G_0} \le[\le( \ov G_1 
- a \ov G_0\ri)\Delta x + \le( 4a_0\ov G_1 - \le(4a_0^2-2b_0^2\ri)\ov G_0\ri) \Delta t \ri] +  \mathcal O(\Delta t^2+\Delta x^2).
\ee
At $(x_0,t_0,\a_0)$ we have $G_1 = \a_0 G_0$ and hence a simplification of the above equation yields 
\bea
\Delta\a^2 = -\frac {1} {G_0 } \le[\Delta x +2 \le(\a_0+a_0\ri) \Delta t \ri] +  \mathcal O(\Delta t^2+\Delta x^2).
\eea
From (\ref{Gjs}) we find $G_0 =-\frac {15i C_1}{ {8} \sqrt{2ib_0}}$ with $C_1$ defined by (\ref{C1}) and, thus, obtain (\ref{Dalph^2}).
\QED

Let us introduce the scaling variables  $x=x_0 + \varepsilon^{\frac 45} X,\ t =t_0 + \varepsilon^{\frac 45} T$ in the neighborhood
$D$ of $(x_0,t_0)$.
Then, from Thm. \ref{thmalphaxt}
\be (\D \a)^2=(\a -\a_0)^2  =  \varepsilon^{\frac 4 5 }
\KK^2
(X + 2(\a_0 +a_0)T)\big(1 + \mathcal O(\varepsilon^\frac 2 5)\big).
\label{alphaXT}
\ee

\bc
\label{thmC0hat}
In terms of $X,T$, we have
\bea
C_0  &\&=-\varepsilon^{\frac 2 5 }\left(\frac{-10iC_1}{3}\right)^\hf (2ib_0)^\qt \sqrt{X + 2(\a_0+a_0)T} (1  + \mathcal O(\ve^\frac 25)),
\label{C0}\\
\tau &\& ={-\frac {2^\frac {17} {10} i^\frac 3 2 (2ib_0)^\qt}{3^\hf (5C_1)^\frac 1 {10}}}
\sqrt{X + 2(\a_0+a_0)T} (1  + \mathcal O(\ve^\frac 25)),
\label{tauXT}
\eea
where  { $(2ib_0)^\qt$ stands for ${\rm e}^{i\pi/8} (2b_0)^\qt$} ($b_0>0$), all the roots are principal and the argument of $C_1$ is determined in such a way that the direction of the main arc is 
\be
\arg(z-\a_{0}) = \pi - \frac 2 5 \arg (C_1),\label{mainarc}
\ee
 so that the main arc is mapped to the negative real $\zeta$ -- axis\footnote{Since $\zeta = \ve^{-\frac 2 5 } C (z-\a_0)(1 + \dots)$ and $C = (5 C_1/4)^\frac 25$, 
the condition for the main arc is $\arg(\zeta) = \pi$, whence the formula (\ref{mainarc}).}.
Moreover (\ref{alphaXT}) can be written using (\ref{tauXT}, \ref{KKdef}) as 
\bea
\Delta \alpha= \ve^{\frac 25} \frac{\tau}{2C},\  \ ~~{\rm~~where} \label{alphatau}\\
C= \le(\frac {5 C_1}4\ri)^\frac 25 = \le(
\frac {\sqrt{2ib_0}}{6\pi} \int_{\gt_m}{{f''(\z)}\over{(\z-\a_0)R(\z)_+}}d\z
\ri)^\frac 25 .
\label{expressC}
\eea
\ec
{\bf Proof.}
It is known from the  modulation  (Whitham) equations \cite{TVZ1} that 
\be
\pa_x h(z;x,t) = R(z)\ ,\qquad \pa_t h(z;x,t) = 2(z+a) R(z).
%R(z):= \sqrt{ (z-\a) (z-\ov \a)}\ ,\qquad \a  = \a(x,t)  = a+ i b\ .
\label{Riemanninv}
\ee
Expanding in series near $z=\a$ and comparing the terms  we find
\bea
&& \a_x C_0  =-\frac {2i \sqrt{2ib}}{3}\ ,\qquad
\a_tC_0 = -\frac {4i\sqrt{2ib} (\a +a)} {3}
\eea
Since $\a_x = \Delta\a_x $, using Thm. \ref{thmalphaxt}
we have 
\be
C_0 = -\frac {2 i\sqrt{2ib}}{3 \a_x} = -\ve^{\frac 25}\frac {4i \sqrt{2ib_0}}{3\sqrt{\frac{8 \,\sqrt{2ib_0}}{15 i  C_1}} } \sqrt{X+ 2(\a+a)T}\, (1+ \mathcal O(\ve^{\frac 25})) 
\ee
and (\ref{tauXT}) follows from the expression (\ref{tau}). Direct calculations  confirm (\ref{alphatau}).
\QED

\begin{figure}[t]
\begin{center}
\resizebox{0.6\textwidth}{!}{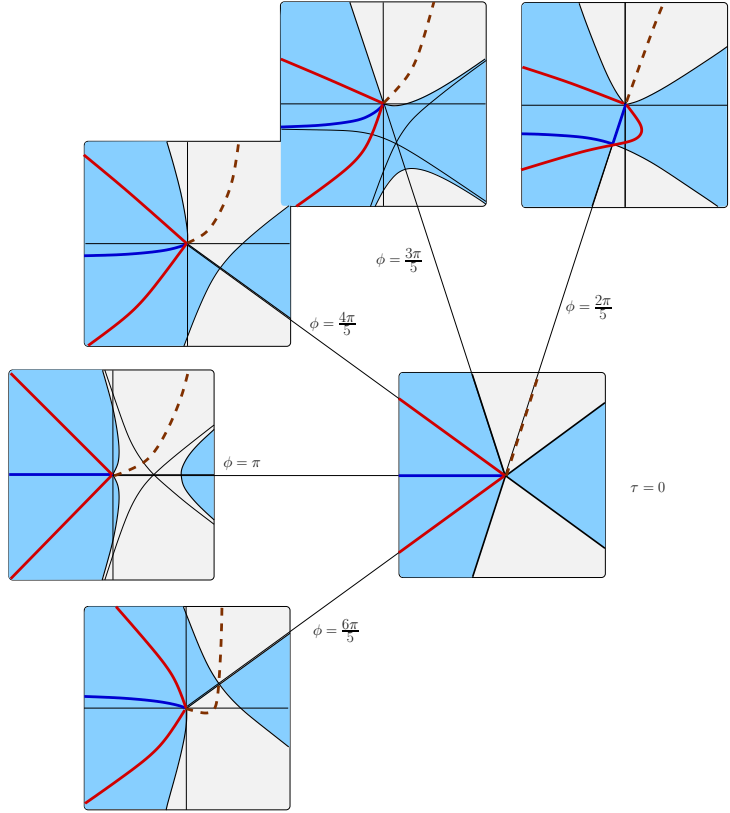}
\caption{
The level lines of the $\Im h$  in the $\zeta$--plane and for different values of 
$\tau$. 
When $\phi:= \arg(\tau) = \frac {2\pi }5$ 
or $\phi = \frac {6\pi}5$  
the complementary arc or one of the rims of the lens is {\em pinched} and 
the solution enters the genus $2$ region. Thus the genus $0$ region corresponds to $\frac {2\pi}5 <\arg(\tau) < \frac {6\pi}5$.
} 
\label{tauphases}
\end{center}
\end{figure}

Corollary \ref{thmC0hat} states that --in the scaling limit-- the map $(x_0 + \varepsilon^{\frac 45} X,t_0 + \varepsilon^{\frac 45} T)\to \t^2$ 
is a map. For definiteness and later purposes we introduce the following definition.

\bd
\label{Painmap}
The map
\be
v(x,t;\varepsilon) = \frac 3 8 \t^2(x,t;\varepsilon)
\ee
will be called the {\bf Painlev\'e\ coordinatization}.
\ed
According to the previous analysis, the function $v(x,t;\ve)$ is a local map in the neighborhood $D$ 
of the point of gradient catastrophe. To this end we formulate the following corollary.
\bc
\label{corvXT}
In terms of the scaling coordinates $X= \frac {x-x_0}{\ve^{4/5}} ,\ T = \frac {t-t_0}{\ve^{4/5}}$ the function $v$ reads, to the leading order as
\be
v =\frac 3 8 \tau^2  =  -i \sqrt{\frac{2ib_{0}}{C}}  \big( X + 2(\a_{0}+a_{0})T\big) (1 + \mathcal O(\ve^{\frac 25})),\label{vXT}
\ee
where $C$ is defined by (\ref{expressC}).
\ec
{\bf Proof}. A simple manipulation from the Def. \ref{Painmap} and eq. (\ref{tauXT}). \QED

\br The complex--valued function $v(x,t;\ve)$ is an approximate linear map from the neighborhood $D$ of $x_0,t_0$ of size $\mathcal O(\ve^\frac 45)$ onto a neighborhood $V$ of the origin $v=0$ uniformly 
bounded (in $\e$); in later sections $v$ will play the role of independent variable for 
the Painlev\'e\  I. 
\er

\subsubsection{The image of the genus two region}

We can now find the opening of the sector $S$ in the $\t^2$-plane (and hence $v$--plane as well) that is the image of the genus two part
of the neighborhood $D$ of the gradient catastrophe point.

The critical value of $\frac{i}\varepsilon h(z;x,t) =\frac 45 \zeta^{\frac 52} +  \tau \zeta^{\frac 32}$ is given by $2 \zeta +\frac 32 \tau =0$, so the critical value $\z_c$ of $\z$ is $\zeta_c = - \frac 3 4  \tau$.
Thus
\be
\frac{i}\varepsilon h(z_c ;x,t)  = \frac 45 \zeta_c^{\frac 52} +  \tau \zeta_c^{\frac 32}  =
\frac 45 \le(- \frac 34 \tau\ri)^\frac 52  + \tau\le(- \frac 34 \tau\ri)^\frac 32 .
\ee 
The breaking curves are determined implicitly by $\Re [ i h(z_{cr};x,t)]=0$ and hence by the condition 
\be
\nu(\t)=\Re\le[ \le(- 3 \tau\ri)^\frac 52  + 5\tau\le(- 3 \tau\ri)^\frac 32 \ri] =2\Re\le[\le(- 3 \ri)^\frac 32 \tau^\frac 52\ri]= 0.
\ee

Care must be exercised due to the presence of the fractional powers: recall that the choice of 
conformal parameter $\zeta$ has been made so that the main arc is mapped to $\zeta<0$; the $\zeta$ image of the critical point,  $\zeta_c$ determines whether we are in the genus zero or two region  as explained presently.

The breaking curves correspond to the first directions where $\nu(\t)=0$ starting from the $\t>0$ or --which is the same--
\be
\cos\le(\frac 5 2 \arg(\zeta_c)\ri)  = 0 \ \ \ \Leftrightarrow \ \ \ 
\varphi = \arg (\zeta_c) = \frac \pi 5 + \frac {2\pi k} 5.
\ee
Thus the two arcs of the breaking curves correspond to  two rays amongst the ones below 
\be
\arg(\tau) \in \le \{ \frac {2\pi k}5, k\in \Z\ri\}\ .
\ee
In order to explain which rays we need to choose we have to consider the topology of the level 
lines of $\Re[ih] = \Re\le[\frac 45 \zeta^{\frac 52} +  \tau \zeta^{\frac 32} \ri]$
for different values of $\tau$. Due to the scale invariance ($\zeta\mapsto \lambda^2 \zeta$, $\tau \mapsto \lambda^3\tau$, \ \  
$\lambda>0$ we can restrict ourselves to studying the argument of $\tau$ only. We will consider $\tau = \frac 43 {\rm e}^{i\phi}$. 
The level lines of the real part  of $i h(z)/\varepsilon$ in the $\zeta$--plane for different values of $\phi$ are plotted in Fig. \ref{tauphases}. 
The transition between the genus $0$ and genus $2$ regions occurs when the connectivity of the complementary arc and/or the rims of the lens needs to change. 
This happens for $\arg\tau = \frac {2\pi}5$ when the complementary arc is pinched between the ``sea'' ($\Re ih<0$) or for $\arg(\tau) = \frac {6\pi}5 $, 
when the main arc is about to break into two arcs.

The above discussion about the directions of the breaking curves can be summarized in the 
following lemma. 
\bl\label{assimD} 
The asymptotic image of the genus zero part of the region $D$ around the point of gradient catastrophe
$(x_0,t_0)$ under the map $v=v(x,t,\e)$ in the limit $\e\ra 0$ is the sector 
\be
\arg(v) \in \le[\frac {4\pi}5 , \frac{ 12 \pi}5\ri]
\ee
in the Painlev\'e\ $v$--plane.
\el
This is so because the argument of $v$ is twice the argument of $\tau$ and from the previous discussion.

The complementary sector of aperture $\frac {2\pi} 5$ is the asymptotic image of the genus--two region in the Painlev\'e\ plane; in the following section we will describe the asymptotics of $q(x,t,\e)$ in terms of the {\em tritronqu\'ee} solution, which has --conjecturally-- poles only in such a sector \cite{DubrovinGravaKlein}.

\br[The angle between the breaking curves in the $(x,t)$--plane]
\label{remvXT}
Since we know now that the breaking curves correspond to the directions $\arg(v) = \frac {2\pi} 5 , \frac {4\pi}5$, we can compute the angle at which the two breaking curves meet at the point of gradient catastrophe. Using  (\ref{vXT}), we calculate 
\be\label{dvxt}
\pa_X v  = \k,~~~~\pa_T v  =2\k(\a + a),\ ,\ \ \ 
\k:= -i \sqrt{\frac{2ib}{C}}=-i\le(\frac{-24\pi b^2}{\int_{\gt_m}{{f''(\z)}\over{(\z-\a_0)R(\z)_+}}d\z}
\ri)^\frac 15. 
\ee
Now we have $v=
\k \le( X+4a T + 2i b T\ri)(1 + \mathcal O(\ve^{\frac 25}))$.

The breaking curves correspond to $\arg(v) = 2\pi/5, 4\pi/5$ in the Painlev\'e\ plane.
Thus, we have $\arg L_1=\frac 2{5} \pi -\arg \k$ and $\arg L_2=\frac {4}{5} \pi -\arg \k    $
respectively. These values of $\arg L$ define the rays
\be 
t = \frac {\tan(\arg L_j)}{ 2b -4a\tan(\arg L_j)} x,~~~~~j=1,2
\ee
on the physical plane that are tangential to the breaking curves at
the point of gradient catastrophe. So,
the angle $\Theta$ in the $(x,t)$--plane between the breaking curves is 
\be\label{wedgeangle}
\tan(\Theta) =\frac{ \frac {\tan(\varphi_2)}{ 2b -4a\tan(\varphi_2)} - \frac {\tan(\varphi_1)}{ 2b -4a\tan(\varphi_1)}  } 
{1 + \ \frac {\tan(\varphi_2)}{ 2b -4a\tan(\varphi_2)} \frac {\tan(\varphi_1)}{ 2b -4a\tan(\varphi_1)} },
\ee
where $\varphi_j=\arg L_j$, $j=1,2$.
\er

\begin{example}\label{examplemapping}
{Let $\hat r_0(z,\e)=\Sscr_\e \hat q(x,0,\e)$ be the reflection coefficient of the initial data  
 (\ref{ourfamily}) for the NLS (\ref{FNLS}), where $\mu\ge 0$, and let $ r_0(z,\e)=e^{\frac {2i}\e f_0(z)\sign(\frac\m 2- z)}$,
where, similarly to Example \ref{exourfamily}, $ f_0(z,\e)=-\lim_{\e\ra 0}\frac{i\e}2\ln \hat r_0(z,\e)$. For solution $q(x,t,\e)$,
defined by the initial data $q(x,0,\e)=\Sscr^{-1}_\e r_0(z,\e)$,}
the point of gradient catastrophe was calculated to be $(x_0,t_0)=\left(0, \frac{1}{2(\m+2)}\right)$, the corresponding value   $\a(x_0,t_0)=
i\sqrt{\m+2}$ and the    slopes of the two breaking curves at the point of gradient catastrophe \footnote{This expression for $m$ is provided in 
Theorem 5.6, \cite{TVZ1}; however, the expression for 
$m$ given in Theorem 1.1, \cite{TVZ1}, should be replaced by its inverse.} - 
\be\label{slopes}
\pm m=\pm\frac{\cot\frac{\pi}{5}}{2\sqrt{\m+2}}~,
\ee
see \cite{TVZ1}. Since the map $v(x,t,\e)$ asymptotically (as $\e\ra 0$) maps breaking curves (of slope $\pm m$) onto the 
rays $\arg v =\frac 85 \pi \pm\frac 45 \pi$
respectively, we have (all angle equations are mod $2\pi$)
\be\label{dirder}
\arg D_{\overrightarrow{u}}=\frac{\k}{\sqrt{1+m^2}}[1\pm2(\a+a)m]=\frac 85 \pi \pm\frac 45 \pi,
\ee
where the vector $\overrightarrow{u}=(1,m)$ and $D_{\overrightarrow{u}}$ denote the derivative in the direction of $\overrightarrow{u}$.
Thus,
\be\label{slopeeq}
\tan^{-1}\frac{2bm}{1+4am}+\arg \k=\frac 25 \pi,~~~~~~~~~~~~\tan^{-1}\frac{2bm}{4am-1}+\arg \k+\pi=\frac {4}{5} \pi.
\ee
Equation (\ref{slopeeq}) defines the slope of the breaking curves at the point of gradient catastrophe $(x_0,t_0)$ in terms of $\a(x_0,t_0)$
and $C_1(x_0,t_0)$. Let us show the slopes (\ref{slopeeq}), found in \cite{TVZ1}, are consistent with Lemma \ref{assimD}.
Substitution of the slopes $\pm m$ from (\ref{slopes}) into  (\ref{slopeeq}) yields
\be \label{kj}
\frac{\pi}{2}\mp\frac{\pi}{5} +\arg \k =\frac 85 \pi \pm\frac 45 \pi.
\ee
We now use (\ref{C1}) to calculate $\k$. Considering for simplicity the solitonless case $\m\geq 2$, we obtain
\be
C_1=\frac{4i\sqrt{2ib}}{15\pi}\int_\R \frac{\Im f''(\z)}{(\z-\a)R(\z)_+}d\z
\ee
where $\Im f'(\z)=\frac{\pi}{2} {\rm sign}\z(1-\chi_{[-T,T]})$, $T=\sqrt{\frac{\m^2}{4}-1}$,
was calculated in \cite{TVZ1}, Sect. 6.4.  {(The choice of branch of $R(z)$  
in (\ref{C1}) and elsewhere in this paper
is  opposite to those used in \cite{TVZ1}. That is why the sign of $\frac{f''(\z)}{R(\z)}$ in (\ref{C1}) is opposite to the one
that would have been calculated according to \cite{TVZ1}). }
Direct calculation of the latter integral yields 
\be
C_1=-\frac{2i\sqrt{2ib}}{15}\le[\frac{1}{(T-\a)R(T)}-\frac{1}{(T+\a)R(-T)}\ri],
\ee
so after some algebra we get $C_1=\frac{32\sqrt{2i}}{15(\m+2)^\frac 94}$. 
Then $\arg C_1=\frac{\pi}{4}+2\pi k$ for some $k\in\Z$.
Taking into account (\ref{alphatau}), we obtain 
\be\label{argK}
\arg \k=-\frac{\pi}{2}+\frac{\pi}{4}-\frac{\pi}{20}\mp\frac{2\pi k}{5}.
\ee
Substitution of (\ref{argK}) into (\ref{kj}) shows that (\ref{kj}) holds with $k=-1$ (mod $5$).  Thus, slopes (\ref{slopeeq}) from \cite{TVZ1} are consistent with Lemma \ref{assimD}. We also conclude that
\be\label{C1mu}
C_1=\frac{32\sqrt{2}}  {15(\m+2)^\frac 94} e^{-\frac 74 i\pi}.
\ee
\end{example}

%\blue
{
\br\label{hgenrem}
The map $v=v(x,t)$ (Painlev\`{e} coordinatization) and the calculation for the image of the breaking curve are valid at any generic
point of gradient catastrophe (when a new main arc emerges from an endpoint of an existing main arc) {\em regardless of the genus} of the  solution,
i.e., regardless of the number of the existing main arcs. Therefore, our analysis can be potentially  extended to points of gradient catastrophe 
where a solution changes genus from $2n$ to $2n+2$ with $n\geq 1$.
\er
}

\subsection{The behavior of the
 phase  $\Phi(x,t)$ near the point of gradient catastrophe}

The genus zero (Whitham) approximation $q_0(x,t;\ve)$ to the semiclassical solution $q(x,t;\ve)$ is the leading approximation and it is valid uniformly in the ``genus zero'' region; its dependence on $x,t$ is determined by the modulation (Whitham) equations \cite{TVZ1}. 

These equations can actually be utilized to extend the definition of $q_0(x,t;\ve)$ beyond the genus zero region where --however-- the actual solution $q(x,t;\ve)$ will have a different behavior (typically of oscillatory nature), {see, for example, \cite{BertolaTovbis1}, where $q_0$ was extended beyond
the breaking curve.}  It will actually turn out that  $q_0(x,t;\ve)$ 
can still be used in a  neighborhood of the point of gradient catastrophe as a ``reference'' for describing the actual behavior of $q(x,t;\ve)$. 
For this reason we briefly analyze $q_0(x,t;\ve)$ near $(x_0,t_0)$.

In the genus zero region, 
the leading order approximation $q_0(x,t,\e)$ of the amplitude and the phase of $q(x,t;\ve)\in\Uscr $ (Def \ref{defUscr}), according to
(\ref{lotg0}), (\ref{phase}), are given by $b(x,t)$ and $4g(\infty;x,t)$ respectively,
where the branch-point $\a(x,t)= a(x,t) + i b(x,t)$ and the $g$-function $g$ was defined by the scalar
RHP (\ref{rhpg}), (\ref{modeq}).
In (\cite{TVZ1} Lemma 4.3, formula (4.43)) it was shown that 
\bea\label{dgxt}
g_x(z;x,t)= \frac 1 2 \le(\sqrt{(z-\a)(z-\ov \a)} -z\ri) \ ,\qquad 
g_t(z;x,t) = (z+a)\sqrt{(z-\a)(z-\ov \a)} - z^2,
\eea
where the determination of the square root is such that they behave like $z$ at infinity\footnote{Note that in \cite{TVZ1} the determination being used is the opposite one.}.
Hence for the phase $\Phi(x,t)=4g(\infty;x,t)$ we have 
\bea\label{diffPhi}
\Phi_x &\&= -2 a(x,t) = -2 \Re(\a), \cr
\Phi_t &\& = -4a^2(x,t) + 2 b^2(x,t) = -2 \Re (\a (\a+a)).
\eea

\bt
\label{thmWhitham} If $q\in\Uscr$ 
{(Def. \ref{defUscr})}
 then the increment of the phase $\Delta \Phi(x,t):= \Phi(x,t)- \Phi(x_0,t_0)$ 
of the modulated plane wave (genus zero) approximation $q_0$ of $q$
 near the point of gradient catastrophe $(x_0,t_0)$ has the expansion
\bea\label{phidelta} 
\Delta \Phi(x,t)   =&\&    -2 \Re \bigg(\a_0 \le[\Delta x   + (\a_0 + a_0) \Delta t \ri] \bigg) -   \Re\le( 
\frac {4 \KK}3 
(\Delta x  + 2(\a_0 +a_0)\Delta t)^{\frac 32}\ri) + \mathcal O(\Delta x^2 + \Delta t^2) \cr
=&\& 
-2a_0\Delta x - 2(2a_0^2 - b_0^2) \Delta t
-   \ve^\frac 65\Re\le( 
\sqrt{\frac {2i}{Cb}} \frac {\tau^3}{8}
\ri) + \mathcal O(\Delta x^2 + \Delta t^2),
\eea
where $\Delta x=x-x_0,~\Delta t =t-t_0$ and $\KK$, $C$ and $\t$ were defined  in (\ref{KKdef}), (\ref{alphatau}) and
in (\ref{tau}, \ref{tauXT}) respectively. 
\et
{\bf Proof}.
If we write $\a = \a_0 + \Delta \a$ we have from (\ref{diffPhi})
\be
\Phi_x = -2 a_0 -2 \Delta a\ ,\qquad 
\Phi_t = -4 a_0^2 + 2b_0^2 - 8 a_0\Delta a + 4 b_0 \Delta b - 4 \Delta a^2 + 2 \Delta b^2.
\ee
In our  problem $\Delta \a =\mathcal O(\ve^\frac 25)$ and hence we can approximate
\be
\Phi_x = -2\Re \a_0 - 2 \Re \Delta\a\ ,\qquad 
\Phi_t = -2 \Re (\a_0 (\a_0+a_0)) - 2\Re\le( { 2(\a_0+a_0) \Delta\a} \ri) + \mathcal O(\ve^\frac 45).
\label{371}
\ee
From (\ref{alphaXT})  {and integration of (\ref{371})} we obtain (\ref{phidelta}).\QED
\br
\label{extendq_0}
The formul\ae\ for $\Phi$ allow us to extend the definition of $q_0(x,t)$ within the genus-2 region using (\ref{lotg0}); 
taking the imaginary part of (\ref{alphatau}) we find that 
\be \label{extendb}
b(x,t) = b_0 +\frac 1 2 \ve ^\frac 25 \Im \le(\frac \tau C\ri) (1+ \mathcal O(\ve^{\frac 25}))
\ee 
and hence 
\bea
 &\& q_0(x,t)=\le( b_0 +\frac 1 2 \ve ^\frac 25 \Im \le(\frac \tau C\ri)\ri)\times \cr
&\& \times \exp \,\frac i\ve \le[ \Phi(x_0,t_0)
 -2 a_0 \Delta x   -2 (2a_0^2-b_0^2)\Delta t -   \ve^\frac 65\Re\le( 
\sqrt{\frac {2i}{Cb_0}} \frac {\tau^3}{8}
\ri) 
\ri](1+ \mathcal O(\Delta x^2 + \Delta t^2))
\eea
In the following we will understand that $q_0(x,t)$, $b(x,t)$, $\a(x,t)$ have been extended as indicated above. It is to be noticed that this extension is discontinuous due to the definition of $\tau$ (\ref{tauXT}) involving a square root. Such ambiguity will not be present in the final formul\ae.
\er

\section{The  Riemann--Hilbert problem for Painlev\'e\ I}\label{sectPRHP}

\begin{wrapfigure}{r}{0.5\textwidth}
\resizebox{0.5\textwidth}{!}{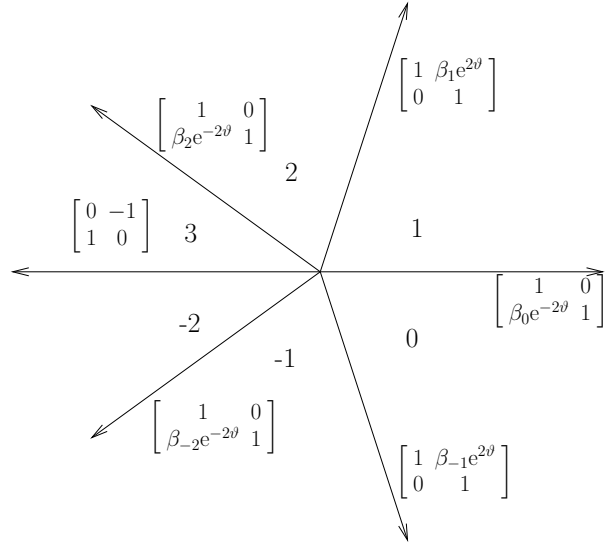}
\caption{The jump matrices for the Painlev\'e\ 1 RHP: here $\vartheta:= \vartheta(\xi;v):= \frac 45 \xi^{\frac 52} - v \xi^{\frac 12}$.}
\label{RHPP1}
\end{wrapfigure}

The heart of the present paper is in the detailed analysis of the ``local parametrix''. This will be constructed in terms of the so--called Psi-function $\Psi(\x,v)$ of the Painlev\'e\ I Lax system,
that depends on the spectral variable $\xi$ and the Painlev\'e\ variable $v$.
The analysis of the Riemann--Hilbert problem for $\Psi(\x,v)$  is contained in a number of papers and books,
see, for example, \cite{KapaevP1,ItsKapaevFokasBook}; this analysis, however, does not cover 
the case when the Painlev\'e\ variable $v$ is at or is approaching a pole $v=v_p$ of the solution to P1 that
is defined through $\Psi(\x,v)$ (as the isomonodromy condition). Furthermore,
it can be shown (\cite{Masoero}) that $\Psi(\x,v)$ has a pole at $v=v_p$.
{\em Analysis of the RHP for $\Psi(\x,v)$ at or close to a pole $v=v_p$ of the tritronqu\'ee} solution (transcendent) $y(v)$ to P1 {\em is a matter of
crucial importance in our study of the height and the shape of the spikes}. 
We start from the summary of the known facts about P1. 
Let the   invertible matrix-function $\P=\P(\z,v)$ be analytic in each sector of the complex $\xi$-plane
shown on Fig. \ref{RHPP1} and satisfy the  multiplicative jump conditions along the oriented boundary of each sector with jump matrices shown on Fig. \ref{RHPP1}. 

The entries of the jump matrices satisfy the following symmetry conditions 
\be\label{jumpsym}
\begin{array}{l}
1+\beta_0\beta_1= - \beta_{-2},\\
1+\beta_0\beta_{-1}= - \beta_{2},\\
1+\beta_{-2}\beta_{-1}=\beta_{1},
\end{array}
\label{betaconditions}
\ee
so that the jump matrices in Fig. \ref{RHPP1} depend, in fact, only on $2$ complex parameters
(that uniquely define a solution to P1).
The matrix function $\P(\z,v)$ is uniquely defined by the following RHP.

\begin{problem}[Painlev\'e\ 1 RHP \cite{KapaevP1}]
\label{P1RHP}
The matrix $\P(\xi;v)$  is locally bounded, admits boundary values on the rays shown in Fig. \ref{RHPP1} and satisfies
\bea
&\& \P_+ = \P_- M,\\
&\& \P(\xi)=\frac{\xi^{\s_3/4}}{\sqrt{2}}
\begin{bmatrix} 1 & -i \cr 1 & i \cr \end{bmatrix}
\left(\1 + O(\xi^{-\hf}) \right), \label{asymppsi}
\eea
where the jump matrices $M=M(\xi;v)$ are the matrices indicated on the corresponding ray in 
Fig. \ref{RHPP1}.
\end{problem}

For any fixed values of the parameters $\beta_k$, Problem \ref{P1RHP} admits a unique solution for generic values of $v$; there are isolated points in the $v$--plane where the solvability of the problem fails as stated and it will need to be modified.

The piecewise analytic function  
\be\label{PsiPain}
\Psi(\xi,v) = \P(\xi,v) {\rm e}^{\vartheta \sigma_3},
\ee
where $\vartheta:= \vartheta(\xi;v)= \frac 45 \xi^{\frac 52} - v \xi^{\frac 12}$,
solves a slightly different RHP with {\em constant} jumps on the same rays. 
The new jump matrices can be obtained from the old ones by replacing the exponential factor in every jump matrix by one.
It then follows that it solves the ODE \cite{KapaevP1}
\be
\frac d {d \xi} \Psi(\xi,v)    = \le[
\begin{array}{cc}
y' & 2\xi^2 + 2 y \xi - v + 2 y^2\\
2 \xi - 2 y & - y'
\end{array}
\ri]\Psi(\xi,v),
\label{P1Lax}
\ee
where $y = y(v)$ solves the Painlev\'e\ I  equation 
\be
y'' = 6 y^ 2 - v.\label{P1ODE}
\ee

Direct computations using the ODE \ref{P1Lax}  and formal algebraic manipulations of series along the lines of \cite{Wasow, JMU1, JMU2, JMU3} show that $\Psi$ admits the following formal solution
\bea
\Psi =&\& \frac{\xi^{\s_3/4}}{\sqrt{2}}
\begin{bmatrix} 1 & -i \cr 1 & i \cr \end{bmatrix}\times \cr
&\& \times
\left(I {-}  \frac {H_I \sigma_3}{\sqrt {\xi}} + \frac { H_I^2\1 + y \sigma_2}{ 2\xi} + 
\frac {( v^2 \, -\, 4 H^3_I - 2y')}{24 \xi^{\frac 32 }}\sigma_3 + \frac {iy' - 2 iH_I y}{4\xi^{\frac 32 }}\sigma_1 + \mathcal O(\xi^{-2})
\right) {\rm e}^{\vartheta \sigma_3},\cr
H_I &\& := \frac 12 (y')^2 + y v - 2 y^3, 
\label{P1expansion}
\eea
where $\mathcal O(\xi^{-2})$ denotes the sum of terms with higher order powers of $\xi^{-1}$.
Such an expansion has to be understood  as representing the asymptotic behavior of an actual solution of the ODE (\ref{P1Lax}) within a sector of angular width smaller than $\frac {4\pi}5$.

\subsection{Failure of the Problem \ref{P1RHP}}
The choice of the parameters $\beta_k$ is (transcendentally)  equivalent to \green{the} choice of Cauchy--initial values for the ODE (\ref{P1ODE}); 
it is known since the original work of Painlev\'e\  that the only (finite) singularities of Eq. (\ref{P1ODE}) are poles and these poles  coincide precisely with the set of exceptional values of $v$ for which Problem \ref{P1RHP} fails to admit a solution. 
From the P1 equation (\ref{P1ODE})  for $y(v)$ one can find the Laurent expansion around any such pole $v=\pole$ to be of the form
\be
y(v) = \frac 1{(v-\pole)^2} + \frac \pole {10} (v-\pole)^2 + \frac 1 6 (v-\pole)^3 + \beta  (v-\pole)^4 + \frac {\pole^2}{300} (v-\pole)^6 + \mathcal O((v-\pole)^7).
\label{TaylorP1}
\ee

We can then proceed as follows \cite{Masoero}:
define the matrix $\wh \Psi(\xi,v)$ via  
\bea
&\&  \Psi(\xi;v ):= (\xi-y)^{ - \sigma_3/2} \le[ 
\begin{array}{cc}
\frac 1 2\le(y' + \frac 1{2(\xi-y)}\ri) & 1\\
1& 0 
\end{array}\ri]\wh \Psi(\xi;v),\cr
&\& \wh  \Psi(\xi;v ):=G(\xi;v) \Psi(\xi;v), \label {Psihat}\\
&\& G(\xi;v ):= \le[ 
\begin{array}{cc}
0 & 1\\
1& -\frac 1 2\le(y' + \frac 1{2(\xi-y)}\ri) 
\end{array}\ri] (\xi-y)^{ \sigma_3/2}.\nonumber  
\eea
It then satisfies the ODE
\bea
\frac {{\rm d}}{{\rm d}\xi} \wh \Psi(\xi;v)
&\&=
 \le[
\begin{array}{cc}
0 & 2 \\
V(\xi;v) & 0
\end{array}
\ri] \wh \Psi(\xi;v)
,\label{hatPsiODE}\\
%:
V(\xi; v)&\& :=2 \xi^3   - v \xi - 2 y^3 + y  v  + \frac 1 2 (y')^2 + \frac {y'}{2(\xi -y)} + \frac 3 {8(\xi-y)^2}.
\eea

It is promptly seen from a direct computation that the function $V(\xi;v)$ admits a limit as $v \to a$  
\bea
V(\xi;v ) &\& = 2\xi^3 - v \xi +\overbrace{ \frac 12 (y')^2 - 2 y ^3 + v y - \frac {y'}{2y}}^{=: \wh H_I}  + \frac {y' \xi}{y(\xi-y)} + \frac 3 {8(\xi-y)^2} =\cr  
&\&= 2\xi^3 - v \xi + \wh H_I  + \frac {y' \xi}{y(\xi-y)} + \frac 3 {8(\xi-y)^2} \to 
2 \xi^3 - \pole \xi - 14 \beta,\\
\wh H_I &\& :=  H_I - \frac {y'}{2 y}  = -14 \beta - \frac \pole 6 (v-\pole)^3 + \mathcal O(v-\pole)^4, \label{hat_Hamiltonian} 
\eea
where the convergence is uniform over compact subsets of the $\xi$ plane \cite{Masoero}.
It was also shown ibidem that $\wh \Psi(\xi;v)$ tends to a finite (holomorphic) matrix $\wh \Psi(\xi; v_p)$ which satisfies the (essentially a scalar ODE)
\be
\frac {{\rm d}}{{\rm d}\xi} \wh \Psi(\xi;\pole) = \le[
\begin{array}{cc}
0 & 2 \\
2 \xi^3 - \pole \xi - 14 \beta   & 0
\end{array}
\ri] \wh \Psi(\xi;\pole) =: \mathcal A(\xi;\pole,\beta)\wh \Psi(\xi;\pole).\label{atpoleODE}
\ee
Most importantly, the
solutions $\wh \Psi(\xi;v)$   to the system (\ref{hatPsiODE}) and    $\wh \Psi(\xi;\pole)$ to the limiting system (\ref{atpoleODE})
have the same Stokes' matrices. In fact, the Stokes' matrices for these solutions are  {\bf the same as those of  $\Psi(\xi,v)$}, except minor changes introduced 
by the obvious nontrivial monodromy of the transformation $G(\xi,v)$ in (\ref{Psihat}).
That follows from the {\bf isomonodromic} property of the equation (\ref{P1Lax}) that defines the P1 equation and the fact that the {\em left} multiplication
by $G(\xi,v)$ does not change the Stokes' phenomenon.

\br
The formal monodromy around $\xi = \infty$  for $\Psi$ is $-i\s_2$ but the one  of $\wh \Psi$ is $i\s_2$ because of the additional monodromy $(-1)$ around $\xi=y$.   Using the  explicit expression (\ref{Psihat}) the reader can also verify that 
\be
\wh \Psi = \frac{\xi^{-\frac 3 4\s_3}}{\sqrt{2}}
\begin{bmatrix} 1 & i \cr 1 & -i \cr \end{bmatrix}
\left(\1 + O(\xi^{-\hf}) \right) {\rm e}^{\vartheta\s_3}\label{normalpsihat}
\ee
\er
\subsection{Analysis in a neighborhood of the pole of PI}

It is essential for our application to  investigate the behavior in which $v\to \pole$ at a certain rate, namely, 
to study how (and in which sense) the limiting expansion of $\wh \Psi(\xi;\pole) $, given by (\ref{psihatform_at_a}) is approached. 
It is proven in Appendix \ref{towards} {(see Theorem \ref{convther} and Corollary \ref{CorFinal})} that 

\bea
\hat\Psi(\xi,v)=\xi^{-\frac 3 4 \s_3}\frac{1}{\sqrt{2}}(\s_1+\s_3)\left(
\begin{bmatrix}1 &0\\0& i\end{bmatrix}
+\mathcal O\left(\xi^{-\hf}, y^{-4},{\rm e}^{-p_2\frac{|y|^{5/2}}{|\xi_0|^{5/2}}}\right)\right)
\le(
\frac {\sqrt{\xi}+ \sqrt{y}}{\sqrt{\xi-y}}
\ri)^{\sigma_3}
{\rm e}^{\vartheta(\xi;v)\sigma_3}
\label{hatpsiasym}
\eea
where $y=y(v)$ satisfies P1 and the term $\mathcal O(\xi^{-1/2})$ is 
{\bf uniform} w.r.t. $v$ in a {\bf finite} neighborhood of the pole $v=\pole$.  { The above expansion has to be properly 
understood under the assumptions stated in Thm. \ref{convther}. In particular, we are going to use it only in the regime where 
$|y|>|\xi|$ and $|y-\xi|$ bounded away from zero;  moreover the above expansion  (\ref{hatpsiasym}) is made on a large 
circle $|\xi|\to \infty$. Indeed, in the application to the construction of the relevant parametrix (Thm. \ref{thmlocalP1}) 
the function $\Psi$ is evaluated on a contour that expands at the rate $\mathcal O(\varepsilon^{-\frac 2 5})$, and the double-scaling is such that $y$ is also growing at the same rate.}

The limiting case of (\ref{hatpsiasym}) for $y=\infty$ (i.e. for $v=\pole$) is 
\bea
\wh \Psi(\xi;\pole) &\&
= \frac{\xi^{-\frac 3 4\s_3}}{\sqrt{2}}
{\begin{bmatrix} -i & -1 \cr -i & 1 \cr \end{bmatrix}}
\left(\1  + 14 \beta  \frac { \sigma_3}{\sqrt {\xi}} + \frac {98 \beta^2 \1 }{ \xi} 
+ \le(\frac {\pole^2}{24} + \frac {1372 \beta ^3}3\ri)
\frac {\sigma_3 }{\xi^{\frac 3 2 } }
+\ri.\nonumber \\
&&+\le.  \le(\frac 7 {12} \pole^2 \beta + \frac {4802}3 \beta^4 \ri)\frac{\1}{\xi^{2}} + \frac \pole {8\xi^2}  \sigma_2 +  \mathcal O(\xi^{-\frac 5 2})
\right){\rm e}^{\vartheta(\xi;v)\sigma_3}
\ .
\label{psihatform_at_a}
\eea
where the expressions for the various coefficients are obtained from the formal solution of the ODE (\ref{atpoleODE}) using the standard techniques in \cite{Wasow}.
\subsubsection{The {\em tritronqu\'ee} transcendent}\label{tritron}

The term {\em tritronqu\'ee} dates back to Boutroux \cite{Boutroux1, Boutroux2}. A generic solution to the ODE (\ref{P1ODE}) has infinitely many poles that accumulate asymptotically for large $|v|$ along the rays $\arg (v) = \frac {2\pi k}5$. Certain one-parameter families (corresponding to the vanishing of one of the Stokes' parameters $\beta_k$ of the associated Riemann--Hilbert problem (\ref{P1RHP}) have the properties that along one of these rays the poles eventually stop appearing as $|v|\to\infty$, or they get {\em truncated}, whence the term {\em tronqu\'ee}. If {\em two} consecutive $\beta_j$'s vanish we have $5$ very special solution for which the poles truncate along {\em three} consecutive rays, whence the naming {\em tritronqu\'ee}.
In fact there are --strictly speaking-- several {\em tritronqu\'ee} solutions: they correspond to the vanishing of the $\beta_j$'s on two {\bf consecutive} rays in Fig. \ref{RHPP1}. 
There are --thus-- $5$ such functions.

However a closer look \cite{KapaevP1}  reveals that the solutions $y(v;\{\beta_j\})$ have the symmetry
\be
y(v;\{\beta_k\}) = {\rm e}^{\frac { 4i  n \pi }5} y\le({\rm e}^{\frac {2i\pi n}5} v; \{\beta _{k+ 2n}\} \ri)\ ,\qquad  n\in \Z\ , \beta_{k+5}:= \beta_{k}
\ee
and hence there is essentially only one {\em tritronqu\'ee} solution. 

\begin{wrapfigure}{r}{0.3\textwidth}
\resizebox{0.3\textwidth}{!}{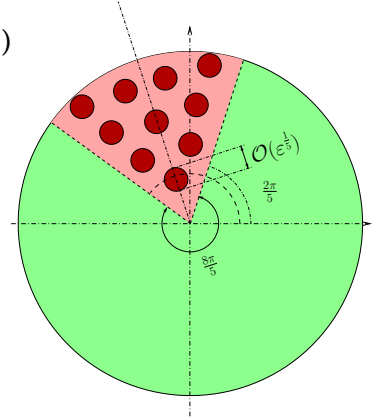}
\caption{The plane of the tritronqu\'ee transcendent, with the poles covered by disks of radius $\varepsilon ^{\frac 1 5}$; 
given that  the $v$--plane is an image of a $\varepsilon^{{\frac 45}}$-size  neighborhood $D$ of the gradient catastrophe point, 
the transversal size of the spikes in the amplitude oscillations of $q(x,t;\varepsilon)$ is $\mathcal O(\varepsilon)$.
%\blue
{It was conjectured in \cite{DubrovinGravaKlein} that all the poles are located in the ``pink'' sector as shown;
this conjecture has not been proven yet, however, the results of our paper do not depend on it.}}
\label{Painplane}
\end{wrapfigure}

Such a solution is characterized by the following theorem.
\bt[\cite{KapaevP1}, Thm. 2.1 and Corollary 2.5 and eqs. (2.72)]\footnote{Note that in \cite{KapaevP1} the independent variable $x$ coincides with our $-v$.}
\label{tritronquee}
There exists a unique solution $y(v)$ corresponding to $\beta_{0} = 0 = \beta_{-1}$ with the asymptotics 
\bea
y = \sqrt{\frac{ {\rm e}^{-i\pi}}6  v} + \mathcal O(v^{-2}) \ ,\ \ \ v\to \infty,\cr 
\arg(v) \in \le[-\frac {6\pi}5+0, \frac{2\pi}5 - 0 \ri].
\eea
Such a solution has no poles for $|v|$ large enough in the above sector (or --equivalently-- has at most a finite number of poles within said sector).
\et
It is conjectured in \cite{DubrovinGravaKlein} that the tritronqu\'ee solution has actually {\bf no poles at all} within said sector: all poles (of which it is known to be infinitely many) lie in the complementary sector, represented in  the shaded area in Fig. \ref{Painplane}.
Such conjecture is so far supported by rather compelling numerical evidence and is consistent with WKB analysis \cite{Masoero}.

We are going to see below that --in fact-- each pole of the {\em tritronqu\'ee} corresponds to a ``spike'' in the asymptotic solution of NLS and such spikes are to be expected only in the region of paroxysmal oscillations (genus $2$).  
{ This correspondence is completely independent of the location of these poles, hence independent of the above-mentioned conjecture}.
While our analysis does not rely in the least on the {\em position} of such poles,  the ``physical intuition'' strongly suggests that indeed they will be confined to the indicated wedge.

\section{Leading order approximation of NLS away from a spike}\label{leadordappr}
As we have seen in Section \ref{mapxt} (Def. \ref{Painmap}), the map $v = v(x,t;\varepsilon)$ maps diffeomorphically a neighborhood $D$ 
of size $\mathcal O(\varepsilon ^{{\frac 45}})$ to the complex-$v$ plane, namely, to the plane of the independent variable of the Painlev\'e\ I {\em tritronqu\'ee} transcendent.

As we have seen in the previous section, the region of the $(x,t)$--plane that corresponds to the genus-$0$ region is mapped to the complement of 
the sector $\arg(v)\in (2\pi /5, 4\pi/5)$.
In the present  section and the following Section \ref{spikepole} 
we shall consider two different  limits in which a point  $(x,t)$ approaches the point of gradient catastrophe 
$(x_0,t_0)$. These limits, express in terms of the map $v(x,t;\varepsilon)$, are:
\begin{itemize}
\item $v(x,t;\varepsilon)$ is in a compact subset  of  the ``swiss--cheese" region $K_\d$, $\d>0$ is constant\footnote{The definition of $K_\delta$ is in Sect. \ref{descreslts}.},  so that $v$ is at least on the distance $\d$ away 
from any  pole of the tritronqu\'ee solution; this is the case considered in this section;
\item $v(x,t;\varepsilon)\in B_\d$, where $B_\d$ is a disk of radius  $\d=O(\e^{\frac 15 +\nu})$, $\nu\geq 0$, centered
at $v=v_p$ - one of the poles  of the tritronqu\'ee  i.e., $v$
can approach a pole $\pole$ of the tritronqu\'ee at a rate  $\varepsilon^{\frac 15}$ or faster; this is the case considered in Section \ref{spikepole}   .
\end{itemize}

Of course, we shall  consider also the case where $v$ is exactly at  a pole (Section \ref{spikepole}), as well as the case when 
$v(x,t;\varepsilon)\ra v_p$ at the rate $O(\e^{\frac 15 -\nu})$, where $\nu\in(0,\frac 15)$ (this section). 
What will transpire from the analysis is the enticing picture sketched in the statements 1 - 5 of Section \ref{descreslts}.

\subsection{Asymptotic behavior  away from the spikes}

In the genus zero region, the leading order solution to the RHP for $Y(z)$, i.e., solution to the model RHP, is (\cite{TVZ1})
\be\label{Psi0}
\Psi_0(z)=
\hf \begin{bmatrix}
-i & -1\cr 1 & i
\end{bmatrix} \left(\frac{z-\a}{z-\bar \a}  \right)^{\frac{\s_3}{4}}
\begin{bmatrix}
i & 1\cr -1 & -i
\end{bmatrix} ~.
\ee
More than its specific form, it is important that near the point $z=\alpha$ it has the behavior
\be\label{leadzeta}
\Psi_0(\z)=\mathcal O(1) \frac{1}{\sqrt{2}}(z-\alpha)^{\frac{s_3}{4}}
\begin{bmatrix}
i & 1\cr 
{-}1 & -i
\end{bmatrix}~, 
\ee
with  the jump matrix $i\s_2= \begin{bmatrix}
0 & 1\cr -1 & 0
\end{bmatrix}$ on the main arc. Here 
$\mathcal O(1)$ denotes a matrix function that is  invertible and analytic in a neighborhood of $\alpha$.

We shall construct an approximation to the matrix $Y(z; x,t,\varepsilon)$ appearing in (\ref{GtoY}) (and henceforth to the matrix $\GGamma$) in the form 
\be
Y(z) = 
\le\{ 
\begin{array}{cc}
\mathcal E(z) \Psi_0(z) & \mbox { for $z$ {\bf outside} of the disks } \mathbb D_\a, \mathbb D_{\ov \a}\\[4pt]
\mathcal E(z) \Psi_0(z)\mathcal P_{\a}(z) & \mbox { for $z$ {\bf inside} of the disk } \mathbb D_\a, \\[4pt]
\mathcal E(z) \Psi_0(z)\mathcal P_{\ov \a}(z) & \mbox { for $z$ {\bf inside} of the disk } {\mathbb D}_{\ov \a},\\[4pt]
 {\mathcal E(z) \Psi_0(z)\mathcal P_{\mu_+}(z)} &  {\mbox { for $z$ {\bf inside} of the disk } {\mathbb D}_{\mu_+}},
\end{array}
\ri.
\label{leadingapprox}
\ee
where $\mathbb D_\a,~\mathbb D_{\ov \a}$ are small  disks centered in $\a$, $\bar \a$ respectively, see Fig. \ref{YRHP}.

\begin{remark}\label{muparametrix}
The existence  of the parametrix  $\mathcal P_{\mu_+}$ in the disk $\mathbb D_{\mu_+}$ centered at the point $z=\mu_+$ together with the uniform estimate
$\1 + \mathcal O(\e)$ on the boundary of $\mathbb D_{\mu_+}$ was established in \cite{TVZ1}. Since $\mathcal P_{\mu_+}$
 does  not affect the accuracy of any of our calculations, we do not discuss it in this paper. 
\end{remark}

Due to the symmetry of the problem in Prop. \ref{propsymmetry} we must have 
\vspace{-10pt}
\be
\mathcal P_{\ov a} (z) = {\le(\mathcal P_\a(\ov z)\ri)^{\star}}^{-1}
\ee
and hence it suffices to consider the construction near the point $\a$ only.
\subsubsection{Local parametrix}\label{locparam}
The local parametrix $\mathcal P_\a(z)  = \mathcal P(z)$ (we understand and suppress the subscript $\a$) must satisfy a certain number of  properties (see Thm. \ref{thmlocalP}),
one of them being the restriction of $\mathcal P(z)$
\be\label{param55}
\mathcal P(z)\bigg|_{z\in \pa\mathbb D} = \1 + o_\varepsilon (1)
\ee
on the boundary of $\mathbb D_\a$,
where $o_\varepsilon (1)$ denotes some infinitesimal of $\varepsilon$, uniformly in $z\in \pa \mathbb D_\a$ and in $(x,t)\in \hat K=v^{-1}(K_\d)$ 
(here a small $\d>0$ is fixed).

If $\mathcal P$ (and the corresponding parametrix near $z=\ov \a$) can be found that satisfy those requirements then the ``error matrix'' $\mathcal E(z)$ is 
seen to satisfy a {\em small--norms} RHP and be uniformly close to the identity. More precisely, the matrix $\mathcal E$ has jumps on:
{\bf (a)} the parts of the lenses and of the complementary arcs 
that lie outside of the disks $\mathbb D_\a, \mathbb D_{\ov \a}$, and; {\bf (b)} on the boundaries of the two disks $\mathbb D_\a, \mathbb D_{\ov \a}$.
The jumps in {\bf (a)} are {\bf exponentially close} to the identity jump in any $L^p$ norm (including $L^{\infty}$) while  on the 
boundary of the disk $\mathbb D_\a$ we have 
\be
\mathcal E_+ (z) = \mathcal E_-(z) \Psi_0(z) \mathcal P(z) \Psi_0^{-1}(z)\bigg|_{z\in \pa \mathcal D_\a} = \mathcal E_- \le(\1 + o_{\varepsilon}(1)\ri)
\label{jumpE}
\ee
Since $\mathcal E(z) = \1 + \mathcal O(z^{-1})$ as $z\to \infty$ it follows  \cite{DKMVZ} that $\|\mathcal E(z)-\1\| \to 0$ (uniformly on the Riemann--sphere) and 
that the rate of convergence is estimated as the same as the $o_\varepsilon(1)$ that appears in (\ref{param55}) as $\e\ra 0$.

Thus, the accuracy of the approximation (i.e. neglecting the term $\mathcal E(z)$) is directly related to the rate of convergence to the identity matrix of the local parametrix $\mathcal P$ on the boundary of the disk(s).

\begin{definition}[Local parametrix away from the spikes.]
\label{deflocalP}
Let $\zeta(z;\varepsilon)$ be the local conformal coordinate introduced in Def. \ref{defzetatau} so that 
\be
\frac i\varepsilon h(z;x,t) = \theta(\zeta;\tau ) = \frac 4 5 \zeta^{\frac 5 2 } + \tau \zeta^{\frac 32}\ . 
\ee
Let $\Psi(\xi;v)$ denote the Psi--function of the Painlev\'e\ I problem with $\beta_0 = 0 = \beta_{-1}$ (and $\beta_{-2} = -1$, $\beta_{2} = -1$),
introduced by (\ref{PsiPain}).
The parametrix $\mathcal P(z)$  is defined by 
\bea
\mathcal P(z) = \frac 1{\sqrt{2}} 
{\begin{bmatrix}
-i & i \\ 1 & 1
\end{bmatrix}}
\zeta^{-\frac {\s_3}4}
\Psi\le ( \zeta +  \frac \t 2; \frac 3 8 \t^2\ri) \begin{bmatrix} 0 & 1\\ -1 & 0 \end{bmatrix}  {\rm e}^{\theta(\zeta;\t) \s_3} \label{localP}.
\eea
\end{definition}

\begin{theorem}\label{thmlocalP}
The matrix $\mathcal P$ satisfies: 
\begin{enumerate}
\item Within $\mathbb D_\a$, the matrix  $\mathcal P(z)$ solves the exact jump conditions on the lenses and on the complementary arc;
\item On the main arc (cut) $\mathcal P(z)$ satisfies 
\be
\mathcal P_+(z) = \sigma_2 \mathcal P_-(z) \s_2\ ,\label{oddjump0}
\ee
so that $\Psi_0 \mathcal P$ within $\mathbb D_\a$ solves the exact jumps on all arcs contained therein  (the left-multiplier in the jump (\ref{oddjump0}) 
cancels against the jump of $\Psi_0$);
\item The product $\Psi_0 (z) \mathcal P(z)$ (and its inverse) are --as  functions of $z$-- bounded within $\mathbb D_\a$, namely the 
matrix $\mathcal P(z)$ cancels the growth of $\Psi_0$ at $z=\a$;
\item The restriction of $\mathcal P(z)$ on the boundary of $\mathbb D_\a$ is 
\be
\mathcal P(z)\bigg|_{z\in \pa\mathbb D_\a} = \1 + \le(H_I + \frac {\t^3}{16}\ri)\frac {\s_3}{\sqrt{\zeta}} 
+ \frac 1 {2\zeta} \le[\le( H_I  + \frac {\tau^3 }{16} \ri)^2\1  + 
{\le(y+\frac \tau 4\ri)} \s_2\ri] + 
\mathcal O(\zeta^{-\frac 3 2}),
\ee
where
\be
v = \frac 3 8 \tau^2,~~~~~~~H_I = \frac 1 2 (y')^2 + y v - 2 y^3 = \int y (s) \d s.
\ee
\end{enumerate}
\end{theorem}

{\bf Proof.}
{\bf (1)} The matrix $\Psi(\xi; v)$ has  {\bf constant jumps} of the same triangularity as the jumps indicated in Fig. \ref{RHPP1} (with $\beta_0 =0=\beta_{-1}$ and $\beta_{-2} = \beta_{2} =-1=-\beta_1$).
{In particular, these jumps can be arbitrarily shifted by any (finite) amount so as they consist of rays originating from $\zeta = 0$ rather than (as it would appear) from $\zeta = - \frac \tau 2$.}
These are altogether of  the opposite triangularities we need, hence the second-last (constant) matrix in (\ref{localP}). 
The last multiplication with ${\rm e}^{\theta(\zeta;t)} = {\rm e}^{\frac i \varepsilon h(z;x,t)}$ gives the exact (non-constant) jumps on the 
parts of the complementary/main arcs and lenses within the disk $\mathbb D_\a$.  On the other hand, the  matrix
\be
\mathcal F(z):= \frac 1{\sqrt{2}} 
{
\begin{bmatrix}
-i & i \\ 1 & 1
\end{bmatrix}}
\zeta(z)^{-\frac {\s_3}4}
\ee
has the jump $\mathcal F_+(z) = i\s_2 \mathcal F_-(z)$ on the {\bf left}, whence the part {\bf(2)}. 

As for part {\bf (3)}, the product $\Psi_0(z) \mathcal P(z)$ is a bounded function of $z$ because the singularities of $\Psi_0(z)$ are canceled by 
those of $\mathcal F(z)$
\bea
\Psi_0(z) \mathcal F(z) = \hf \begin{bmatrix}
-i & -1\cr 1 & i
\end{bmatrix} \left(\frac{z-\a}{z-\bar \a}  \right)^{\frac{\s_3}{4}}
\begin{bmatrix}
i & 1\cr -1 & -i
\end{bmatrix}  \frac 1{\sqrt{2}}
{
\begin{bmatrix}
-i & i \\ 1 & 1
\end{bmatrix}}
\zeta(z)^{-\frac {\s_3}4} = \cr
= \hf 
{
\begin{bmatrix}
-i & i \\ 1 & 1
\end{bmatrix}}
\left(\frac{z-\a}{z-\bar \a}  \right)^{\frac{\s_3}{4}}
\zeta(z)^{-\frac {\s_3}4} = \mathcal O(1)
\eea
since $\zeta(z) =\mathcal O(z-\a)$. In fact,  we see that the product is actually analytic.

Finally, part {\bf (4)} follows from the asymptotics of $\Psi(\xi;v)$. Indeed, for $z\in \pa\mathbb D_\a$ the conformal coordinate $\zeta(z)$ grows 
(homothetically) as $\varepsilon ^{-\frac 2 5 }$ and hence we can use the expansion (\ref{P1expansion}) for $\Psi(\xi;v)$ near infinity. 
To see how it works let us recall the notation 
\be
\vartheta(\xi;v):= \frac 45 \xi^{\frac 5 2} - v \xi^{\frac 1 2}\ ,\ \ \ v=   \frac 3 8 \t^2,
\ee
so that we can write
\bea
\mathcal P(z) =  \frac 1{{2}}
{
\begin{bmatrix}
-i & i \\ 1 & 1
\end{bmatrix}}
\zeta^{-\frac {\s_3}4} 
\le(\zeta + \frac \t 2 \ri)^{\sigma_3/4} 
\begin{bmatrix}
1 & -i\\ 1& i 
\end{bmatrix} 
\le( \1 - \frac {H_I\s_3}{\sqrt{\zeta+ \frac \t 2}} +\frac{H_I^2\1+ y\s_2}{2\zeta+\tau}  + \mathcal O(\zeta^{-\frac 32}) \ri)\times \cr
\times 
{\rm e}^{\vartheta(\zeta+\frac \t 2; v)\s_3} \begin{bmatrix} 0 & 1\\ -1 & 0 \end{bmatrix}  {\rm e}^{\theta(\zeta;\t) \s_3}
\\
=  \underbrace{\frac 1{{2}}
{
\begin{bmatrix}
-i & i \\ 1 & 1
\end{bmatrix}}
\le(
\frac{\zeta + \frac \tau 2 }{\zeta} \ri)^{\sigma_3/4}
{
\begin{bmatrix}
i & 1 \\ -i & 1
\end{bmatrix}}
} _{\ds =\1 +\frac \tau {8\zeta} \s_2 + \mathcal O(\zeta^{-2})}  
\le( \1 + \frac {H_I\s_3}{\sqrt{\zeta+ \frac \t 2}} + \frac{H_I^2\1+ y\s_2}{2\zeta+\tau}  + \mathcal O(\zeta^{-\frac 32})  \ri) {\rm e}^{\le(\theta(\zeta;\t) - \vartheta(\zeta+\frac \t 2; v)\ri) \s_3}=\cr
= 
\le( \1 +  \frac {H_I\s_3}{\sqrt{\zeta}} +\frac{H_I^2\1+ (y+\tau/4)\s_2}{2\zeta}  + \mathcal O(\zeta^{-\frac 32})  \ri) {\rm e}^{\le(\theta(\zeta;\t) - 
\vartheta(\zeta+\frac \t 2; v)\ri) \s_3}.\label{517}
\eea
We also have 
\bea
\theta(\zeta;\t) - \vartheta\le(\zeta+\frac \t 2; -\frac 3 8 \t^2\ri)= \frac 45 \zeta^{\frac 52} + \t \zeta^{\frac 32} - \frac 45 \le(\zeta  + \frac   \t 2\ri)^{\frac 52} + \frac 38 \t^2 
\le(\zeta+\frac \t 2\ri )^\frac 1 2 = \frac {\t^3}{16\sqrt{\zeta}} + \mathcal O(\zeta^{-\frac 32}),
\label{expasy}
\eea
so that --continuing from (\ref{517})-- we have 
\bea
\mathcal P(z)\bigg|_{z\in\pa\mathbb D_\a} &\& = \1 + \le(H_I + \frac {\t^3}{16}\ri)\frac {\s_3}{\sqrt{\zeta}} 
+ \frac 1 {2\zeta} \le[\le( H_I^2 + \frac {\tau^6}{256}  + \frac {\tau^3 H_I}8 \ri)\1  + \le(y+\frac \tau 4\ri) \s_2\ri] + 
\mathcal O(\zeta^{-\frac 3 2}) =\nonumber \\
 &\& = 1 + \le(H_I + \frac {\t^3}{16}\ri)\frac {\s_3}{\sqrt{\zeta}} 
+ \frac 1 {2\zeta} \le[\le( H_I  + \frac {\tau^3 }{16} \ri)^2\1  + \le(y+\frac \tau 4\ri) \s_2\ri] + 
\mathcal O(\zeta^{-\frac 3 2}).
\eea
\QED

At this point we already know that the error term $\mathcal E(z)$ in (\ref{leadingapprox}) is within $\mathcal O(\varepsilon^{\frac 1 5})$ from the identity; 
if we simply ignore it, we would get the leading order approximation to $Y$ and --in turn-- to $\GGamma$, which would produce the leading order approximation of the 
NLS solution $q(x,t,\e)$. Next, we will find the first subleading approximation by solving the first step in the iterative approximation of the error term itself.

\subsection{Subleading correction}

\bt
\label{mainthmgen0}
The behavior  of  a solution $q\in \Uscr$  to the focusing NLS (\ref{FNLS}) in the domain $D$ of the point of gradient catastrophe (scaled like $\e^\frac 45$) is given by 
\bea
q(x,t,\e) &\& =
b\le[1 -2 \ve^{\frac 25} \Im \le(\frac {y(v)}{C b}\ri)
+\mathcal O(\varepsilon^{\frac 35})\ri]\times\cr
&\& \times
\exp  \frac{i}{\varepsilon}\le[ \Phi(x_0,t_0) - {2} \le( a \Delta x +(2a^2-b^2) \Delta t \ri) 
+ 2 \ve^\frac 65\Re\le(\sqrt{\frac{2i}{C b }}H_I(v)   \ri) \ri],\label{620}
\eea
where $\mathcal O(\varepsilon^{\frac 35})$ term is uniform ``away from  spikes'', i.e., is uniform in $(x,t)\in\hat K_\d=v^{-1}(K_\d)$
with some fixed $\d>0$, namely, as long as $v(x,t,\epsilon)$ remains uniformly bounded away from all the poles of the tritronqu\'ee solution (wherever these might be).
Here $\a(x_0,t_0)=a+ib$, $C = \le( \frac {5 C_1}4\ri)^\frac 25$ given by (\ref{expressC}),
\be
H_I = \frac 1 2 (y'(v))^2 + v y(v) - 2y^3(v)
\ee
is the Hamiltonian of the Painlev\'e\ I equation, evaluated along the  {\em tritronqu\'ee} solution appearing in Theorem. \ref{tritronquee},
while $v$ can be expressed (Corollary \ref{corvXT}) as 
\bea
v &\&=  -i\e^{-\frac 45} \sqrt{\frac {2ib}{C}} \bigg(\Delta x  +2(\a+a)\Delta t \bigg)(1 + \mathcal O(\varepsilon^\frac 1 5))
\eea
uniformly in $D$.
\et
%\red{[ I suggest to replace $\Phi_0$ with $\Phi(x_0,t_0)$]}
\br
The class of solution $\Uscr$, for which Theorem \ref{mainthmgen0} is valid, can be extended according to Remark \ref{rem2}.
\er
\br[Comparison with the Conjecture from \cite{DubrovinGravaKlein}]
\label{remarkConjecture}
The approximation formula (\ref{620}) ``away from the spikes'' 
is consistent (but not a proof since our initial data are different) with the conjecture from \cite{DubrovinGravaKlein} about the behavior of the amplitude and the phase of the genus zero (modulated plane wave)
approximation $q_0(x,t,\e)$ in the genus zero (non-oscillatory) part of the neighborhood $D$ of the point of gradient catastrophe.
This conjecture can be written as 
\be\label{dubconj}
U+ i \sqrt{U_0} V  = U_0+ i \sqrt{U_0} V_0 + \ve^{\frac 25} {K}  y(v) + \mathcal O(\ve^\frac 45),
\ee
where $U= |q|^2,\ \ V=\ve \pa_x \arg(q)$, with $K$ being a (complex)  constant.
Using our expression (\ref{620}), $U_0=b^2$,  and the fact that $H'_I(v) = y(v)$ we find 
\be
U = b
^2 - 4 \ve^{\frac 2 5} \Im \le(\frac {b y(v) }{C}\ri) + \mathcal O(\ve^\frac 35),  \ \qquad
V = -2a +  4\ve^{\frac 25} \Re \le(\frac { y(v)}{C} \ri),
\ee
so that 
\bea
U+ib V &\& = b^2 - 4 \ve^{\frac 2 5}b  \Im \le(\frac {y(v) }{C}\ri)   -2iab +  
4i\ve^{\frac 25} b\Re \le(\frac {  y(v)}{C} \ri)+ \mathcal O(\ve^\frac 35)=\nonumber \\
&\& = b^2 -2iab + \ve^\frac 25 \frac {4ib} C y(v) + \mathcal O(\ve^\frac 35).\label{ourconj}
\eea
%\red{[there is no need to give a number for each line of this formula.]}

To replace the error estimate $\mathcal O(\ve^\frac 35)$ in (\ref{ourconj}) with  the  estimate $\mathcal O(\ve^\frac 45)$
from the conjecture (\ref{dubconj}), if correct, would require calculation of the higher order corrections to the RHP (\ref{GtoY}).  
It may be true that the approximation of $q$  is in powers of $\ve^{\frac 25}$ rather than in powers of $\ve^{\frac 15}$, but
that, again,  would require additional analysis. 
The situation here {may resemble the} analogous statements in random matrix 
theory in regards to the expansion of the partition function in {\em even} powers  $1/N$. We also did not dwell into 
the notation of \cite{DubrovinGravaKlein} to compare all the constants used.
Finally, we omitted the term proportional to $(t-t_0)^2$ in (\ref{dubconj}) because, in our scaling, it is of 
order $\ve^\frac 45$ and hence not ``visible'' at this order of approximation.
\er

{\bf Proof of Thm. \ref{mainthmgen0}}.
%\blue
{During the proof, it will be convenient to use the notation $\a_0=a_0+ib_0=\a(x_0,t_0)$, while reserving the notation
$\a=a+ib=\a(x,t)$ for $x,t$ in a vicinity of $x_0,t_0$.}
In a RHP of the form 
\bea
\mathcal E(z) = \1 + \mathcal O(z^{-1}) \ ,\ \ \ \ z\to \infty,\\
\mathcal E_+(z) = \mathcal E_-(z) \bigg(\1 + \Delta M(z)\bigg)
\eea
the solution (if it exists) can be written as 
\be
\mathcal E(z) = \1 + \frac 1{2i\pi}\int \frac {\mathcal E_-(s) \Delta M(s) {\rm d}s}{s-z},
\ee
where the integration extends over all contours supporting the jumps (it is a simple exercise using Sokhotskii--Plemelji formula to 
verify that the above singular integral equation is equivalent to the Riemann--Hilbert formulation). If --in addition-- the term $\Delta M(z)$ 
is sufficiently small in the appropriate norms (at least in $L^\infty$ and $L^2$ of the contours) then the above formula can be used in 
an iterative approximation approach, where 
\bea
\mathcal E^{(0)}(z)&\& \equiv \1,\\
\mathcal E^{(j+1)}(z) &\& =  \1 + \frac 1{2i\pi}\int \frac {\mathcal E_-^{(j)} (s) \Delta M(s) {\rm d}s}{s-z}\ ,\qquad j=0,1,\dots,
\label{resE}
\eea
which can be shown to converge to the desired solution. 

In the RHP for $\mathcal E(z)$, stated in Section \ref{locparam},  we have  $\Delta M$ exponentially small (in $\e$) in any $L^p$ norm 
on all parts of the contour outside the disks $\mathbb D_\a$ and $\mathbb D_{\bar\a}$,
and approaching zero in the $L^\infty$ norm  on the boundaries $\partial\mathbb D_\a, \ \partial\mathbb D_{\ov \a}$.
The latter estimate is valid  in any  $L^p$ norm  due to compactness.  We shall thus find the first correction term in the above approximation procedure. 

As noted in  Theorem \ref{thmlocalP}, part 4,  and in (\ref{jumpE}), the jump of $\mathcal E$ on $\partial\mathbb D_\a$ is 
(note that $\Psi_0$ commutes with $\s_2$)
\bea
\mathcal E_{+}(z) =&\& \mathcal E_{-}(z)\Psi_{0}(z) \mathcal P_\a^{-1}(z) \Psi_{0}^{-1}(z),
\eea
where
\bea
\Psi_{0} \mathcal P_\a^{-1} \Psi_{0}^{-1} = &\& 
\1 -   \le(H_{I}+\frac {\t^{3}}{16}\ri)\frac 1 2 
\le(\sqrt{\frac {p}\zeta} \begin{bmatrix}
1 & i\\
i & -1
\end{bmatrix} +
\frac 1 {\sqrt{p \zeta}} \begin{bmatrix}
1 & -i\\
-i & -1
\end{bmatrix}
\ri)+\nonumber \\
&\&  
+\frac 1{2\zeta} \le[\le(H_I+ \frac {\tau^3}{16} \ri)^2\1 - \le(y + \frac \tau 4\ri) \s_2 \ri]
+ \mathcal O(\varepsilon^{\frac 3 5} )=
\label{534}\\
&\& 
= \1 - \frac {u_1}2 \le(\sqrt{\frac p\zeta} N_1^\star + \frac 1{\sqrt{p\zeta}} N_1\ri) + \frac 1{2\zeta} \le(u_1^2 \1 - u_2 \s_2\ri),
\label{DONE2}\\
&& p:= \frac {z-\a}{z-\ov\a},\qquad 
{ N_1:= \begin{bmatrix} 1 & -i\\ -i &-1\end{bmatrix}},\\
&&
{u_1 := H_I + \frac {\tau^3}{16}\ ,\qquad u_2:=  
{\le(y + \frac \tau 4\ri)}}.
\eea

Due to the symmetry, the jump  on the disk around $z=\ov \a$ is 
\bea
\Psi_0 \mathcal P_{\ov \a}^{-1} \Psi_0^{-1} &\& =
 {\1 + \frac {\ov u_1}2 \le(\sqrt{\frac {p^\dagger}{\zeta^\dagger} } N_1 + \frac 1{\sqrt{p^\dagger\zeta^\dagger}} N_1^\star \ri) + 
\frac 1{2\zeta^\dagger} \le(\ov u_1^2 \1 + \ov u_2\s_2\ri) + \mathcal O(\ve^\frac 35)},
\\
&&\zeta^\dagger(z) := \ov{\zeta(\ov z)}\ ,\ \ p^\dagger = \ov{p(\ov z)} = \frac 1 p.\ 
\eea

We now proceed to the computation of the first-order correction to $\mathcal E(z)$ according to the formula (\ref{resE}). 
In that formula the integral should extend to all the jumps of $\mathcal E$, which include the lenses, the complementary arcs and the disk around $\mu_{+}$; 
the former are exponentially small and the latter is of order $\mathcal O(\varepsilon)$, thus they can be neglected altogether to within this order.

When $z$ is outside the disks this residue computation annihilates the analytic term  with $\sqrt{p/\zeta}$ in (\ref{DONE2}) and  yields  (here $\mathcal C_{\a}, \mathcal C_{\ov \a}$ denote two counterclockwise small circles of radius $\delta>0$ around $\alpha, \ov \alpha$, respectively) 
\bea
&\& \mathcal E^{(1)}(z) =\1  -  
\frac {u_1}{4i\pi}  
N_1
\oint_{\mathcal C_\alpha} \sqrt{\frac {t-\ov \a} {\zeta(t) (t-\a)}}\frac {d\, t}{t-z} +
\frac {\ov  u_1}{4i\pi} 
N_1^\star 
\oint_{\mathcal C_{\ov \a}} \sqrt{\frac {t- \a} {\wh \zeta(t) (t-\ov \a)}}\frac {d\, t}{t-z} +\nonumber \\
&\&  + \le(\frac{ u_1^2\1 -
{u_2} \s_2}2  \oint_{\mathcal C_{\a}} \frac 1{\zeta(t)(t-z)}\frac{\d t}{2i\pi}  +\frac{\ov u_1^2  +  
{\ov u_2}\s_2}{2}  \oint_{\mathcal C_{\ov \a}} \frac 1{\zeta^\dagger (t)(t-z)}\frac{\d t}{2i\pi}\ri)
 +\mathcal O(\varepsilon^{\frac 35}) =\nonumber \\
 =&\& \1 +  \frac {\varepsilon^{\frac 15} u_1N_1 }{2(z-\a)} \sqrt{\frac {\a-\ov\a}{C}}
    - \frac {\varepsilon^{\frac 15}\ov u_1 N_1^\star  }{2(z-\ov  \a)} \sqrt{\frac {\ov\a-\a}{\ov C}}
{-\ve^\frac 25  \le(\frac { u_1^2\1 - 
{u_2}\s_2}{2C(z-\a)} + \frac {\ov u_1^2\1 + 
{\ov u_2}\s_2 }{2\ov C(z-\ov \a)} \ri) +}
\mathcal O(\varepsilon^{\frac 35}).
\eea
So,
\bea
\mathcal E^{(1)}(z)=&\&  \1  +
{\frac{u_1}2\sqrt{\frac {2ib}C} }
 \frac {\varepsilon^{\frac  15} \,N_1}{z-\alpha} - 
 { \frac{\ov u_1}2\sqrt{\frac {-2ib}{\ov C}}}    \frac {\varepsilon^{\frac  15} \, N_1^\star }{z-\ov \alpha}  
  {-\ve^\frac 25  \le(\frac { u_1^2\1 - 
  {u_2} \s_2}{2C(z-\a)} + \frac {\ov u_1^2\1 + 
  {\ov u_2}\s_2 }{2\ov C(z-\ov \a)} \ri) } + 
 \mathcal O(\varepsilon^{\frac 35}),\label{DONE3}
\eea
where --by definition-- $C$ appears as $\zeta(z) = \ve^{\frac 25} C(z-\a)(1 + \dots)$.
We now need to use once more (\ref{resE}) with the expression (\ref{DONE3} and retain only the terms up to $\ve^\frac 25$): 
\bea
\mathcal E^{(2)}(z)  = \mathcal E^{(1)}(z)  + 
\frac {\ve ^{\frac 25} }{4} \le(\frac {u_1^2 N_1N_1^\star}{C(z-\a)} + \frac {\ov u_1^2 N_1^\star N_1}{\ov C(z-\ov a)}\ri)+ \frac {i |u_1|^2 \ve^{\frac 25} }
{4|C|}\le(\frac {N_1^\star N_1} {z-\a}   - \frac {N_1N_1^\star} {z-\ov \a}\ri).
\eea

Hence the approximation of the solution $Y(z)$ is 
\bea
Y(z) &\&(z) =\mathcal E(z) \Psi_0(z) = 
\mathcal E(z)
\hf \begin{bmatrix}
-i & -1\cr 1 & i
\end{bmatrix} \left(\frac{z-\a}{z-\bar \a}  \right)^{\frac{\s_3}{4}}
\begin{bmatrix}
i & 1\cr -1 & -i
\end{bmatrix} ~.
\eea
Writing $\mathcal E(z) = \1 +\frac{ \mathcal E_1}z + \mathcal O(z^{-2})$ near $z=\infty$, we have 
\bea
Y =  \mathcal E(z)\le(\1 - \frac {\alpha-\ov \alpha}{4z} \le[\begin{array}{cc}
0&  -i\\
i&0
\end{array}\ri] + \mathcal O(z^{-2} )\ri) =\nonumber \\
=  \1 +   \frac {\mathcal E_1} z +\frac {1}{2 z} \le[\begin{array}{cc}
0&  -b\\
b&0
\end{array}\ri]+ \mathcal O(z^{-2}).
\eea
The correction comes from the $(1,2)$ entry  of $\mathcal E_1$; we use the second iteration $\mathcal E^{(2)}$ and 
\bea
\le[
{  \frac{u_1\ve ^\frac 1 5 }2\sqrt{\frac {2ib}C} }N_1 - 
{ \frac{\ov u_1\ve ^\frac 1 5 }2\sqrt{\frac {-2ib}{\ov C}}}  N_1^\star
 {-\ve^\frac 25  \le(\frac { u_1^2\1 - 
 {u_2}\s_2}{2C} 
+ \frac {\ov u_1^2\1 + 
{\ov u_2}\s_2 }{2\ov C} \ri) } +\ri.   \cr 
\le.+\frac {\ve ^{\frac 25} }{4} \le(\frac {u_1^2 N_1N_1^\star}{C} 
+ \frac {\ov u_1^2 N_1^\star N_1}{\ov C}\ri)
+ \frac {i |u_1|^2 \ve^{\frac 25} }{4|C|}\le(
\underbrace{{N_1^\star N_1}  -  {N_1N_1^\star}}_{=4\s_2}\ri)
\ri]_{12} =
\cr
=  - i \varepsilon^{\frac 1 5} \frac {u_1}2 \sqrt{\frac {2ib}{C}} 
-  i \varepsilon^{\frac 1 5} \frac {\ov u_1}2 \sqrt{\frac {-2ib}{\ov C}} 
+ 
{\ve^{\frac 25} \Im \le(\frac {
{u_2}}C\ri)     
+\ve^{\frac 25} \frac {i u_1^2}{2C} -\ve^{\frac 25}  \frac {i \ov u_1^2}{2\ov C}  + \ve^{\frac 25} \frac {|u_1|^2}{|C|}
}
=
\\= -i\varepsilon^{\frac 15 }
\Re\le(\sqrt{\frac{2ib}C}u_1  \ri)  +
{\frac {\ve^{\frac 25}}4 \le(\frac {2i u_1^2}{C} -\frac {2i \ov u_1^2}{\ov C}  + \frac {4|u_1|^2}{|C|}\ri)  + \ve^{\frac 25} \Im \le(\frac {
{u_2}}C\ri)
}=\nonumber \\
= - i b \varepsilon^{\frac 15 }
\Re\le(\sqrt{\frac{2i}{C\, b}}u_1  \ri)  +
{b \ve^{\frac 25} \Re\le(\sqrt{\frac{2i}{C\,b }}u_1  \ri)^2  + \ve^{\frac 25} \Im \le(\frac {
{u_2}}C\ri).
}
\eea
We thus have
\bea
q(x,t,\e)&\&= -2 \lim_{z\to\infty} z\GGamma_{1,2}(z) =-2  {\rm e}^{\frac{4i}\varepsilon g(\infty;x,t)} \lim_{z\to\infty} z \le(\mathcal E \Psi_{0}\ri)_{12} = \cr
&\& =
{\rm e}^{\frac{i}\varepsilon \Phi (x,t)} \le(b(x,t) 
+ 2  i b \varepsilon^{\frac 15 }
\Re\le(\sqrt{\frac{2i}{C b}}u_1  \ri)  
-2
{b \ve^{\frac 25} \Re\le(\sqrt{\frac{2i}{Cb }}u_1  \ri)^2  
-2 \ve^{\frac 25} \Im \le(\frac {
{y + \frac \tau 4}}C\ri)}
+
\mathcal O(\varepsilon^{\frac 35})\ri)=\cr
&\&= \exp\le[\frac{i}\varepsilon \Phi (x,t)+ 2  i  \varepsilon^{\frac 15 }
\Re\le(\sqrt{\frac{2i}{C b}}u_1  \ri) \ri]  \le(b(x,t) 
-2 \ve^{\frac 25} \Im \le(\frac {
{y + \frac \tau 4}}C\ri)
+
\mathcal O(\varepsilon^{\frac 35})\ri).\label{miracle}
\eea
Now care must be exercised before factoring  $b(x,t)$ out: indeed from (\ref{alphatau}) we see that 
\be
b(x,t) = b_0 +\frac 1 2 \ve ^\frac 25 \Im \le(\frac \tau C\ri)
\ee 
and thus the $\tau$ in (\ref{miracle}) cancels out and we obtain 
\bea
q(x,t,\e) &\& =
b_0
\exp\le[ {\frac{i}\varepsilon \Phi (x,t)} + 2i \ve^\frac 15\Re\le(\sqrt{\frac{2i}{C b }}u_1  \ri) \ri]\le[1 
 {-2 \ve^{\frac 25} \Im \le(\frac {y}{C b}\ri)}
+
\mathcal O(\varepsilon^{\frac 35})\ri] =\nonumber \\
=
b_0&\&
\exp\le[ {\frac{i}\varepsilon \Phi (x,t)} + 2i \ve^\frac 15\Re\le(\sqrt{\frac{2i}{C b }}\le(H_I +  \frac {\tau^3}{16}\ri)  \ri) \ri]\le[1 
 {-2 \ve^{\frac 25} \Im \le(\frac {y}{C b}\ri)}
+
\mathcal O(\varepsilon^{\frac 35})\ri]=\nonumber \\
=
b_0&\&
\exp\le[ {\frac{i}\varepsilon \Phi (x,t)}+ i\ve^\frac 15 \Re \le(\sqrt{\frac{2i}{C b }}\frac {\tau^3}8 \ri) + 2i \ve^\frac 15\Re\le(\sqrt{\frac{2i}{C b }}H_I   \ri) \ri]\le[1 
{-2 \ve^{\frac 25} \Im \le(\frac {y}{C b}\ri)}
+
\mathcal O(\varepsilon^{\frac 35})\ri].\label{almost}
\eea
Note that we could also replace the remaining occurrences of $b$ by $b_0$  in (\ref{almost}) since it would affect the result by terms of order $\mathcal O(\ve^\frac 35)$.
Recall here the expression (\ref{phidelta}) for the increment of  phase near the point of gradient catastrophe: 
\be
\Delta \Phi = 
-2a_0 \Delta x   -2 (2a_0^2-b_0^2) \Delta t  -   \ve^\frac 65\Re\le( 
\sqrt{\frac {2i}{Cb}} \frac {\tau^3}{8}
\ri) + \mathcal O(\Delta x^2 + \Delta t^2)
\ee
The reader may notice that the discontinuous term containing  $\tau^3$ cancels and we complete the proof.
\QED

\br
From the  expression (\ref{620}) it is clear that the approximation cannot hold in proximity of a pole of the {\em tritronqu\'ee} $y(v)$, for the Hamiltonian $H_I$ 
has a simple pole there and $y$ has a double pole.

The reader could verify that the above analysis holds as long as we approach a pole $v=v_p$ but {\em not too quickly}; 
\be
v-v_p = \mathcal O(\varepsilon ^{\frac 1 5 - \nu })\ ,\ \ \ \ \frac 1 5 > \nu> 0
\ee
In this case the formula in the above theorem is still correct but with the error term of order $\mathcal O(\varepsilon^{\frac 35  - 3\nu})$, 
and then the leading correction has --in fact-- order $\mathcal O(\varepsilon^{\frac 1 5 - \nu})$. 
Indeed the term $\mathcal O(\zeta^{-\frac 32})$ contains terms with {\em triple} poles (see (\ref{P1expansion})), and --in general-- the term $\zeta^{-\frac k2}$ has a coefficient with a pole of order $k$ in the Painlev\'e\ variable $v$. Hence the estimate $\mathcal O(\ve^{\frac 35})$ in (\ref{534} and following would have to be replaced throughout by $\mathcal O(\ve^{\frac 35-3\nu})$. 
\er

It appears that something awry is occurring when we approach a pole too fast, and  a different approximation parametrix need to be constructed.
This is the goal for the rest of the paper.

\section{Approximation near a spike/pole of the {\em tritronqu\'ee}} \label{spikepole}

With the preparatory material covered in Section \ref{sectPRHP} we shall now address the approximation of $q(x,t;\varepsilon)$ near a spike or --which is 
essentially the same-- in a (shrinking) neighborhood of a pole of the {\em tritronqu\'ee } solution.

In order to motivate the  construction used below we illustrate the difficulties in constructing the leading approximation to the solution $Y(z)$ of the RHP: 
it should appear that the local parametrix $\mathcal P$ needs to be expressed now in terms of the modified Psi function $\wh \Psi$ (\ref{Psihat}). 

Looking at the asymptotic expansion for large $\xi$ of $\wh \Psi$ (\ref{hatpsiasym}) it appears that the first modification  we need to make in order 
to match the boundary behavior of the $\wh \Psi$ with the outer parametrix is to replace the solution  $\Psi_0$  to the model RHP,
(see (\ref{Psi0})),  by the 
solution
\be
\Psi_1(z):= \frac 1  2 
\begin{bmatrix}
-i & -1\\
1 & i
\end{bmatrix}
\le(\frac{z-\a}{z-\ov \a}\ri)^{-\frac 34 \sigma_3} 
\begin{bmatrix}
i & 1 \\
-1& -i
\end{bmatrix}.
\label{Psi1}
\ee
The difference between $\Psi_1$ and  $\Psi_0$  is simply in the power growth near the endpoints $\a,\ov \a$ of the main arc. The transformation that links $\Psi_0$ to $\Psi_1$ is called {\bf (discrete) Schlesinger isomonodromic transformation}. In fact the two matrices are simply related one to another as seen below
\bl[Schlesinger chain]
The matrices  
\be
\Psi_K(z):=  \frac 1  2 
\begin{bmatrix}
-i & -1\\
1 & i
\end{bmatrix}
\le(\frac{z-\a}{z-\ov \a}\ri)^{\le(\frac 14-K\ri) \sigma_3} 
\begin{bmatrix}
i & 1 \\
-1& -i
\end{bmatrix}\ ,\qquad K\in \Z,
\ee
are related by a left-multiplication by a rational matrix 
\be
\Psi_K(z) =R_K(z) \Psi_0(z),
\ee
where 
\be
R_K (z) = \begin{bmatrix}
\frac {p^K + p^{-K}}2 &  i\frac {p^K -p^{-K}}2\\
- i\frac {p^K -p^{-K}}2 & \frac {p^K + p^{-K}}2
\end{bmatrix}\ ,\ \ p:= \frac {z-\a}{z-\ov \a}.
\ee
\el
{\bf Proof.}
The expression of $R_K$ follows from straightforward computations; we only point out that the {\em existence} of such a rational left multiplier follows 
from the fact that all $\Psi_K$  solve the same RHP (jump conditions and normalization at infinity), but have different growth behaviors at the points $\a,\ov \a$.    \QED

Mimicking the previous case of Definition \ref{deflocalP}, we shall state the following  new definition.
\begin{definition}[Local parametrix near the spikes.]
\label{deflocalPspike}
Let $\zeta(z;\varepsilon)$ be the local conformal coordinate in $D$, introduced in Definition \ref{defzetatau}, so that 
\be
\frac i\varepsilon h(z;x,t) = \theta(\zeta;\tau ) = \frac 4 5 \zeta^{\frac 5 2 } + \tau \zeta^{\frac 32}\ . 
\ee
Let $\wh \Psi(\xi;v)$ denote the modified Psi--function (\ref{Psihat}) of the Painlev\'e\ I problem with $\beta_0 = 0 = \beta_{-1}$ (and $\beta_{-2} = -1$, $\beta_{2} = -1$).
Then we define the parametrix
\bea
\mathcal P_{1;\a}(z) =
\frac 1{\sqrt{2}} 
{ \begin{bmatrix}
1&-1\\
i & i
\end{bmatrix}}
\zeta^{\frac 3 4 {\s_3}} 
\wh \Psi\le ( \zeta +  \frac \t 2; \frac 3 8 \t^2\ri) \begin{bmatrix} 0 & 1\\ -1 & 0 \end{bmatrix}  {\rm e}^{\theta(\zeta;\t) \s_3}, \label{localPspike}
\eea
and we set
\be
\mathcal P_{1;\ov\a}(z) :={ \le(\mathcal P_{1;\a}(\ov z)\ri)^\star}^{-1}.
\ee
\end{definition}

For brevity we will write simply $\mathcal P_1 = \mathcal P_{1;\a}$. 
We can then formulate the statement corresponding to Thm. \ref{thmlocalP} for the new local parametrix.
\bt
\label{thmlocalP1}
The matrix $\mathcal P_1$ satisfies: 
\begin{enumerate}
\item Within $\mathbb D_\a$, the matrix  $\mathcal P_1(z)$ solves the exact jump conditions on the lenses and on the complementary arc;
\item On the main arc (cut) $\mathcal P_1(z)$ satisfies 
\be
\mathcal P_{1+}(z) = \sigma_2 \mathcal P_{1-}(z) \s_2\ ,\label{oddjump}
\ee
so that $\Psi_0 \mathcal P_1$ within $\mathbb D_\a$ solves the exact jumps on all arcs contained therein  (the left-multiplier in the jump (\ref{oddjump}) 
cancels against the jump of $\Psi_0$);
\item The product $\Psi_0 (z) \mathcal P_1(z)$ (and its inverse) are --as  functions of $z$-- bounded within $\mathbb D_\a$, namely the 
matrix $\mathcal P_1(z)$ cancels the growth of $\Psi_0$ at $z=\a$;
\item The restriction of $\mathcal P_1(z)$ on the boundary of $\mathbb D_\a$ is 
\be
\mathcal P_1(z)\bigg|_{z\in \pa\mathbb D_\a} = \le(\1 + 
\mathcal O(\zeta^{-\frac 1 2})\ri) \le( \frac{ \sqrt{1-\zeta/y}}{ 1+ \sqrt{\zeta/y}} \ri)^{\s_3}\label{611} 
\ee
where $\mathcal O(\zeta^{-\frac 12})$ is uniform w.r.t. $v$ in a neighborhood of a pole $\pole$ not containing any zero of $y(v)$.
\end{enumerate}
\et
{\bf Proof.} 
For the first three points the  proceeds exactly as in Thm. \ref{thmlocalP} and hence is omitted. \\
{\bf (4)}  Due to the asymptotic expansion (\ref{hatpsiasym}) for $\wh \Psi$, when restricted to
the boundary $\partial\D_\a$,
we have (following the same computations as in Theorem \ref{thmlocalP})
\bea
\mathcal P_1(\zeta)\bigg|_{\pa \mathbb D_\a}=&\& \frac{1}{\sqrt{2}}
 \begin{bmatrix}
1& -1 \\
i & i
\end{bmatrix}
\z^{\frac 3 4\s_3}
\wh \Psi\le (\zeta + \frac \tau 2; \frac 38 \tau^2\ri )\begin{bmatrix} 0& 1 \cr -1 &0 \end{bmatrix} {\rm e}^{\theta(\zeta)\sigma_3} = 
\cr
=&\& \frac{1}{{2}}
 \begin{bmatrix}
1 &-1\\
i & i
\end{bmatrix}
\le(\frac {\z}{\zeta + \frac \tau 2 } \ri)^{\frac 3 4\s_3} 
\begin{bmatrix}
i &1\\
-i &1
\end{bmatrix}
\le(\1 + \mathcal O(\zeta^{-\frac 12 }) 
\ri)
 \le(\frac {\sqrt{ \zeta +\tau/2-y} } {  \sqrt{\zeta + \tau/2} + \sqrt{y} } \ri)^{  \s_3  }
=\cr
=&\& (i)^{\s_3}
\le( \1 + \mathcal O(\zeta^{-\frac 1 2 })
\ri)
 \le(\frac {\sqrt{\zeta/y-1} }{ 1 + \sqrt{\zeta /y}} \ri)^{ \s_3  } =
\le( \1 + \mathcal O(\zeta^{-\frac 1 2 })
\ri)
\le(\frac {\sqrt{1-\zeta/y} }{ 1 + \sqrt{\zeta /y}} \ri)^{ \s_3  }.    
\eea
\QED

Corresponding to  this new local parametrix, we  setup the approximation of the solution as 
\be
Y(z) = \le\{ 
\begin{array}{cc}
\ds\mathcal E(z) \Psi_1(z) & \mbox { for $z$ {\bf outside} of the disks } \mathbb D_\a, \mathbb D_{\ov \a},\\[4pt]
\ds \mathcal E(z) \Psi_1(z)\mathcal P_{1;\a}(z) & \mbox { for $z$ {\bf inside} of the disk } \mathbb D_\a, \\[4pt]
\ds \mathcal E(z) \Psi_1(z)\mathcal P_{1;\ov \a}(z) & \mbox { for $z$ {\bf inside} of the disk } \mathbb D_{\ov \a}, \\[4pt]
 {\ds \mathcal E(z) \Psi_1(z)\mathcal P_{\mu_+}(z)} &  {\mbox { for $z$ {\bf inside} of the disk } \mathbb D_{\mu_+}}.
\end{array}
\ri.
\label{leadingapproxspike}
\ee
 { where the parametrix $\mathcal P_{\mu_+}$ is the same  used in (\ref{leadingapprox}) (see Remark \ref{muparametrix}).}
Recall that we are considering the regime $v -v_p = \mathcal O(\varepsilon^{\frac {1}5 +\nu })$ (and hence $y = \mathcal O(\ve^{-\frac 2 5 -2\nu})$)
where  $v = \frac 38 \t^2$ and $\nu\geq 0$. The boundaries of both 
disks $\mathbb D_\a, \wh{\mathbb D_{\bar\a}}$ are mapped by the conformal changes of coordinates $\zeta,\wh \zeta$ on some closed curves in the respective planes, 
that expand homothetically with a scale factor  $\varepsilon ^{-\frac 25}$.
Then the behavior of the jump in (\ref{Ejumpspike}) is determined by the behavior of the local parametrix $\mathcal P_1=\mathcal P_{1;\a}$ on the boundary of 
$\mathbb D_\a$.
The jump of $\mathcal E$ is 
\be
\mathcal E_+(z) = \mathcal E_-(z) \Psi_1(z) \mathcal P_{1;\a}^{-1} \Psi_1^{-1}(z)\ ,\ \ \ z\in \pa \mathbb D_\a\ .\label{Ejumpspike}
\ee

From eq. (\ref{611})  it is clear that the rate of approach of $\mathcal P_1$ to the identity on the boundary is seriously 
impeded by the last factor in (\ref{611}), which fails to converge to $\1$ when $\nu=0$, namely, when $y = \mathcal O(\varepsilon^{-\frac 25}))$.
More precisely we have from (\ref{Ejumpspike}) (since $\zeta/y = \mathcal O(\ve ^{2\nu})$)
\be
\mathcal E_+(z) = \mathcal E_-(z) \le(\1 + \mathcal O(\ve ^{\min(\frac 1 5 , 2\nu) }\ri).
\ee
Before tackling the general problem $\nu=0$, we shall see what happens when $v=v_p$ (namely $y=\infty$) and we are exactly  at the ``top of a spike''

\subsection{The top of the spike: amplitude}

When $v=v_p$ is exactly a  pole of the {\em tritronqu\'ee} we can use the expansion (\ref{psihatform_at_a}) for the expansion of the  
local parametrix on the boundary of the disks $\mathbb D_\a, \mathbb D_{\ov \a}$. Since the first term after the identity is of order $\mathcal O(\zeta^{-\frac 12}) = \mathcal O(\ve^{\frac 15})$ when restricted on the boundary, the error term in (\ref{leadingapproxspike}) is then  of  the form
\be\label{E1}
\mathcal E(z) = \1 + \mathcal O(\varepsilon^{\frac 15}, z^{-1}). 
\ee
Near $z=\infty$ we can write $\mathcal O(\ve^\frac 15, z^{-1})$ as $\frac {\mathcal O(\ve^{\frac 15})}z$ and so we have
\bea
Y(z) =  \mathcal E(z) \Psi_1(z) = \le(  \1 +\frac 1 z \mathcal O(\varepsilon^{\frac 15})
\ri)\le(\1 +\frac 3 2  \frac {b}{z} \le[\begin{array}{cc}
0&  1\\
-1&0
\end{array}\ri] + \mathcal O(z^{-2} )\ri).
\eea
We thus have
\bea\label{topspikeexp}
{q(x,t,\e)=}-2\lim_{z\to\infty} z\GGamma_{1,2}(z) =&\&-2 {\rm e}^{\frac{4i}\varepsilon g (\infty;x,t)} \lim_{z\to\infty} z \le(\mathcal E \Psi_{1}\ri)_{12} = 
{\rm e}^{\frac{i}\ve \Phi (x,t)} \le(-3 b(x,t)+ \mathcal O(\varepsilon^{\frac 15})\ri) =\nonumber \\
=  &\&{\rm e}^{\frac{i}\ve\le ( \Phi (x,t)- \varepsilon  \pi \ri)} 3 b(1 + \mathcal O(\varepsilon^{\frac 1 5 })),\nonumber
\eea
where 
$q\in\Uscr$ and $(x,t)=v^{-1}(v_p)$ corresponds to the top of the spike. Since the map
$v=v(x,t,\e)$ is scaled as  $\e^\frac 45$, see Corollary \ref{corvXT}, the corresponding to $v_p$ spike 
will approach the point of gradient catastrophe $(x_0,t_0)$ at $O(\e^\frac 45)$ rate. Here $b$ can be taken 
as the value at $(x_0,t_0)$ because the difference between the values at $(x_p,t_p)$ and $(x_0,t_0)$ is of order $\mathcal O(\ve^{\frac 25})$ 
(see (\ref{alphatau}) or (\ref{alphaXT})). Thus, we have proved the following theorem;

\bt
\label{ThmAmplitude}
The asymptotic amplitude of a spike near the gradient--catastrophe point $(x_0,t_0)$ is 
(up to  $\mathcal O(\varepsilon^{\frac 1 5 })$ accuracy)
{\bf three times} the amplitude predicted by  the Whitham  modulation equations at $(x_0,t_0)$.
\et

This result is a bit unexpected in that it entails a very simple {\bf  universality}; the three-fold amplitude of the first spikes appears to be entirely independent of the initial data.

In fact the factor of $3$ (together with the phase-shift, i.e., the minus sign) can be traced to the exponent $-\frac 34 \sigma_3$ of the outer parametrix $\Psi_1$, compared with the exponent $\frac 1 4 \sigma_3$ of $\Psi_0$. 
In turn, this exponent is determined by the asymptotic behavior of the modified $\Psi$-function for the P1 problem (\ref{psihatform_at_a}). The latter is due to the shearing transformation for the cubic potential $V(\xi;a)$ appearing in (\ref{atpoleODE}). 

It seems clear that, were we to study a non-generic gradient catastrophe 
(i.e., more than one new main arc emerging from the endpoint of an already existing main arc),  
we would have to replace the P1 problem by a higher member of the Painlev\'e\ I hierarchy, which are characterized by exponents $\frac 72, \frac 92$ etc. 
Thus, we conjecture that the amplitude of the first spikes after a non-generic gradient catastrophe will be 
$5, 7, 9, \dots $ times the amplitude at the gradient catastrophe $(x_0,t_0)$, depending on the degree of degeneracy.
\begin{conjecture}
\label{P1conjecture}
The asymptotic amplitude of the spikes in the neighborhoods of the points of gradient catastrophe are odd multiples of the amplitude at the point itself. 
\end{conjecture}
We reserve to verify this in a subsequent publication. 
The possible  shape of these spikes in the case of a degenerate gradient catastrophe are discussed in Remark \ref{rembreathe} below.

\subsection{The shape of the spike}
Due to the rightmost term in (\ref{611}), the rate of approach of $\mathcal P_1$ to the identity on the boundary of $\z(\mathbb D_\a)$ in the regime
\be
v-\pole = \mathcal O(\varepsilon^{\frac 15 + \nu}) \ \ \Leftrightarrow y = \mathcal O(\varepsilon ^{-\frac 2 5 - 2 \nu})
\ee
becomes slower  as $\nu$ approaches the critical value $\nu=0$, at which point the parametrix $\mathcal P_1$ does not tend to the identity at all.
Indeed, since $\zeta = \mathcal  O (\e^{-\frac 25})$, 
we see that on the boundary of $\z(\mathbb D_\a)$
\be
\mathcal P_1(\zeta) \sim  \le(\1 + \mathcal O (\varepsilon^{\frac 15}) \ri) 
\underbrace{
\le(\frac {\sqrt{1-\zeta/y}}{1+\sqrt{\zeta/y}}\ri)^{\s_3}
}_{=:Q(z)} = \1 + \mathcal O(\varepsilon^{\min\le( \frac {2\nu}{5(1-\nu)}, \frac 1 5\ri)}),\label{624}
\ee
so that the jumps of $\mathcal E$ on the circles $\mathbb D_a,~\mathbb D_{\bar\a}$ are 
\be
\mathcal E_+(z)=\mathcal E_-(z)\le( \1 + \mathcal O(\varepsilon^{\min\le( \frac {2\nu}{5(1-\nu)}, \frac 1 5\ri)})\ri).
\ee
From the standard approximation theorems for Riemann--Hilbert problems, it is seen that $\mathcal E$ converges (uniformly) to the identity only up to  the same rate of convergence of the jumps, in this case $\mathcal O(\varepsilon^{\min\le( \frac {2\nu}{5(1-\nu)}, \frac 1 5\ri)})$; in particular the ``error'' becomes worse and worse as $\nu\to 0$.

As we shall presently see, it is possible (and necessary)  to modify the outer parametrix $\Psi_1$  in such a  way that the troublesome factor $Q(z)$ above is exactly taken care of.
To account for the term $Q(z)$ we shall seek an {\bf exact} solution of the Riemann--Hilbert problem described hereafter:
let $\zeta(z),  \zeta^\star (z) = \ov \zeta(\ov z)$ be the local conformal scaling coordinates  in the neighborhoods of $\a, \ov \a$ respectively of the form
\bea
\zeta(z) =\varepsilon^{-\frac 2 5 } C(z-\a)(1 + \mathcal O(z-\a))\ ,\qquad
\wh \zeta(z)= \varepsilon^{-\frac 2 5 }\ov C (z-\ov \a) (1  + \mathcal O(z-\ov \a),
\eea
where $C = \le(\frac {5}4  C_1\ri)^\frac 25\neq 0 $. Let us assume that 
the circles $\partial \mathbb D_\a$ and $\partial \mathbb D_{\bar\a}$ (oriented counterclockwise)
have some small radius $r$, so that $|\zeta(z)/y|<1$ and $| \zeta^\star (z) / \ov y |<1$ respectively  on these two circles. 
\begin{problem}
\label{Eproblem}
Find a piecewise analytic matrix $E(z)$ on the complement of the two circles described above and such that 
\bea
E(z) &\&= \1 + \mathcal O(z^{-1})~~~~~~~~{{\rm as}~~z\ra\infty},\\
E_+(z) &\&= E_-(z) \Psi_1(z) Q_\a^{-1}(z) \Psi_1^{-1}(z) =E_- F  M_\a(z) F^{-1}\ ,\ \ |z-\a| = r,\\
E_+(z) &\& = E_-(z)  \Psi_1(z) Q_{\ov a}^{-1}(z) \Psi_1^{-1}(z) = E_-  F M_{\ov \a}(z) F^{-1} \ ,\ \ |z-\ov \a|=r,\\
M_\a(z)
&\&=\frac 1{\sqrt{1-\zeta /y}}
\begin{bmatrix}
1 & -i \sqrt{\frac {\zeta(z) (z-\ov \a)^3}{ y (z-\a)^3}}\\
i \sqrt{\frac {\zeta(z) (z- \a)^3}{ y (z-\ov \a)^3}} & 1
\end{bmatrix},
\label{623}\\
M_{\ov\a}(z) 
&\&=\frac 1{\sqrt{1- \zeta^\star/\ov y }}
\begin{bmatrix}
1 & i \sqrt{\frac {\zeta^\star (z) (z-\ov \a)^3}{ \ov y (z-\a)^3}}\\
-i \sqrt{\frac { \zeta^\star(z) (z- \a)^3}{\ov y(z-\ov \a)^3}} & 1
\end{bmatrix},
\label{6233}\\
F&\& := \frac 1{\sqrt{2}} \begin{bmatrix}
-i & -1 \\
1 & i
\end{bmatrix}.
\eea
\end{problem}

Before solving Problem \ref{Eproblem} we show how its solution will be used.
If $E(z)$ solves Problem \ref{Eproblem} then we re-define the approximation of $Y$ as 
\bea
Y(z) = \le\{
\begin{array}{cc}
\mathcal E(z) E(z) \Psi_1(z) & \mbox { for $z$ {\bf outside} of the disks}~ \mathbb D_\a, \mathbb D_{\bar\a},\\[10pt]
\mathcal E(z) E(z) \Psi_1(z) \mathcal P_1(z) &  \mbox { for $z$ {\bf inside} of the disks}~
\mathbb D_\a, \mathbb D_{\bar\a}.
\end{array}
\ri. 
\eea
Then, according to (\ref{624}), the jump of $\mathcal E(z)$ will  be 
\bea\label{627}
\mathcal E_+ = \mathcal E_- E_- \Psi_1 \mathcal P_1 \Psi_1^{-1} E_{+}^{-1} =
\mathcal E_- E_- \Psi_1 \mathcal P_1 Q \Psi_1^{-1} E_{-}^{-1} 
\cr
= \mathcal E_-E_- \le(\1 + \mathcal O(\varepsilon^{\frac 15}) \ri) E_{-}^{-1} = \mathcal E_- \le(\1 + \mathcal O(\varepsilon^{\frac 15}) \ri),
\eea
where the last equality holds provided that $E_-(z), E_-^{-1}(z)$ are  {\bf bounded uniformly in $\varepsilon$} on the boundaries (which will be the case indeed). 

\subsubsection{Solution to Problem \ref{Eproblem}}
The problem has an {\bf explicit solution}.
For the sake of simpler computations we will conjugate $E(z)$ by the constant matrix $F$, so that $\wh E(z) := F ^{-1} E (z)F$  has the matrices $M_\a, M_{\ov \a}$ (\ref{623}, \ref{6233})  for jumps.

It is known that the solution $\wh E(z)$ (if it exists) must satisfy the integral equation (here $\mathcal C_{\a}, \mathcal C_{\ov \a}$ are the counterclockwise boundaries of the disks $\mathbb D_\a,\mathbb D_{\ov \a}$) 
\be
\wh E(z) = \1 + \oint_{\mathcal C_\a} \!\!\!\! \frac {\wh E_-(s) (M_\a(s)-\1)}{s-z} \frac{{\rm d}s}{2i\pi}+ \oint_{\mathcal C_{\ov \a}} \!\!\!\! \frac {\wh E_-(s) (M_{\ov \a}(s)-\1)}{s-z} \frac{{\rm d}s}{2i\pi}, \label{ERHP}
\ee
where $\wh E_-(s)$ is the (analytic continuation of the) solution from the outside of the circles
$\mathbb D_\a, \mathbb D_{\bar\a}$.
It is crucial that the jump matrices $M_{\a, \ov \a}$ admit a simple decomposition of the form (recall  that (\ref{expressC}) $C:= \e^\frac 25 \zeta'(\a)$)
\bea
M_\a(z) -\1 &\&=  \mathcal O_\a(z) + 
\frac {-i \varepsilon^{-\frac 1  5} \sqrt{C/y (\a-\ov \a)^3} }{z-\a}
\begin{bmatrix} 0 & 1 \\ 0  & 0 
\end{bmatrix} =: \mathcal O_\a(z) + \frac {n_{\a}}{z-\a}\s_+,\cr
M_{\ov\a}(z) -\1 &\&=  \mathcal O_{\ov \a}(z) + 
\frac {-i \varepsilon^{-\frac 15}\sqrt{\ov C/\ov y  (\ov \a- \a)^3} }{z-\ov\a}
\begin{bmatrix} 0 & 0 \\ 1  & 0 
\end{bmatrix} =: \mathcal O_{\ov\a}(z) + \frac {n_{\ov\a}}{z-\ov \a}\s_-,  
\eea
where $\mathcal O_\a(z)$ denote some {\bf locally analytic} matrices in the respective neighborhoods (whose expression the reader can evince 
from the above formul\ae\ but which has no bearings in the considerations to follow). 
What is essential in the following is that when evaluated at $z=\a$  and $z=\bar\a$, these local analytic matrices will be multiples of $\s_+$ and
$\s_-$ respectively.

Consider the Ansatz
\be
\wh E_-(z) =\1 + \frac \A{z-\a} + \frac {\wh \A}{z-\ov a}.\label{outhatE}
\ee
The expression of $E(z)$ in the inside of the disks $\mathbb D_\a, \mathbb D_{\bar\a}$ (i.e. $E_+(z)$) has no particular interest for us and can be simply obtained from the jump condition.
Using  (\ref{outhatE}), the  formula (\ref{ERHP}) becomes
\bea
\frac{\A}{z-\a} + \frac {\wh \A}{z-\ov \a} &\& = 
\frac{\A \mathcal O_{\a}(\a)}{(\a-z)} - \frac {\A\,\s_+ n_{\a}}{(z-\a)^2} + \frac {n_{\a}\s_+}{\a-z} + \frac {\wh \A\, \s_+ n_{\a}}{(\a-z)(\a-\ov \a)} + \cr
&\& + \frac{\wh \A \mathcal O_{\ov \a}(\ov \a)}{(\ov \a-z)} - \frac {\wh \A \,\s_- n_{\ov \a}}{(z-\ov \a)^2} + \frac {n_{\ov \a}\s_-}{\ov \a-z} + \frac {\A\, \s_-n_{\ov \a}}{(\ov \a-z)(\ov\a- \a)}.
\eea
We thus have the system 
\bea
\A \s_+ =0,  &&
\wh \A \s_-  = 0,\\
\A  + \frac {n_{\a}}{ \a-\ov \a} \wh \A \s_+ = 
-n_{\a} \s_+, & & 
\wh \A + \frac {n_{\ov \a}}{\ov\a-\a} \A \s_-  =
-n_{\ov \a} \s_-.
\eea
The solution is given by 
\bea
\A &\& = \frac 1{
1+ \frac {n_\a n_{\ov \a}}{(\a-\ov \a)^2}
}
\begin{bmatrix}
     0& 
     -n_{\a}  \\  0 & \frac{
     -n_\a n_{\ov \a}}{{\ov \a- \a}}
    \end{bmatrix}  =  \frac 1 {1+ \frac{2b |C|}{\varepsilon^{\frac 25} |y|}  }
\begin{bmatrix}
0 &  
 i \ds \sqrt { \frac{  (2ib)^3 C}{\varepsilon ^{\frac 25}y}}\\[10pt]
0 & 
i\ds \frac{ |C| 4 b^2}{\varepsilon ^{\frac 25} |y| }
\end{bmatrix},
\cr
\wh \A  &\& =  \frac 1{
1+ \frac {n_\a n_{\ov \a}}{(\a-\ov \a)^2}
} \begin{bmatrix}
  \frac{
  -n_\a n_{\ov \a}}{\a-\ov \a} & 0 \\ 
  n_{\ov \a} & 0
 \end{bmatrix} =  \frac 1 {1+\frac{ 2b |C|}{\varepsilon^{\frac 2 5 } |y|}  }
\begin{bmatrix}
\ds 
{-}  i \frac{|C| 4 b^2}{\varepsilon^{\frac 2 5}|y|} & 0\\[15pt]
 i \sqrt{ \frac { (-2i b)^3 \ov C}{\varepsilon^\frac 2 5 \ov y}} & 0 
\end{bmatrix}.
\label{AAhat}
\eea
Thus the solution of the problem for $\wh E$ has the form (\ref{outhatE}),
(\ref{AAhat}) in the region outside the disks. 

\subsubsection{Partial Schlesinger transformation and improved leading order asymptotics: shape of the spike}
\bt[Shape of the spikes]
\label{thmpeakshape}
The spikes around the point of gradient catastrophe for $q(x,t,\e)\in\Uscr$ are in one-to-one correspondence with the poles of the {\em tritronqu\'ee} solution $y(v)$ of the Painlev\'e\ I equation
\be
y'' =6y^2 -v
\ee
and have the following  universal shape: for a given pole $v=v_p$ of the {\em tritronqu\'ee} solution, the shape
of the spike is described by
\be
\frac{q(x,t,\e)}{q_0(x_0,t_0, \e)} = \frac {1 + (s+2i)(\ov s+ 2i)}{1+|s|^2}
\le(1 + \mathcal O(\varepsilon^{\frac 1 5})\ri) = \frac {|s|^2 - 3 + 4i \Re (s) }{1+|s|^2}
 \le(1 +        \mathcal O(\varepsilon^{\frac 15})     \ri),
\label{toothshape}
\ee
where  $q_0(x,t,\e)$ is the genus zero approximation (see  (\ref{lotg0}), (\ref{phase}), and Remark \ref{extendq_0})
the  variable $s$ is defined in terms of the {\em tritronqu\'ee} solution $y(v)$ as
\bea
\ds s^2 &\&: =- \le(\frac {5C_1}4\ri)^{\frac 25}\frac {2ib}{\varepsilon^{\frac 2 5} y} =  
-\le(\frac {5C_1}4\ri)^{\frac 25}\frac{2ib }{\varepsilon^\frac 25}(v-\pole)^2 \le(1 +\mathcal O(v-\pole)^2\ri)\cr
&\&=-\frac {4 b^2 }{\varepsilon^{\frac 25}} \bigg( \frac{x-x_p}{\e^\frac 45} + 2(2a+ib)
\frac{t-t_p}{\e^\frac 45} \bigg)^2 \le(1 + \mathcal O(\varepsilon^\frac 25)\ri),\label{sdef}
\eea
and the map $v=v(x,t,\e)$ is given in Corollary \ref{corvXT}. The formula and the error term are valid uniformly for $(x,t)$ in a  $\mathcal O(\ve)$--neighborhood of the center of the spike $(x_p,t_p)=v^{-1}(v_p)$, or --which is the same-- for $v(x,t,\ve)- \pole = \mathcal O(\ve^{\frac 15})$.
\et
\br  The class of solution $\Uscr$, for which Theorems \ref{ThmAmplitude} and \ref{thmpeakshape} are valid, 
can be extended according to Remark \ref{rem2}.
\er
\br
 We stress once more that the above result holds in the vicinity of any pole, {\em wherever this might occur}. In particular, the validity of the above description holds even if the pole should be outside of the sector conjectured by Dubrovin et al. \cite{DubrovinGravaKlein}.
\er
{\bf Proof}.
It is apparent that both $E_-(z)$ and $E_-^{-1}(z)$ are uniformly bounded on the boundaries of the disks 
$\mathbb D_\a,\mathbb D_{\bar\a}$ as long as $y = \mathcal O(\varepsilon ^{-\frac 25 - \nu})$ for some $\nu\geq 0$. Thus,
according to (\ref{627}), the matrix $\mathcal E(z)$ will be $\1 + \mathcal O(\varepsilon^{\frac 1 5})$. 
Then the leading order approximation is thus given by  
\bea
\wt \Psi_1=  E(z) \Psi_1(z)= F \wh E(z) \le(\frac {z-\a}{z-\ov a}\ri)^{-\frac 34 \sigma_3} F^{-1},
\eea
where
\be
F=
{\frac {\s_2+\s_3}{\sqrt{2}}}. 
\ee
From this expression we find the behavior of the amplitude 
\bea
\wt \Psi_1(z) &\&= F \le(\1 + \frac {\A}{z-\a} + \frac {\wh \A}{z-\ov \a} \ri)  \le(\frac {z-\a}{z-\ov \a}\ri)^{-\frac 34 \sigma_3} F^{-1} = \1 + F\le( \frac {\A + \wh \A}{z} + \frac {3(\a-\ov \a)} 4 \frac{\sigma_3}z\ri)F^{-1}=\cr
&\& = \1 
%\replace{-}
{+} \  \frac {3i b}{2z}   \begin{bmatrix}
0 & -i\\
i&0
\end{bmatrix} + \frac 1 {1+\frac{ 2b |C|}{\varepsilon^{\frac 2 5 } |y|}  } \frac 1 z  F
 \begin{bmatrix}
-i \frac{|C| 4 b^2}{\varepsilon^{\frac 2 5}|y|}  &        
  i\varepsilon^{-\frac 1 5} \sqrt { - 8i C b^3/y}    \\
 i \varepsilon^{-\frac 1 5} \sqrt{8i \ov C b^3/\ov y} &    
 i\frac{ |C| 4 b^2}{\varepsilon ^{\frac 25} |y| }   
\end{bmatrix}
F^{-1} + \mathcal O(z^{-2}). 
\eea
Since $q = -2 \lim_{z\to\infty} z \GGamma_{12}(z) $, where
\be
\GGamma(z) = {\rm e}^{\frac {2i} \varepsilon g(\infty) \sigma_3} \mathcal E(z) \wt \Psi_1(z)  {\rm e}^{ - \frac {2i} \varepsilon g(z) \sigma_3}, 
\ee
we have
\bea
q(x,t,\e) &\& =  {\rm e}^{\frac {4i}\varepsilon g(\infty)}\le(-3b 
%\replace{+}
{-} 2 (F (\A + \wh \A)F^{-1})_{12} + \mathcal O(\varepsilon^\frac 15) \ri) =  \\
&\& =
{\rm e}^{\frac {i}\varepsilon \Phi}\le(-3 b +
\frac {2} {{\varepsilon^{\frac 2 5 } |y|}+{ 2b |C|}  }
\le(
{|C| 4 b^2} + i\varepsilon^{\frac 1 5}  \Re\le(\sqrt {-8i C b^3 \ov y } \ri)\ri)  + \mathcal O(\varepsilon^ \frac 1 5)\ri)= \nonumber \\
&\& =- \frac 1{{1 + \frac {2b|C|}{\varepsilon^{\frac 2 5 } |y|} }} {\rm e}^{\frac {i}\varepsilon \Phi} b \le(3   - \frac {2b|C|}{\varepsilon^{\frac 2 5 } |y|}     - 4   i \Re \le(
\sqrt{ \frac {-2ib C }{\varepsilon^{\frac 2 5} y }}
\ri)  \ri). \label{749}
\eea
Since the variable $v$ is a map on $D$, which is a size $\varepsilon^{\frac 4 5 }$  neighborhood of  the point of gradient catastrophe, then the variable $s$ defined in (\ref{sdef})  is a map on a neighborhood of size $\mathcal O(\varepsilon)$ around the spike. In terms of $s$ we can rewrite (\ref{749}) as  
\be
q(x,t,\e)  =-  {\rm e}^{\frac {i}\varepsilon \Phi} b\frac{\le( 3 - |s|^2 - 4i \Re(s)\ri)}{1+|s|^2}\big (1 + \mathcal O(\varepsilon^{\frac 1 5})\big) =  {\rm e}^{\frac {i}\varepsilon \Phi} b\, \frac { 1  + (s+2i)(\ov s+2i)}{1+|s|^2}\big (1 + \mathcal O(\varepsilon^{\frac 1 5})\big).
\label{peakshape}
\ee
\QED

We remark in the following corollary that from (\ref{toothshape}) it follows immediately that each spike has two zeroes (nodes), one on each side.
\bc 
\label{isochronousnodes}
For each spike near the gradient catastrophe point there are two zeroes (nodes)  and they occur
at the time $t=t_p$  (asymptotically as $\varepsilon\to 0$)  and with $x-x_p = \pm \varepsilon \frac 1{2b} \sqrt{3}$.
\ec

\begin{wrapfigure}{r}{0.3\textwidth}
\vspace{-30pt}
\resizebox{0.3\textwidth}{!}{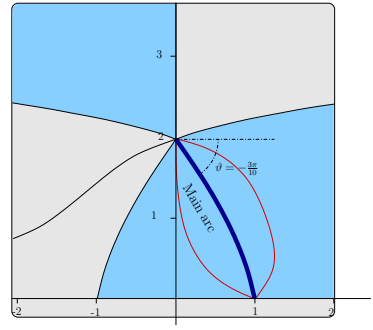}
\caption{A numerically accurate representation of the main arc (thick blue), the complementary arc (thin black),
the lenses (red)  and level lines of $\Im(h)$ for Example \ref{Exmu2}.}
\label{figmu2}
\end{wrapfigure}

\bx
\label{Exmu2}
 {Using the same example as in Ex. \ref{examplemapping} with $\mu=2$  (see Fig. \ref{figmu2}) one can verify that the main arc  forms an angle  $\vartheta = -\frac {3\pi}{10} =\frac  \pi 2 - \frac {4\pi }5$ with the horizontal. 
Using the  formula (\ref{mainarc}) that relates the direction of the main arc and the argument of $C_1$, and then using 
(\ref{dvxt}, \ref{expressC}) one verifies that the images $v_{node}$ of the  nodes in the Painlev\'e\ plane  are aligned  in the directions $\arg(v_{node}- \pole) = \frac {\pi}{10} +k \pi$, namely,  aligned {\em perpendicularly} to the bisecant of the sector $\arg(v)\in (2\pi/5, 4\pi/5)$.}
\ex

\subsubsection{The spike as a rational breather}
\label{breather}
To gain a bit more insight into the  formula (\ref{toothshape})  
we use  the second expression in (\ref{sdef}).
The phase can be expanded near the spike using  (\ref{phidelta}) and retaining just the linear term;
indeed the scale of the spike is $\mathcal O(\ve)$ and hence $ x-x_p=  \mathcal O(\ve)$
and $t-t_p=  \mathcal O(\ve)$.
We thus find 
\bea
s &\& = 2i b \bigg(\xi + 4 a \eta + 2ib \eta\bigg) = - 4b^2 \eta   + 2ib (\xi + 4a \eta),\\
&\& \xi := \frac {x-x_p}\ve\ ,\qquad \eta:= \frac {t-t_p}\ve.
\eea 
So,
\be
q(x,t,\e)  ={\rm e}^{\frac i \ve \Phi(x_p,t_p)} Q_{br}(\xi,\eta)
(1 + \mathcal O(\ve^\frac 15)),
\ee
where
\bea
Q_{br}(\xi,\eta)&\& := {\rm e}^{-2i \le(a \xi + (2a^2 - b^2) \eta\ri)} 
b\, \le( 1 - 4\frac {1+ 4ib^2 \eta }{1 +  4b^2 (\xi  +4 a \eta)^2 + 16 b^4 \eta^2} \ri).
\eea

\begin{figure}[th]
\centerline{\includegraphics[width=0.67\textwidth]{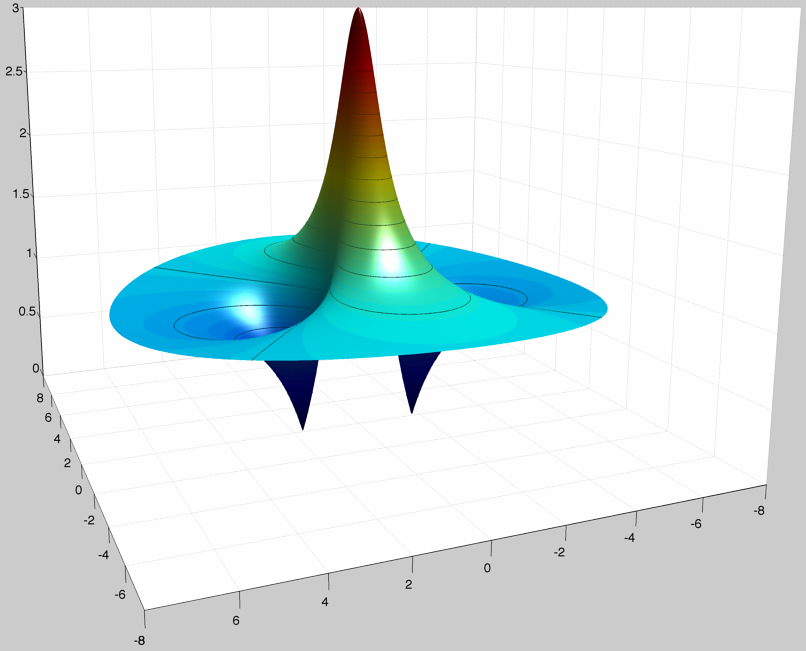}}
\caption{The theoretical shape of the spike as a function of $\ds s =i \sqrt{\frac{2i b C} {y \varepsilon ^{  \frac 2 5}}}$ as given by Eq. (\ref{peakshape}). This  also represents the exact single-breather solution $Q_{br}$ where $s = 2i b(\xi + 4a \eta +2ib \eta)$.}
\label{NLSpeak}
\end{figure}

The expression $Q_{br}(\xi,\eta)$ is well known in the literature \cite{Peregrine84}; it is called the {\bf rational breather solution} for NLS. 
Indeed it solves exactly the NLS equation 
\be
i \ve\pa_t Q_{br} + \ve^2\pa_{x}^2 Q_{br} + 2\le|  Q_{br} \ri| ^ 2 Q_{br}=0,
\ee
where
\be
\xi = \frac {x-x_p} \ve\ ,\qquad \eta  = \frac {t-t_p}\ve\ .
\ee
In our case it is obtained from the ``stationary'' breather (depicted in Fig. \ref{NLSpeak})
\be
Q_{br}^0 (\xi,\eta)={\rm e}^{2i\eta}\le( 1 - 4\frac {1+4i\eta}{1 + 4 \xi^2 + 16 \eta^2}\ri)
\label{656}
\ee
by applying the transformations (mapping solutions into solutions)
\be
\wt Q(\xi, \eta)  =  \lambda Q(\lambda\xi,  \lambda^2 \eta),~~~~~~~~~~
\wh Q(\xi, \eta)  = {\rm e}^{i(k x - k^2\eta)} Q(\xi -2 k \eta, \eta).
\ee

The breather  has  the maximum equal to $3$ at $\xi=0=\eta$
and tends to $1$ as $|s|\to \infty$; therefore it smoothly interpolates the regime near the spike with the regime ``far'' from it.

\br
\label{rembreathe}
While completing the preparation of this paper we came across \cite{Ankiewicz010}, 
where a hierarchy of rational breathers generalizing the  standard one (\ref{656}) is investigated. Interestingly, 
the maximum amplitude of these breathers is also an odd integer and hence this prompts the speculation  that they should 
play the same r\^ole  of (\ref{656}) in the cases of gradient catastrophes with higher degeneracy (compare with Conjecture \ref{P1conjecture}).
\er

\br[Consistency Check]
Suppose $y$ remains fixed and $\varepsilon\to 0$ in the formul\ae\ for $\A, \wh \A$ (\ref{AAhat}); we then obtain the limit
\bea
\wt \Psi_1 (z)&\& = F\le(\1 + \begin{bmatrix} 
\frac {\ov \a - \a} {z-\ov \a} & 0 \\
0 &  \frac { \a - \ov \a} {z- \a}
\end{bmatrix}
\ri)\le(\frac {z-\a}{z-\ov \a}\ri)^{-\frac 34 \sigma_3} F^{-1} =\nonumber \\
&\& =
 F \begin{bmatrix} 
\frac {z - \a} {z-\ov \a} & 0 \\
0 &  \frac { z - \ov \a} {z- \a}
\end{bmatrix}
\le(\frac {z-\a}{z-\ov \a}\ri)^{-\frac 34 \sigma_3} F^{-1} 
= F\le(\frac {z-\a}{z-\ov \a}\ri)^{\frac 14 \sigma_3} F^{-1} = \Psi_0(z).
\eea
\er

\begin{figure}[t]
\resizebox{0.99\textwidth}{!}{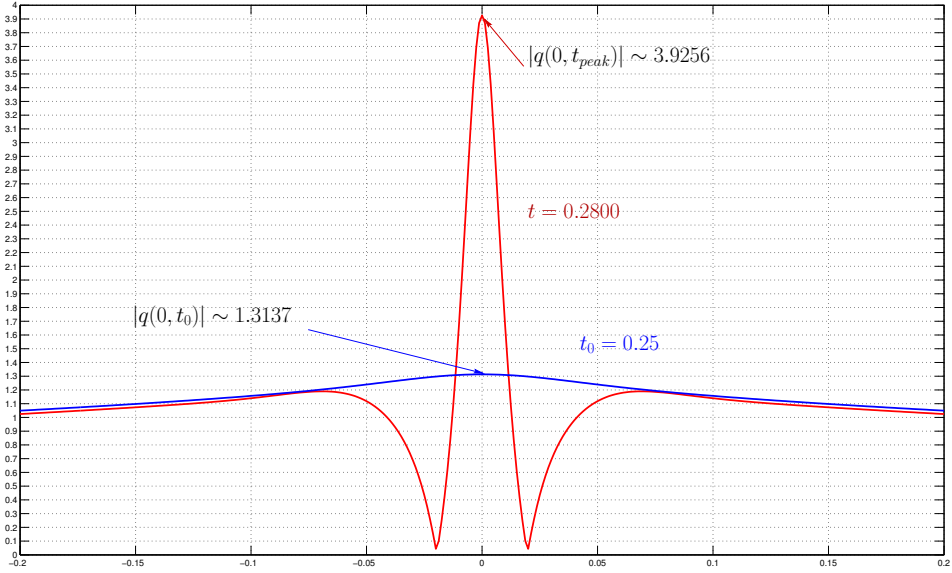}
\caption{Some simple numerics (using Fast Fourier Transform) for the initial data $q(x,0) = \frac 1{\cosh(x)}$ and $\ve = \frac 1 {33}$; note that the amplitude at the first spike (numerically determined) is $|q_{peak}| = 3.9256$ and $|q_0|=1.3137$. This is almost exactly three times, $3|q_0| = 3.9411$. Moreover the localization of the spike at $t\simeq 0.2800$ is in excellent agreement with the theoretical value $t=0.27912...$ obtained from the numerical localization of the first  pole of the {\em tritronqu\'ee} (after \cite{JoshiKitaev01}). In this case $\mu=0$ and $t_0 = \frac 14$ as per Example \ref{examplemapping} . }
\label{numericpeak}
\end{figure}
\bx\label{firstspike}
According to numerical studies by \cite{JoshiKitaev01}, the closest pole to the origin of the {\em tritronqu\'ee} on the bisecant of 
the sector where the poles are asymptotically confined  is at distance of about $|\pole|\sim 2.38$. According to our Theorem \ref{thmpeakshape}, this is the first spike with $x=x_p$ after the 
gradient catastrophe and its 
location is thus determined by the map $v=v(x,t,\e)$.

In the case of the initial data of Example \ref{examplemapping} with $\mu=0$, the solution is symmetric ($x\mapsto -x$),
hence the first spike occurs at $x=x_0 = 0$; the time of the first spike is then estimated at 
\be
t_{spike} = t_0 + \ve^\frac 45 \le(\frac {5|C_1|}4 \ri)^\frac 15 \frac 1{(2 b_0)^\frac 32} |\pole| \le(1+ \mathcal O(\ve^\frac 25)\ri).
\ee
For the case of the initial data of Example \ref{examplemapping} ($\mu=0$), 
the exact values of $b_0=b(x_0,t_0)=\sqrt 2 $,  $t_0=\frac 1 4$  and $|C_1|=\frac 8{15} 2^{\frac 14}$  
were found in \cite{TVZ1}; this gives 
\be
t_{spike} \sim \frac 1 4+ \ve^{\frac 45} 0.4776312294~. \label{tpeak}
\ee
In Fig. \ref{numericpeak} we report on a simple numerical example with initial data $q(x,0)=sech(x)$, $\ve = \frac 1{33}$; plotted are the profiles of $|q(x,t)|$ for $t=t_0=1/4$ which is the theoretical time of gradient catastrophe; the second curve is the time where numerically the maximum amplitude is achieved at $x=0$. The numerics indicates $t_{spike} =  0.28$ whereas the formula (\ref{tpeak}) predicts $t_{spike} = 0.2791260482$, which --considering that $\ve$ is not that small-- is an amazingly close estimate.
The results on \cite {JoshiKitaev01} do not exclude that there are closest poles to the origin: however this is the one that occurs on the same $x=x_0$ as the gradient catastrophe. Other numerical indications using Pad\'e\ approximations localize the poles in a triangular pattern \cite{Novokshenov} like the one used as illustration in Fig. \ref{Painplane}.
 
Our numerics was produced by integrating directly the NLS equation using FFT in the $x$-direction; it should be noted that the initial data we have chosen correspond to a pure soliton situation, and hence one could use a better numerical approach based on exact  linear algebra as explained in \cite{LyngMiller}.

\ex

\appendix
\renewcommand{\theequation}{\Alph{section}.\arabic{equation}}

\section{Estimate of the parametrix on a circle of large radius uniformly for large $y$}
\label{towards}

This section is rather technical; its goal is to prove that the sectorial analytic matrix function
$\wh \Psi(\xi;v)$, that is related to the solution $\P(\xi,v)$ of the RHP (\ref{P1RHP}) through
$\wh \Psi(\xi;v)=G(\xi,v)\P(\xi,v)e^{\vartheta(\xi,v)\s_3}$, has asymptotic behavior 
(\ref{hatpsiasym}) as $\xi\ra\infty$ uniformly with respect to $v-v_p=O(\e^\frac 15)$, where 
$v_p$ is one of the poles of the { tritronqu\'ee} solution $y(v)$ of P1, or, equivalently,  
uniformly with respect to $\parallel (x,t)-(x_p,t_p)\parallel =O(\e)$.

In fact, we will prove a somewhat stronger result, stated in Theorem \ref{convther} below. Corollary \ref{CorFinal} of this theorem
has been used in the main body of the paper. 

\begin{figure}
\resizebox{0.7\textwidth}{!}{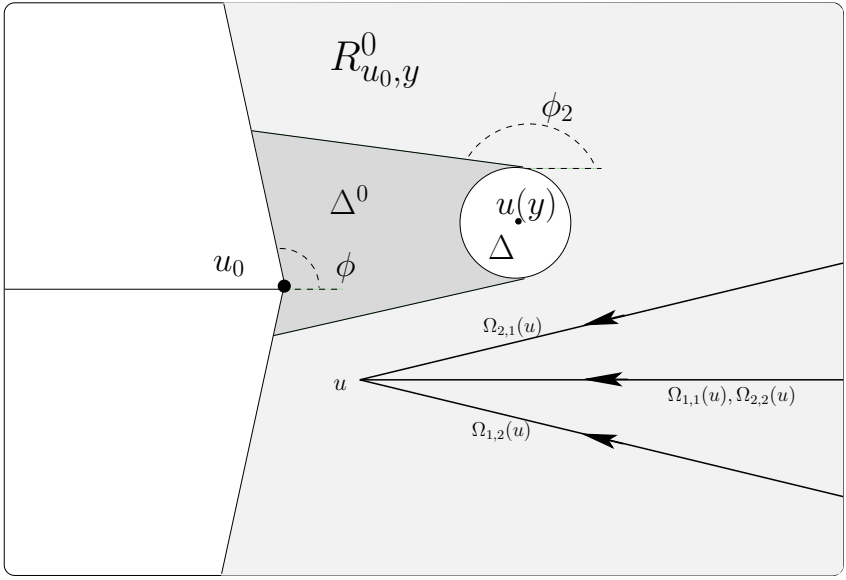}
\caption{Region $R_{u_0,y}$ is the complement of $\D$ and the union of $\D^0$ (darker) and $R_{u_0,y}^0$ (lighter). 
Shown in the picture are the contours of integration $\Omega_{i,j}(u)$, $u\in R_{u_0,y}^0$, for the $(i,j)$ entry of the Volterra operator (\ref{Volterra1}).}
\label{Sigma}
\end{figure}

\begin{theorem}\label{convther}
Let $S$ be one of the sectors shown on Figure \ref{RHPP1} and $\hat\Psi(\xi,v)=G(\xi,v)\P(\xi,v)e^{\vartheta(\xi,v)\s_3}$ be a solution of the ODE (\ref{hatPsiODE}), 
obtained from  the sectorial solutions $\P(\xi,v)$ of the Problem \ref{P1RHP} in the sector $S$. Let $\L$ be a constant nonsingular diagonal matrix  and
$\Dscr_y$ be a disk of fixed radius $r>0$ centered at $\xi=y$.
Then, there exist some constants $\xi_*>0$, $p_2>0$, both independent of $y$ and of each other, such that $\hat\Psi(\xi,v)$ has the representation 
\be\label{asshatPsi2}
\hat\Psi(\xi,v)=2^{\s_3/2}(4\xi^3-2\xi v)^{-\s_3/4}\frac{1}{\sqrt{2}}(\s_1+\s_3)T
\L\left(\1 +O\left(\xi^{-\hf},{ e}^{-p_2 |\frac{y}{\xi_*}|^{5/2}}\right)\right)e^{\q(\xi,v)\s_3}
\ee
where
\bea
&\&\q(\xi,v)=\vartheta(\xi,v)+\frac{y'}{4y}\int_\infty^\xi\frac{\le(s^\frac 32-\frac{v}{2\sqrt{s}}\ri)ds}{(s-y)s^2},   \label{qu} \\
T = \le (\1-\frac{i\s_2}{2} A_1 \ri)\le(\1- \frac{\s_1}{2} A_2 \ri)&\&~~~~~{\rm with}~~~~~A_1=\frac{6\xi^2-v}{2(4\xi^3-2\xi v)^{3/2}}\ ,\qquad A_2=\frac{y'\xi}{8\xi^3y(\xi-y)},
\label{A4}
\eea
uniformly for  $\xi\in S(\xi_*)= S\cap\{\xi:~|\xi|\geq \xi_*\}\setminus \Dscr_y$ and for all sufficiently large $y$. Here $\vartheta(\xi,v) =\frac 45 \xi^\frac 52 -v\sqrt{\xi}$ and
the contour of integration in (\ref{qu}) is a ray  in the direction of the bisector of $S$. In the case when $y\not\in S_{\xi_*}$ the $O$ term in (\ref{asshatPsi2}) is independent of $\frac{|y|}{\xi_*}$,
i.e., it is $O(\xi^{-\hf})$. The same $O$ term  is valid even if $y\in S_{\xi_*}$  
everywhere in $S_{\xi_*}$ except a certain``shadow'' of $\Dscr_y$ region, see Fig. \ref{Sigma} 
and description of this region below.

\end{theorem}

Proof of this theorem constitutes the bulk of this Appendix. Let us  consider sector $S$  that is 
not adjacent to  $\R_-$. This sector is bisected by the Stokes' ray $\ell$ where $\Re \xi^{\frac 5 2} =0$.
Let us extend $S$ by some angle $\d\in(0,\frac 25 \pi)$ in both directions.
For definiteness, we shall focus on the sector $S=\{\arg(\xi)\in (-\delta,\frac {2 \pi} 5+\delta)\}$, the remaining being treated similarly.
It is more convenient to work with variable $\vartheta(\xi,v)$ instead of $\xi$; therefore, we introduce a new variable
\be\label{umap}
u=u(\xi)=-i\vartheta(\xi;v) =-i\left[\frac{4}{5}\xi^{5/2}-v\xi^{\hf}\right]~.
\ee
We will consider $u(\xi)$ as an analytic function (with the principal determinations of the roots) on sectors  of the $\xi$--plane of width up  to 
$4\pi/5$ and for $|\xi|$ sufficiently large: therefore in any such region the map $u(\xi)$ is holomorphically {\em invertible} because
$v$ will be chosen uniformly bounded (in fact- in a neighborhood of $\pole$), and we will denote its inverse by $\xi(u)$. We will write for brevity $u$ or $\xi$ with the understanding that they may be viewed as functions of the other variable. 
A direct calculation shows that 
\be\label{ximap}
\xi=\xi(u)=\left(\frac{5i}{4}\right)^{2/5}u^{2/5}
\left[1+\frac{v{\rm e}^{-i\pi/5}}{5^{2/5}2^{1/5}} u^{-4/5}+O(u^{-8/5}) \right],
\ee
We will let $v-\pole$ be bounded above, say, by $2$, and,  since the branch-points of $u(\xi)$ are at $\pm \sqrt{v/2}$,  
the map $u(\xi)$ is invertible in any sectorial domain $\{|\xi|>|\sqrt\pole+1|$,  
as long as the opening of this domain is less than $2\pi$.

The image of $S$ under the map (\ref{umap}) will be denoted by $\S$. Then, asymptotically for large $u$,
\be\label{Sigmasec}
\S= \{-\phi<\arg (u)<\phi \}\ ,\ \ \ \frac 1 2\pi < \phi < \pi. 
\ee  
It will be convenient for us to consider (\ref{Sigmasec}) as the definition for $\S$, whereas $S$ is the preimage of $\S$, i.e., $S=u^{-1}(\S)$.
By $\S_{u_0}$ we denote sector $\S$ shifted from the origin to some $u_0\in \C$, i.e., $\S_{u_0}=\S+u_0$; for convenience we shall take $u_0\in \R_+$.
Without any loss of generality we assume $u_0$ to be  sufficiently large.
Our first step in proving Theorem \ref{convther} is to ``pre-normalize'' the ODE (\ref{hatPsiODE}), satisfied by $\hat\Psi$, written with respect 
to the independent variable $u$.

\bl
\label{lemmaWform}
Let $W(u)$ be defined by 
\be\label{A5}
\hat\Psi(\xi) = 2^{\s_3/2}(4\xi^3-2\xi v)^{-\s_3/4}\frac{1}{\sqrt{2}}(\s_1+\s_3) T(\xi)W(\xi)
\ee
where the matrix $T$ is defined by (\ref{A4}).
Then $W$ satisfies the ODE in the variable $u$
\be
W_u  = \le[\le(i + \frac {iy'\xi}{8\xi^3 y(\xi-y)}\ri)\s_3 + B(u)\ri] W,    \label{WODE}
\ee 
where $B(u) = \mathcal O(u^{-6/5})$ uniformly with respect to sufficiently large values of $y$ in $u\in R_{u_0,y}=\S_{u_0}\setminus \D$. 
Here $\D=u(\Dscr_y)$. 
\el
{\bf Proof.}
The transformation 
\be
\hat\Psi=2^{\s_3/2}(4\xi^3-2\xi v)^{-\s_3/4}\frac{1}{\sqrt{2}}(\s_1+\s_3) X
\ee
reduces the system (\ref{hatPsiODE})
to
\be\label{XODE}
X_\xi= \left[\sqrt{4\xi^3-2\xi v}\s_3 + \frac{6\xi^2-v}{2(4\xi^3-2\xi v)}\s_1 +\frac{M}{2\sqrt{4\xi^3-2\xi v}}(\s_3+i\s_2)\right]X~,
\ee
where 
\be\label{M}
M=2\wh H_I+M_2+M_3=\left(2H-\frac{y'}{y}\right)+\frac{y'\xi}{y(\xi-y)}+\frac{3}{4(\xi-y)^2}.
\ee
The three terms in $M$ have uniform bounds in $S_{\xi_0,y}=u^{-1}(R_{u_0,y})$, where $u_0=u(\xi_0)$:
\begin{itemize}
\item Since $\xi\not\in D$ we have $|M_3|<\frac 1 {r^2}$;
\item  According to (\ref{TaylorP1}) and the fact that $H'(v)=y(v)$, we have 
\be\label{Hterm}
2 \wh H_I=2H(v)-\frac{y'(v)}{y(v)}= 28 \beta + O(v-\pole);
\ee
\item According to (\ref{TaylorP1}), $(\ln y)'=-2\sqrt{y}(1+O(v-v_p)^4)$ for all sufficiently large $y$, so we now need to estimate $\frac{y^\hf\xi}{\xi-y}$.
Note that 
$\le|\frac{\sqrt{\frac{y}{\xi}}}{\frac{y}{ \xi} -1}\ri|\leq \frac{2}{d_0}$ 
outside the domain $|\xi-y|<d_0|\xi|$. Thus,
$M_2=O(\xi^\hf)$ outside this domain.
Inside  the latter domain but outside $\Dscr_y$ we have $\frac{y^\hf\xi}{\xi-y}=O(\xi^\frac 32)$, so that 
\be\label{M2n}
M_2=O(\xi^\frac 32)
\ee
for all $\xi\in S_{\xi_0,y}$ and uniformly in all sufficiently large $y$.
\end{itemize}

Rewriting (\ref{XODE}) in the variable $u$ we obtain 
\bea\label{XODEu}
X_u=i\frac {2\sqrt{\xi}}{4\xi^2-v}
\left[\sqrt{4\xi^3-2\xi v}\s_3 + \frac{6\xi^2-v}{2(4\xi^3-2\xi v)}\s_1 +\frac{M}{2\sqrt{4\xi^3-2\xi v}}(\s_3+i\s_2)\right]X = \cr
=i\le[\s_3 + \frac{6\xi^2-v}{2(4\xi^3-2\xi v)^{3/2}}\s_1 +\frac{M}{2(4\xi^3-2\xi v)}(\s_3+i\s_2) + \mathcal O(u^{-8/5})\ri]X.
\eea
Dropping all the terms in $M$  except $M_2$, the previous  equation   becomes (see the bulleted list above)
\bea\label{XODEu2}
X_u=i\le[\s_3 + \frac{6\xi^2-v}{2(4\xi^3-2\xi v)^{3/2}}\s_1 +\frac{y' \xi (\s_3+i\s_2)}{2y(\xi-y)(4\xi^3-2\xi v )} + \mathcal O(u^{-6/5})\ri]X.
\eea
Using (\ref{M2n}) we have 
\be
\frac{y'\xi }{2y(\xi-y)(4\xi^3-2\xi v )}  = \frac{y' }{8y\,\xi^2(\xi-y)} (1+ \mathcal O(\xi^{-2})) = \frac{y' }{8y\,\xi^2(\xi-y)}  + \mathcal O(u^{-\frac 75})
\ee
and hence 
\bea\label{XODEu3}
X_u=i\le[\s_3 + \frac{6\xi^2-v}{2(4\xi^3-2\xi v)^{3/2}}\s_1 +\frac{y'  (\s_3+i\s_2)}{8\xi^2y(\xi-y)} + \mathcal O(u^{-6/5})\ri]X
\eea
for $u\in R_{u_0,y}$ uniformly in $y$ (for sufficiently large $y$).
Setting $X = TW$ with $T$ as indicated in (\ref{A4}, \ref{A5}),   simplifying and keeping track of the orders already estimated,  yields the equation (\ref{WODE}). \QED

We will use the ODE (\ref{WODE}) in order to setup a recursive approximation scheme within each sector; the key is that the estimate for $B(u)$ in (\ref{WODE}) is uniform for $v$ in a neighborhood of the pole $\pole$ (hence,  for large values of $y$) and $|\xi - y|$ bounded below.

The domain  $\D = u(D_y)$ 
in the $u$ plane is of size $\mathcal O(|u(y)|^{\frac 35})$. For simplicity (and without real loss of generality) we modify $D$ to be the 
preimage of the exact disk $\D$ in the $u$-plane centered at  $u(y)$ of the radius $d_0|u(y)|^\frac 35$.

\bl
\label{lemmaVolterradiff}
Any solution $W(u)$ to (\ref{WODE}) can be written in the form $
W  = \Phi {\rm e}^{\q (u) \s_3}$ where 
\be
\frac{d}{d u} \q (u) = \le(i + \frac {iy'\xi}{8\xi^3 y(\xi-y)}\ri) \label{voltdiff}
\ee
and $\Phi(u)$ solves 
\be
\Phi' = \q '  \big[\s_3,\Phi\big] + B\,\Phi 
\ee
with $B=B(u)$ as in Lemma \ref{lemmaWform} (eq. (\ref{WODE})).
\el
{\bf Proof.} A direct substitution of the proposed expression into (\ref{WODE}). \QED

The differential equation (\ref{WODE}) is equivalent to the following {\bf Volterra} integral equation
\bea\label{Volterra1}
\Phi(u)=\L+e^{\q(u)}\left(\int_{\O (u)}e^{-\q(\eta)}B\Phi e^{\q(\eta)}d\eta\right)e^{-\q(u)\s_3} = \L +\Iscr[\Phi](u),
\eea
where $q$ from (\ref{voltdiff}) can be written as
\be
\q (u):=iu+\frac{iy'}{8y}\int^u_\infty \frac{du}{(\xi(u)-y)\xi^2(u)}.\label{A21}
\ee
The symbol $\int_{\O(u)}$ denotes the integration along a set of contours originating at $u$ and extending to $u=\infty$, 
with a different direction of contour $\O_{i,j}(u)$ for different entries of the matrix, see Fig. \ref{Sigma} for the
choice of the contours (the exponential of the integrand should decrease in the direction of the contour, traversed from $u$ to $\infty$). 
It is promptly seen from the fundamental theorem of calculus 
is equivalent to (\ref{voltdiff}).
The matrix $\L$ is a constant of integration, which, at this point, we choose  to be a {\bf diagonal} and nonsingular matrix, but otherwise undetermined.

Changing variable $u=u(\xi)$ in the integral (\ref{A21}) and using $\frac{du}{d\xi}=-2i\xi^{3/2}+\frac{iv}{\sqrt{\xi}}$ and
\be\label{inteta}
\int_\infty^\xi\frac{dt}{\sqrt{t}(t-y)}=\frac{1}{\sqrt{y}}\ln\frac{\sqrt{\xi}-\sqrt{y}}{\sqrt{\xi}+\sqrt{y}},
\ee
we find
\be\label{Q}
\q(u)=iu+\frac{y'}{4y^{3/2}}\ln\frac{{\xi}-{y}}{(\sqrt{\xi}+\sqrt{y})^2} +O(u^{-3/5})
\ee
in the region $R_{u_0,y}$ uniformly in $y$. Note that 
\bea
\frac {y'}{4 y^{\frac 32  } } =  - \frac 12  + \frac a {8}(v-\pole)^4 + \mathcal O((v-\pole)^5) = -\frac 1  2+ \mathcal O(y^{-2}) 
\label{A24}
\eea 
and hence 
\be\label{W0}
e^{\q (u)\s_3}=e^{\le(\frac 4 5 \xi^3 - v \xi^\frac 1 2\ri)\s_3}\left( \frac{\sqrt{\xi}+\sqrt{y}}{\sqrt{\xi-y}}\right)^{\s_3}(\1 + O(u^{-3/5}, y^{-2}))~
\ee
in the region $R_{u_0,y}$ uniformly for large $y$; the cut of the logarithm is taken from $y$ to $\xi\infty$ parallel to the Stokes line (bisecant to the sector).

The strategy is to show that the integral operator $\Iscr$, defined in (\ref{Volterra1})  is  contractive in a suitable space, 
so that a solution can be sought through an iteration scheme; the analysis depends on the region $u$ belongs to.
For practical reasons we state the following lemma.
\bl
\label{lemmaQ}
The function $\left( \frac{\sqrt{\xi}+\sqrt{y}}{\sqrt{\xi-y}}\right)$ and its inverse   satisfy the inequalities 
\be
\le|\frac{\sqrt{\xi}\pm\sqrt{y}}{\sqrt{\xi-y}} \ri| \leq  K_0 |\xi|^\frac 1 2 \label{Kest}
\ee
outside the circle $|\xi-y|\geq r$,
with $K_0$ a constant independent of $y$ (but depending on $r$).
\el
{\bf Proof.}
The function $\left( \frac{\sqrt{\xi}+\sqrt{y}}{\sqrt{\xi-y}}\right)$ is continuous (in fact smooth except at $\xi=0$ and $\xi=y$) on $\C\setminus \{y\}$ 
and has limit $1$ at $\xi=\infty$, being unbounded in any neighborhood of $\xi=y$. It can be also seen by calculus that it has no critical values in $\C\setminus\{0,y\}$ 
and hence the maximum is attained on a disk around $y$. The value there is easily estimated as in (\ref{Kest}). Similar arguments work for the inverse function.
\QED

\subsection{Convergence of iterations in $R_{u_0,y}^0$}
The region $R_{u_0,y}^0$ is the lighter shaded region in Fig. \ref{Sigma}; it is obtained by excising from $R_{u_0,y}$ the ``cone of shadow'' 
of the disk $\D$ with the angle $\phi_2\in (\phi,\pi)$ indicated in Fig. \ref{Sigma} chosen arbitrarily and fixed once and for all. 
The region $\D^0$ will be such cone of shadow.
For a given $u\in R^0_{u_0,y}$, the contours of integration $\O_{i,j}(u)$ are also shown in Fig. \ref{Sigma}. It is important that the collection of contours
$\O(u)\subset R_{u_0,y}^0$ for any $u\in R_{u_0,y}^0$.

\begin{lemma}\label{14.2a}
If a matrix-function $\chi(u)$ satisfies $\Vert  \chi(u)\Vert \leq c|u|^{-m}$ in $R^0_{u_0,y}$ for some $c>0$ and some $m\geq 0$ 
then $\Vert \Iscr_1 \chi(u)\Vert \leq cK_1|u|^{-m-1/5}$, where $\Iscr_1$ denoted the Volterra integral operator (\ref{Volterra1}) with the contours $\O(u)$
specified above. The constant $K_1$ does not depend on $\chi$, $u_0\in\R^+$ and $y\in\C$, but depend on $m$ (it was assumed above that $u_0$ and $|y|$ are 
large).  \end{lemma}
{\bf Proof.}
The diagonal entries of $B\chi$ are unaffected by the conjugation by ${\rm e}^{\q\s_3}$ and are of order $u^{-m-\frac 6 5}$: after integration they 
become of order $u^{-m-\frac 15}$. Regarding the  off--diagonal entries of $B\chi$, let $F(u)$ denote one of them;
then, according to (\ref{Volterra1}), $|F(u)|\leq cL |u|^{-m-\frac 6 5}$ on $ R_{u_0,y}$, where the constants $L>0$ and $m\geq 0$ do not depend on $y$ 
and  we have (from (\ref{W0}))
\be
{\rm e}^{\pm 2\q(u)}F(u)={\rm e}^{\pm 2i u}F(u)\left( \frac{\sqrt{\xi}+\sqrt{y}}{\sqrt{\xi-y}}\right)^2(\1 + O(u^{-3/5}, y^{-2})),  
\ee 
so that 
\be
\le|F(u)\left( \frac{\sqrt{\xi}+\sqrt{y}}{\sqrt{\xi-y}}\right)(\1 + O(u^{-3/5}, y^{-2})) \ri|\leq cL K_0  \sqrt{|\xi|} |u|^{-m} =   c LK_0  |u|^{-m-1}. 
\ee
Due to the choice of the contours $\O(u)$, the integration along the corresponding $\O_{i,j}(u)$ can change the previous estimate 
only by a constant that does not depend on $u_0$ and $y$ (but depends on $\phi_2$).
\QED

Let us consider successive iterations $\Phi_n=\Iscr_1 \Phi_{n-1}$, $n\geq 1$, of equation  (\ref{Volterra1}), where $\Phi_0=0$.
Let $\D \Phi_n=\Phi_n-\Phi_{n-1}$. Then $\Phi_1=\D \Phi_1=\L$ and $\D \Phi_n=\Iscr_1 \D \Phi_{n-1}$.  
We are now going to prove uniform in $y$ convergence of the series $\sum_1^\infty \D \Phi_n$ in different subregions of $R_{u_0,y}$, provided
that $u_0$, which is independent of $y$, is large enough. In all the analysis below $y$ is assumed to be large.

Let $\Vert \L\Vert =c$. Then, according to Lemma \ref{14.2a},  $\Vert \D\Phi_2\Vert\leq cK_1 |u|^{-1/5}$ in $R^0_{u_0,y}$ uniformly in $u_0$ and in $y$.
We can choose $u_0\in\R^+$ so large that $K_1 |u|^{-1/5}<\hf$ for all $u\in R^0_{u_0,y}$.
This choice of $u_0$ guarantees convergence of the series 
$\sum_1^\infty \D \Phi_n$ in  $R^0_{u_0,y}$  uniformly in $y$. Moreover, we obtain the estimate
\begin{equation}\label{estS0}
\Phi(u)=\L[\1 + O(u^{-1/5})]
\end{equation}
in  $R^0_{u_0,y}$  uniformly in $y$. 

Let $\Rscr_{\xi_0,y}$ denote the preimage of $R^0_{u_0,y}$
under the map $u=u(\xi)$, where $u_0=u(\xi_0)$.
Going back to the system (\ref{hatPsiODE}), we obtain the following result.

\begin{lemma}\label{convlem1}
The solution $\hat\Psi(\xi,v)$ to the system (\ref{hatPsiODE}) has behavior
\be\label{asshatPsi1}
\hat\Psi(\xi,v)=2^{\s_3/2}(4\xi^3-2\xi v)^{-\s_3/4}\frac{1}{\sqrt{2}}(\s_1+\s_3)T\L(\1+O(\xi^{-\hf}))e^{\q(u(\xi))}
\ee
in $\Rscr_{\xi_0,y}$ uniformly in $y$  provided $\xi_0$ is large enough. Moreover, for a fixed diagonal $\L$, condition (\ref{asshatPsi1}) \it{uniquely}
determines the solution $\hat\Psi(\xi,v)$ of (\ref{hatPsiODE}).
\end{lemma}

Comparison of (\ref{asshatPsi1}) and (\ref{Psihat}) together with the asymptotics of $\Psi$ (\ref{P1expansion}) and (\ref{normalpsihat})  yields 
\be\label{matC}
\L={\rm diag}(1,i).
\ee

\subsection{Convergence of iterations in $\D^0$ }

If $y$ does not belong to the sector $S$  we are considering, we can always arrange that $\Dscr_y\cap S=\emptyset$,
so that $\Rscr_{\xi_0,y}$ coincides with the preimage of $\S_{u_0}$ and, thus,  
the estimate of Lemma \ref{convlem1} holds throughout that sector.
We want to extend the statement of 
Lemma \ref{convlem1} into the preimage $\Dscr^0$ of the region $\D^0$ for the case that $y\in S$.
Because of the construction of $S$ we can assume,  without any loss of generality, that $y\in \wh S$,
where $\wh S$ is any proper subsector of $S$.

Consider  the tangent line $\l$ to the disk $\D$ (which is centered at $u(y)$) that is parallel to $e^{-i\phi_1}$
and  located above $\D$, i.e., if $u_*$ is the point of tangency, then $\arg u_*>\arg u(y)$. Here $\phi_1\in(\phi,\phi_2)$.
It is clear that
$\l$ divides $\S_{u_0}$ into two regions. Let $R^1_{u_0,y}$ denote one of these regions, namely, the one
that does not contain $\D$. We want to extend the statement of 
Lemma \ref{convlem1} into the region $R^1_{u_0,y}$. To this end, we construct a solution $\tilde\Phi$
of the integral equation (\ref{Volterra1}) in $R^1_{u_0,y}$ by successive iterations.
Let $u_1$ denote the point of intersection
of $\partial\S_{u_0}$ and $\l$.
The collection of contours of integration $\O(u)$   in the case of the region $R^1_{u_0,y}$ is similar
to the case of the region $R^0_{u_0,y}$, considered above, except that the contour of integration
for the entry $(1,2)$ is the segment $[u_1, u]$, provided that $[u_1, u]\subset R^1_{u_0,y}$.
For $u\in R^1_{u_0,y}$ that do not satisfy 
the latter condition, the contour of integration is the union of $[u_1,u_0]\cup[u_0,u]$.

Let $\Iscr_2$ denote the integral operator in the Volterra equation (\ref{Volterra1}) in the region $R^1_{u_0,y}$ 
with the contours defined above. We can repeat  the previous estimates of integrals  to extend the statement of 
Lemma \ref{14.2a} to the integral operator  $\Iscr_2$ in the  region $R^1_{u_0,y}$: the only difference is the 
finite contour of integration, where the desired estimate comes from Lemma 14.2, \cite{Wasow}. 
The solution $\tilde\Phi$ to (\ref{Volterra1}) thus  satisfies  the same estimate of Lemma \ref{convlem1}. 
We also have  
\begin{equation}\label{estS1}
\tilde\Phi(u)=\L\le(\1+O(u^{-1/5})\ri)
\end{equation}
in  $R^1_{u_0,y}$  uniformly in $y$. Here $\L$ is given by (\ref{matC}).

\begin{figure}
\resizebox{0.7\textwidth}{!}{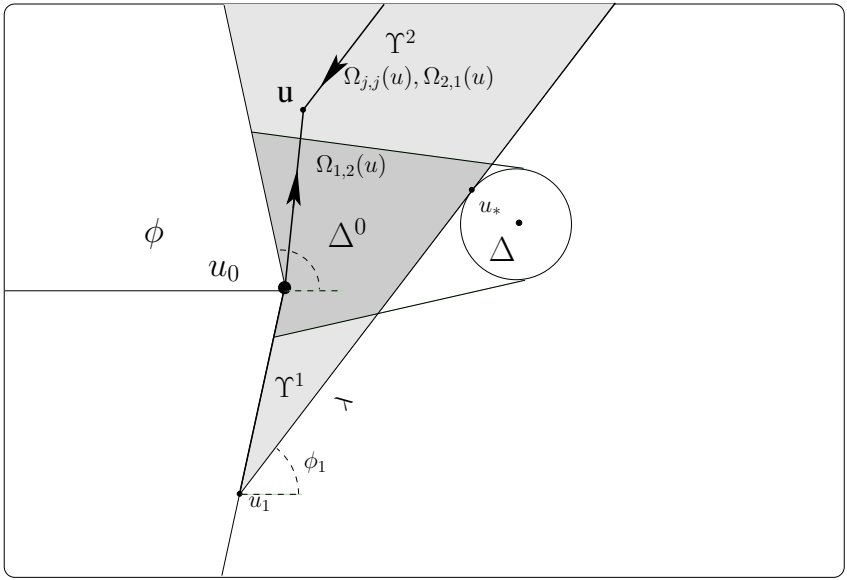}
\caption{The region $R^1_{u_0,y}$ (shaded)  and its two subregions $\Upsilon^{1}_{u_0,y}$, $\Upsilon^2_{u_0,y}$ (lighter shading).
The goal is to extend the estimate (\ref{asshatPsi1}) into the remaining third (darker shading) subregion of $R^1_{u_0,y}$.} 
\label{Sigma0}
\end{figure}

Let $u_0\in \R_+$ be fixed:  for  any sufficiently large $|u(y)|$   the set $R^0_{u_0,y}\cap R^1_{u_0,y}$ consists of two disjoint 
components $\Upsilon^{1,2}_{u_0,y}$ shown in Fig. \ref{Sigma0} (otherwise, we can extend $\D$ in such a way that it will intersect the boundary
of $\S_{u_0}$ and the corresponding $\D^0$ disappear).
Let  $S(y)$ denote the Stokes matrix connecting solutions of (\ref{WODE})  $\tilde\Phi{\rm e}^{\q\s_3}$
and the previously constructed one $\Phi{\rm  e}^{\q\s_3}$. Then 
\be\label{St}
\Phi(u,y)=\tilde\Phi(u,y)e^{\q(u,y)\s_3}S(y)e^{-\q(u,y)\s_3},\qquad S(y)=\begin{pmatrix}
                   1& s(y)\\
                   0& 1
                  \end{pmatrix},
\ee
where the triangularity of $S(y)$ follows from the fact that  $\Phi, \tilde\Phi\to \L$ as $u\ra\infty$, $u\in\Upsilon^{2}_{u_0,y}$. 
Writing $\q=\q(u,y)$ in (\ref{St}), we emphasize that $\q$  depends on $y$.

\paragraph{Estimate for $\mathbf {s(y)}$.} We now want to obtain an estimate for $s(y)$ by comparing $\Phi$ and $\tilde\Phi$ in the region $\Upsilon_{u_0,y}^{1}$.  
Let us denote by $\wh \Phi$ the analytic continuation of $\Phi$ from $\Upsilon^2_{u_0,y}$ throughout $R^1_{u_0,y}$ (throughout the dark shaded region)
into $\Upsilon^1_{u_0,y}$. 
According to (\ref{Psihat}), $\hat\Psi(\xi,y)$ has monodromy $-\1$ as $\xi$ goes around $\xi=y$. Because of (\ref{A5}), 
$W = \Phi\, {\rm e}^{\q \s_3}$ has the same  monodromy. It follows from (\ref{Q}) 
that ${\rm e}^{\q\s_3}$ has monodromy $Mm_y:= \exp\le[2i\pi  \frac {y'}{4 y^\frac 32}\s_3\ri]$ around $\xi=y$.
Therefore we have 
\be
\wh \Phi(u;y) = - \Phi(u;y)m_y^{-1}\ ,\qquad u\in \Upsilon_{u_0,y}^1.
\ee
The matrix $m_y$ (which is constant in $u$!) has the behavior $m_y = -\1 + \mathcal O(y^{-2})$ on account of (\ref{A24}). 
According to (\ref{estS0} and the fact that $m_y$ is uniformly bounded (for all large $y$), we have 
\be
\|\wh \Phi-\L\|\leq K_2|u|^{-\frac 15},~~\| \Phi-\L\|\leq K_2|u|^{-\frac 15}\ \hbox{ in } \Upsilon^1_{u_0,y},
\ee
with a constant $K_2$ independent of $y$. But, according to   (\ref{estS1}),  $\wt\Phi$ too satisfies the same estimate.
Substituting these estimates into (\ref{St}), we obtain
\be
s(y){\rm e}^{2 \q (u)} =O(u^{-\frac 15})  \label{Cs1}
\ee
in $\Upsilon^1_{u_0,y}$ uniformly in $y$. In particular, according to (\ref{Q}),  $|s(y)|\leq K_3 |u_1|^\frac 15 e^{2\Im u_1}$ 
for some $K_3>0$ that is independent of $y$. Therefore, in the region $\D^0\cup R^1_{u_0,y}$ (darker shade on Fig. \ref{Sigma0}),
we have 
\be
\le|s(y){\rm e}^{2 \q (u)}\ri| \leq K_3 |u_1|^\frac 15 e^{2\Im( u_1-u)}  \label{Cs1a}
\ee
Given the geometry of regions and the fact that $y\in \wh S$, we have 
\be\label{Cs1b}
\Im(u-u_1) \geq K_4 |u(y)-u_1|\geq K_5|u(y)-u_0| ,\ \ \forall u\in \D^0,
\ee
where $K_4,K_5$ denote some positive constants independent of $u_0,y$. Thus, the matrix $e^{\q(u,y)\s_3}S(y)e^{-\q(u,y)\s_3}$ in (\ref{St}) is exponentially
close to $\1$ as $y-\xi_0\ra\infty$, where $u(\xi_0)=u_0$. So, according to (\ref{St}), 
the estimate of Lemma \ref{convlem1} (with a modified $O$ term) can be extended  to $\D^0$ (more precisely, to the shaded part of it $\D^0\cap R^{1}_{u_0,y}$), provided $y-\x_0\ra \infty$.
Because of the construction of $\wh S$, there exists $\mu>0$, such
that $\wh S$ contains $S(\mu |\xi_0|)$, where by $S$ now we mean the original sector of 
the RHP (\ref{RHPP1}), see Theorem (\ref{convther}). We denote $\xi_*=\mu|\xi_0|$.
Existence of the positive constant $p_2$ in the estimate (\ref{asshatPsi2}) follows from (\ref{St}) (\ref{Cs1a}) and  (\ref{Cs1b}). 
The case of sectors $S$ adjacent to the jump contour $\R_-$ on Fig. \ref{RHPP1} can be considered similarly since we can rotate  this
contour in both directions  by  angle up to $\frac{\pi}{5}$. 

The  proof of Theorem \ref{convther} is completed.

According to (\ref{A4}) and (\ref{M2n}), the matrix $T$ from (\ref{asshatPsi2}) can be absorbed into $\1+O(\xi^{-\hf})$ term.
To within the same estimate we can replace also $\sqrt{4\xi^3 - v\xi} $ by $2 \xi^\frac 3 2 $ and recast the theorem into 

\bc
\label{CorFinal}%
Under the same assumptions and in the same notations of Theorem \ref{convther} we have 
\bea
\hat\Psi(\xi,v)=\xi^{-\frac 3 4 \s_3}\frac{1}{\sqrt{2}}(\s_1+\s_3)\left(\L+O\left(\xi^{-\hf}, y^{-4},{ e}^{-p_2 |\frac{y}{\xi_*}|^{5/2}}\right)\right)
\le(\frac {\sqrt{\xi}+ \sqrt{y}}{\sqrt{\xi-y}}\ri)^{\sigma_3}{\rm e}^{\vartheta(\xi;v)\sigma_3}\eea
{with the understanding that  $|\xi|\to\infty$ and $|y|\to \infty$ and $|\xi-y|$ is bounded uniformly away from zero.}
\ec
Theorem \ref{convther} together with the convergence of iterations imply the following corollary.

\begin{coroll}
For any sufficiently large $\xi$ we have $\lim_{v\ra\pole}\hat\Psi(\xi,v)=\hat\Psi(\xi,\pole)$. 
\end{coroll}

\paragraph{Acknowledgments.} The authors are grateful to Percy Deift
and to Peter Miller for useful discussions. A. T. thanks the hospitality and support of the  Mathematical Physics Laboratory at the {\em Centre de recherches math\'ematiques} where part of the work was carried out. 
M. B. was supported by the NSERC Discovery Grant "Exact and asymptotic methods in Random Matrix Theory and Integrable Systems".

%

%\red{[We either use initial or full names for all the bibitems in the same way.] }

%\bibliographystyle{plain}
%\bibliography{NLS-Painleve-Universality.bbl}
\end{document}

%% file: Strip.pdf_t
\begin{picture}(0,0)%
\includegraphics{Strip.pdf}%
\end{picture}%
\setlength{\unitlength}{3947sp}%
\begingroup\makeatletter\ifx\SetFigFont\undefined%
\gdef\SetFigFont#1#2#3#4#5{%
  \reset@font\fontsize{#1}{#2pt}%
  \fontfamily{#3}\fontseries{#4}\fontshape{#5}%
  \selectfont}%
\fi\endgroup%
\begin{picture}(12105,11925)(2326,-13186)
\put(8701,-5011){\makebox(0,0)[lb]{\smash{{\SetFigFont{25}{30.0}{\familydefault}{\mddefault}{\updefault}{\color[rgb]{.69,0,0}$\mathcal O(\ve^{\frac 45})$}%
}}}}
\put(7801,-9286){\makebox(0,0)[lb]{\smash{{\SetFigFont{25}{30.0}{\familydefault}{\mddefault}{\updefault}{\color[rgb]{0,0,0}$\mathcal O(\epsilon)$}%
}}}}
\put(6751,-4111){\makebox(0,0)[lb]{\smash{{\SetFigFont{25}{30.0}{\familydefault}{\mddefault}{\updefault}{\color[rgb]{0,0,0}Umbilical grad catastrophe}%
}}}}
\put(9601,-9061){\makebox(0,0)[lb]{\smash{{\SetFigFont{25}{30.0}{\familydefault}{\mddefault}{\updefault}{\color[rgb]{.69,0,0}$\mathcal O(\epsilon)$}%
}}}}
\put(8926,-8311){\makebox(0,0)[lb]{\smash{{\SetFigFont{25}{30.0}{\familydefault}{\mddefault}{\updefault}{\color[rgb]{.69,0,0}$\mathcal O(\epsilon\ln(\epsilon))$}%
}}}}
\end{picture}%

%% file: wgraph.pdf_t
\begin{picture}(0,0)%
\includegraphics{wgraph.pdf}%
\end{picture}%
\setlength{\unitlength}{3947sp}%
\begingroup\makeatletter\ifx\SetFigFont\undefined%
\gdef\SetFigFont#1#2#3#4#5{%
  \reset@font\fontsize{#1}{#2pt}%
  \fontfamily{#3}\fontseries{#4}\fontshape{#5}%
  \selectfont}%
\fi\endgroup%
\begin{picture}(6024,1836)(1189,-4883)
\put(4201,-4336){\makebox(0,0)[lb]{\smash{{\SetFigFont{17}{20.4}{\familydefault}{\mddefault}{\updefault}{\color[rgb]{0,0,0}0}%
}}}}
\put(5026,-4336){\makebox(0,0)[lb]{\smash{{\SetFigFont{17}{20.4}{\familydefault}{\mddefault}{\updefault}{\color[rgb]{0,0,0}$\mu_+$}%
}}}}
\put(2251,-4336){\makebox(0,0)[lb]{\smash{{\SetFigFont{17}{20.4}{\familydefault}{\mddefault}{\updefault}{\color[rgb]{0,0,0}$\mu_-$}%
}}}}
\end{picture}%

%% file: PhaseDiagram.pdf_t
\begin{picture}(0,0)%
\includegraphics{PhaseDiagram.pdf}%
\end{picture}%
\setlength{\unitlength}{3947sp}%
\begingroup\makeatletter\ifx\SetFigFont\undefined%
\gdef\SetFigFont#1#2#3#4#5{%
  \reset@font\fontsize{#1}{#2pt}%
  \fontfamily{#3}\fontseries{#4}\fontshape{#5}%
  \selectfont}%
\fi\endgroup%
\begin{picture}(18270,12310)(1651,-8191)
\end{picture}%

%% file: Y_RHP.pdf_t
\begin{picture}(0,0)%
\includegraphics{Y_RHP.pdf}%
\end{picture}%
\setlength{\unitlength}{3947sp}%
\begingroup\makeatletter\ifx\SetFigFont\undefined%
\gdef\SetFigFont#1#2#3#4#5{%
  \reset@font\fontsize{#1}{#2pt}%
  \fontfamily{#3}\fontseries{#4}\fontshape{#5}%
  \selectfont}%
\fi\endgroup%
\begin{picture}(7813,5219)(707,-4583)
\put(4651,-886){\makebox(0,0)[lb]{\smash{{\SetFigFont{20}{24.0}{\familydefault}{\mddefault}{\updefault}{\color[rgb]{0,0,0}$\alpha$}%
}}}}
\put(5776,-2386){\makebox(0,0)[lb]{\smash{{\SetFigFont{14}{16.8}{\familydefault}{\mddefault}{\updefault}{\color[rgb]{0,0,.56}$\begin{bmatrix} 0&1\\-1 & 0\end{bmatrix}$}%
}}}}
\put(6451,-1636){\makebox(0,0)[lb]{\smash{{\SetFigFont{14}{16.8}{\familydefault}{\mddefault}{\updefault}{\color[rgb]{.56,0,0}$\begin{bmatrix}1 &  - {\rm e}^{-\frac {2i}\epsilon h} \\  0 & 1 \end{bmatrix} $ }%
}}}}
\put(3826,-3286){\makebox(0,0)[lb]{\smash{{\SetFigFont{14}{16.8}{\familydefault}{\mddefault}{\updefault}{\color[rgb]{.56,0,0}$\begin{bmatrix}1 &  - {\rm e}^{-\frac {2i}\epsilon h} \\  0 & 1 \end{bmatrix} $ }%
}}}}
\put(1366,-3451){\makebox(0,0)[lb]{\smash{{\SetFigFont{14}{16.8}{\familydefault}{\mddefault}{\updefault}{\color[rgb]{0,0,0}$\begin{bmatrix}1&0\\ - {\rm e}^{\frac {2i}\epsilon h} & 1 \end{bmatrix} $ }%
}}}}
\put(4516,-2071){\makebox(0,0)[lb]{\smash{{\SetFigFont{14}{16.8}{\familydefault}{\mddefault}{\updefault}{\color[rgb]{0,.56,0}$\1$}%
}}}}
\end{picture}%

%% file: BreakingCusp.pdf_t
\begin{picture}(0,0)%
\includegraphics{BreakingCusp.pdf}%
\end{picture}%
\setlength{\unitlength}{3947sp}%
\begingroup\makeatletter\ifx\SetFigFont\undefined%
\gdef\SetFigFont#1#2#3#4#5{%
  \reset@font\fontsize{#1}{#2pt}%
  \fontfamily{#3}\fontseries{#4}\fontshape{#5}%
  \selectfont}%
\fi\endgroup%
\begin{picture}(19425,21852)(-2078,-14501)
\put(14701,-5911){\makebox(0,0)[lb]{\smash{{\SetFigFont{41}{49.2}{\familydefault}{\mddefault}{\updefault}{\color[rgb]{0,0,0}$\tau=0$}%
}}}}
\put(6162,-9743){\makebox(0,0)[lb]{\smash{{\SetFigFont{41}{49.2}{\familydefault}{\mddefault}{\updefault}{\color[rgb]{0,0,0}$\phi=\frac {6\pi}5$}%
}}}}
\put(3751,-5236){\makebox(0,0)[lb]{\smash{{\SetFigFont{41}{49.2}{\familydefault}{\mddefault}{\updefault}{\color[rgb]{0,0,0}$\phi=\pi$}%
}}}}
\put(6076,-1486){\makebox(0,0)[lb]{\smash{{\SetFigFont{41}{49.2}{\familydefault}{\mddefault}{\updefault}{\color[rgb]{0,0,0}$\phi= \frac {4\pi}5$}%
}}}}
\put(7862,222){\makebox(0,0)[lb]{\smash{{\SetFigFont{41}{49.2}{\familydefault}{\mddefault}{\updefault}{\color[rgb]{0,0,0}$\phi=\frac {3\pi }5$}%
}}}}
\put(12962,-1044){\makebox(0,0)[lb]{\smash{{\SetFigFont{41}{49.2}{\familydefault}{\mddefault}{\updefault}{\color[rgb]{0,0,0}$\phi = \frac {2\pi }5$}%
}}}}
\end{picture}%

%% file: RHPP1.pdf_t
\begin{picture}(0,0)%
\includegraphics{RHPP1.pdf}%
\end{picture}%
\setlength{\unitlength}{3947sp}%
\begingroup\makeatletter\ifx\SetFigFont\undefined%
\gdef\SetFigFont#1#2#3#4#5{%
  \reset@font\fontsize{#1}{#2pt}%
  \fontfamily{#3}\fontseries{#4}\fontshape{#5}%
  \selectfont}%
\fi\endgroup%
\begin{picture}(8969,8176)(129,-8049)
\put(2326,-1861){\makebox(0,0)[lb]{\smash{{\SetFigFont{20}{24.0}{\familydefault}{\mddefault}{\updefault}{\color[rgb]{0,0,0}$\begin{bmatrix} 1&0 \cr \beta_2 {\rm e}^{- 2\vartheta}&1\end{bmatrix}  $}%
}}}}
\put(2776,-3511){\makebox(0,0)[lb]{\smash{{\SetFigFont{29}{34.8}{\rmdefault}{\mddefault}{\updefault}{\color[rgb]{0,0,0}3}%
}}}}
\put(4276,-2611){\makebox(0,0)[lb]{\smash{{\SetFigFont{29}{34.8}{\rmdefault}{\mddefault}{\updefault}{\color[rgb]{0,0,0}2}%
}}}}
\put(6151,-3436){\makebox(0,0)[lb]{\smash{{\SetFigFont{29}{34.8}{\rmdefault}{\mddefault}{\updefault}{\color[rgb]{0,0,0}1}%
}}}}
\put(6076,-5086){\makebox(0,0)[lb]{\smash{{\SetFigFont{29}{34.8}{\rmdefault}{\mddefault}{\updefault}{\color[rgb]{0,0,0}0}%
}}}}
\put(4126,-5386){\makebox(0,0)[lb]{\smash{{\SetFigFont{29}{34.8}{\rmdefault}{\mddefault}{\updefault}{\color[rgb]{0,0,0}-1}%
}}}}
\put(2701,-4861){\makebox(0,0)[lb]{\smash{{\SetFigFont{29}{34.8}{\rmdefault}{\mddefault}{\updefault}{\color[rgb]{0,0,0}-2}%
}}}}
\put(1051,-3361){\makebox(0,0)[lb]{\smash{{\SetFigFont{20}{24.0}{\familydefault}{\mddefault}{\updefault}{\color[rgb]{0,0,0}$\begin{bmatrix}0&-1\cr 1&0\end{bmatrix}$}%
}}}}
\put(2251,-6361){\makebox(0,0)[lb]{\smash{{\SetFigFont{20}{24.0}{\familydefault}{\mddefault}{\updefault}{\color[rgb]{0,0,0}$\begin{bmatrix}1& 0  \cr \beta_{-2} {\rm e}^{ - 2\vartheta} &1\end{bmatrix}$}%
}}}}
\put(5926,-1261){\makebox(0,0)[lb]{\smash{{\SetFigFont{20}{24.0}{\familydefault}{\mddefault}{\updefault}{\color[rgb]{0,0,0}$\begin{bmatrix}1&\beta_1{\rm e}^{ 2\vartheta} \cr  0 &1\end{bmatrix}$}%
}}}}
\put(7351,-4486){\makebox(0,0)[lb]{\smash{{\SetFigFont{20}{24.0}{\familydefault}{\mddefault}{\updefault}{\color[rgb]{0,0,0}$\begin{bmatrix}1&0 \cr \beta_0 {\rm e}^{ - 2\vartheta}&1\end{bmatrix}$}%
}}}}
\put(5926,-7036){\makebox(0,0)[lb]{\smash{{\SetFigFont{20}{24.0}{\familydefault}{\mddefault}{\updefault}{\color[rgb]{0,0,0}$\begin{bmatrix}1& \beta_{-1}{\rm e}^{2\vartheta} \cr 0  &1\end{bmatrix}$}%
}}}}
\end{picture}%

%% file: PainPlane.pdf_t
\begin{picture}(0,0)%
\includegraphics{PainPlane.pdf}%
\end{picture}%
\setlength{\unitlength}{3947sp}%
\begingroup\makeatletter\ifx\SetFigFont\undefined%
\gdef\SetFigFont#1#2#3#4#5{%
  \reset@font\fontsize{#1}{#2pt}%
  \fontfamily{#3}\fontseries{#4}\fontshape{#5}%
  \selectfont}%
\fi\endgroup%
\begin{picture}(6965,8030)(1356,-7661)
\put(6045,-2765){\rotatebox{18.0}{\makebox(0,0)[lb]{\smash{{\SetFigFont{25}{30.0}{\familydefault}{\mddefault}{\updefault}{\color[rgb]{0,0,0}$\mathcal O(\varepsilon^{\frac 1 5})$}%
}}}}}
\put(5101,-4861){\rotatebox{18.0}{\makebox(0,0)[lb]{\smash{{\SetFigFont{20}{24.0}{\familydefault}{\mddefault}{\updefault}{\color[rgb]{0,0,0}$\frac {8\pi}5$ }%
}}}}}
\put(6212,-3393){\makebox(0,0)[lb]{\smash{{\SetFigFont{20}{24.0}{\familydefault}{\mddefault}{\updefault}{\color[rgb]{0,0,0}$\frac{2\pi}5$}%
}}}}
\end{picture}%

%% file: Crit_mu2.pdf_t
\begin{picture}(0,0)%
\includegraphics{Crit_mu2.pdf}%
\end{picture}%
\setlength{\unitlength}{3947sp}%
\begingroup\makeatletter\ifx\SetFigFont\undefined%
\gdef\SetFigFont#1#2#3#4#5{%
  \reset@font\fontsize{#1}{#2pt}%
  \fontfamily{#3}\fontseries{#4}\fontshape{#5}%
  \selectfont}%
\fi\endgroup%
\begin{picture}(6114,5489)(1849,-7348)
\put(4760,-4989){\rotatebox{300.0}{\makebox(0,0)[lb]{\smash{{\SetFigFont{14}{16.8}{\familydefault}{\mddefault}{\updefault}{\color[rgb]{0,0,0}Main arc}%
}}}}}
\put(5318,-4732){\makebox(0,0)[lb]{\smash{{\SetFigFont{12}{14.4}{\familydefault}{\mddefault}{\updefault}{\color[rgb]{0,0,0}$\vartheta =-\frac{3\pi}{10}$}%
}}}}
\put(4231,-5536){\makebox(0,0)[lb]{\smash{{\SetFigFont{12}{14.4}{\familydefault}{\mddefault}{\updefault}{\color[rgb]{0,0,0}1}%
}}}}
\put(4349,-4204){\makebox(0,0)[lb]{\smash{{\SetFigFont{12}{14.4}{\familydefault}{\mddefault}{\updefault}{\color[rgb]{0,0,0}2}%
}}}}
\put(4322,-2817){\makebox(0,0)[lb]{\smash{{\SetFigFont{12}{14.4}{\familydefault}{\mddefault}{\updefault}{\color[rgb]{0,0,0}3}%
}}}}
\put(3210,-7177){\makebox(0,0)[lb]{\smash{{\SetFigFont{12}{14.4}{\familydefault}{\mddefault}{\updefault}{\color[rgb]{0,0,0}-1}%
}}}}
\put(1876,-7171){\makebox(0,0)[lb]{\smash{{\SetFigFont{12}{14.4}{\familydefault}{\mddefault}{\updefault}{\color[rgb]{0,0,0}-2}%
}}}}
\put(5944,-7167){\makebox(0,0)[lb]{\smash{{\SetFigFont{12}{14.4}{\familydefault}{\mddefault}{\updefault}{\color[rgb]{0,0,0}1}%
}}}}
\put(7231,-7182){\makebox(0,0)[lb]{\smash{{\SetFigFont{12}{14.4}{\familydefault}{\mddefault}{\updefault}{\color[rgb]{0,0,0}2}%
}}}}
\end{picture}%

%% file: Peak.pdf_t
\begin{picture}(0,0)%
\includegraphics{Peak.pdf}%
\end{picture}%
\setlength{\unitlength}{3947sp}%
\begingroup\makeatletter\ifx\SetFigFont\undefined%
\gdef\SetFigFont#1#2#3#4#5{%
  \reset@font\fontsize{#1}{#2pt}%
  \fontfamily{#3}\fontseries{#4}\fontshape{#5}%
  \selectfont}%
\fi\endgroup%
\begin{picture}(22383,12683)(1276,-12594)
\put(14701,-7336){\makebox(0,0)[lb]{\smash{{\SetFigFont{25}{30.0}{\familydefault}{\mddefault}{\updefault}{\color[rgb]{0,0,1}$t_0=0.25$}%
}}}}
\put(13726,-4861){\makebox(0,0)[lb]{\smash{{\SetFigFont{25}{30.0}{\familydefault}{\mddefault}{\updefault}{\color[rgb]{.69,0,0}$t=0.2800$}%
}}}}
\put(13726,-1936){\makebox(0,0)[lb]{\smash{{\SetFigFont{25}{30.0}{\familydefault}{\mddefault}{\updefault}{\color[rgb]{0,0,0}$|q(0,t_{peak})|\sim 3.9256$}%
}}}}
\put(6301,-6811){\makebox(0,0)[lb]{\smash{{\SetFigFont{25}{30.0}{\familydefault}{\mddefault}{\updefault}{\color[rgb]{0,0,0}$|q(0,t_0)|\sim 1.3137$}%
}}}}
\end{picture}%

%% file: Sigma.pdf_t
\begin{picture}(0,0)%
\includegraphics{Sigma.pdf}%
\end{picture}%
\setlength{\unitlength}{3947sp}%
\begingroup\makeatletter\ifx\SetFigFont\undefined%
\gdef\SetFigFont#1#2#3#4#5{%
  \reset@font\fontsize{#1}{#2pt}%
  \fontfamily{#3}\fontseries{#4}\fontshape{#5}%
  \selectfont}%
\fi\endgroup%
\begin{picture}(8364,5714)(809,-5767)
\put(4088,-3909){\makebox(0,0)[lb]{\smash{{\SetFigFont{14}{16.8}{\familydefault}{\mddefault}{\updefault}{\color[rgb]{0,0,0}$u$}%
}}}}
\put(7409,-4024){\makebox(0,0)[lb]{\smash{{\SetFigFont{12}{14.4}{\familydefault}{\mddefault}{\updefault}{\color[rgb]{0,0,0}$\O_{1,1}(u), \O_{2,2}(u)$}%
}}}}
\put(4111,-2776){\makebox(0,0)[lb]{\smash{{\SetFigFont{20}{24.0}{\familydefault}{\mddefault}{\updefault}{\color[rgb]{0,0,0}$\phi$}%
}}}}
\put(6151,-1186){\makebox(0,0)[lb]{\smash{{\SetFigFont{20}{24.0}{\familydefault}{\mddefault}{\updefault}{\color[rgb]{0,0,0}$\phi_2$}%
}}}}
\put(4051,-2161){\makebox(0,0)[lb]{\smash{{\SetFigFont{20}{24.0}{\familydefault}{\mddefault}{\updefault}{\color[rgb]{0,0,0}$\D^0$}%
}}}}
\put(5626,-2611){\makebox(0,0)[lb]{\smash{{\SetFigFont{20}{24.0}{\familydefault}{\mddefault}{\updefault}{\color[rgb]{0,0,0}$\D$}%
}}}}
\put(2896,-2731){\makebox(0,0)[lb]{\smash{{\SetFigFont{20}{24.0}{\familydefault}{\mddefault}{\updefault}{\color[rgb]{0,0,0}$u_0$}%
}}}}
\put(5701,-2161){\makebox(0,0)[lb]{\smash{{\SetFigFont{20}{24.0}{\familydefault}{\mddefault}{\updefault}{\color[rgb]{0,0,0}$u(y)$}%
}}}}
\put(4051,-736){\makebox(0,0)[lb]{\smash{{\SetFigFont{29}{34.8}{\familydefault}{\mddefault}{\updefault}{\color[rgb]{0,0,0}$R^0_{u_0,y}$}%
}}}}
\put(5576,-3337){\makebox(0,0)[lb]{\smash{{\SetFigFont{12}{14.4}{\familydefault}{\mddefault}{\updefault}{\color[rgb]{0,0,0}$\O_{2,1}(u)$}%
}}}}
\put(5463,-4371){\makebox(0,0)[lb]{\smash{{\SetFigFont{12}{14.4}{\familydefault}{\mddefault}{\updefault}{\color[rgb]{0,0,0}$\O_{1,2}(u)$}%
}}}}
\end{picture}%

%% file: Sigma0.pdf_t
\begin{picture}(0,0)%
\includegraphics{Sigma0.pdf}%
\end{picture}%
\setlength{\unitlength}{3947sp}%
\begingroup\makeatletter\ifx\SetFigFontNFSS\undefined%
\gdef\SetFigFontNFSS#1#2#3#4#5{%
  \reset@font\fontsize{#1}{#2pt}%
  \fontfamily{#3}\fontseries{#4}\fontshape{#5}%
  \selectfont}%
\fi\endgroup%
\begin{picture}(8354,5726)(809,-5767)
\put(3243,-5058){\makebox(0,0)[lb]{\smash{{\SetFigFontNFSS{14}{16.8}{\familydefault}{\mddefault}{\updefault}{\color[rgb]{0,0,0}$u_1$}%
}}}}
\put(5626,-2611){\makebox(0,0)[lb]{\smash{{\SetFigFontNFSS{20}{24.0}{\familydefault}{\mddefault}{\updefault}{\color[rgb]{0,0,0}$\D$}%
}}}}
\put(4206,-4117){\rotatebox{55.0}{\makebox(0,0)[lb]{\smash{{\SetFigFontNFSS{14}{16.8}{\familydefault}{\mddefault}{\updefault}{\color[rgb]{0,0,0}$\lambda$}%
}}}}}
\put(2204,-2429){\makebox(0,0)[lb]{\smash{{\SetFigFontNFSS{20}{24.0}{\familydefault}{\mddefault}{\updefault}{\color[rgb]{0,0,0}$\phi$}%
}}}}
\put(2842,-2764){\makebox(0,0)[lb]{\smash{{\SetFigFontNFSS{20}{24.0}{\familydefault}{\mddefault}{\updefault}{\color[rgb]{0,0,0}$u_0$}%
}}}}
\put(5528,-2124){\makebox(0,0)[lb]{\smash{{\SetFigFontNFSS{14}{16.8}{\familydefault}{\mddefault}{\updefault}{\color[rgb]{0,0,0}$u_*$}%
}}}}
\put(3793,-4669){\makebox(0,0)[lb]{\smash{{\SetFigFontNFSS{14}{16.8}{\familydefault}{\mddefault}{\updefault}{\color[rgb]{0,0,0}$\phi_1$}%
}}}}
\put(3496,-3962){\makebox(0,0)[lb]{\smash{{\SetFigFontNFSS{17}{20.4}{\familydefault}{\mddefault}{\updefault}{\color[rgb]{0,0,0}$\Upsilon^{1}$}%
}}}}
\put(4596,-553){\makebox(0,0)[lb]{\smash{{\SetFigFontNFSS{17}{20.4}{\familydefault}{\mddefault}{\updefault}{\color[rgb]{0,0,0}$\Upsilon^2$}%
}}}}
\put(3913,-1735){\makebox(0,0)[lb]{\smash{{\SetFigFontNFSS{14}{16.8}{\familydefault}{\mddefault}{\updefault}{\color[rgb]{0,0,0}$\Omega_{1,2}(u)$}%
}}}}
\put(4188,-856){\makebox(0,0)[lb]{\smash{{\SetFigFontNFSS{14}{16.8}{\familydefault}{\mddefault}{\updefault}{\color[rgb]{0,0,0}$\O_{j,j}(u), \O_{2,1}(u)$}%
}}}}
\put(4028,-2476){\makebox(0,0)[lb]{\smash{{\SetFigFontNFSS{20}{24.0}{\familydefault}{\mddefault}{\updefault}{\color[rgb]{0,0,0}$\D^0$}%
}}}}
\end{picture}%